\documentclass[twocolumn]{aastex631}
\received{2021 September 17}
\revised{2021 November 12}
\accepted{2021 November 17}
\published{YYYY MM DD}
\submitjournal{ApJ}
\shorttitle{Oscillatory reconnection in a hot coronal plasma}
\shortauthors{Karampelas et al.}
\graphicspath{{./}{figures/}}
\begin{document}

\title{Oscillatory Reconnection of a 2D X-point in a hot coronal plasma}
% title to include thermal conduction?

\correspondingauthor{Konstantinos Karampelas}
\email{konstantinos.karampelas@northumbria.ac.uk}

\author[0000-0001-5507-1891]{Konstantinos Karampelas}
\affiliation{Department of Mathematics, Physics and Electrical Engineering, Northumbria University,\\ Newcastle upon Tyne, NE1 8ST, UK}
\affiliation{Centre for mathematical Plasma Astrophysics, Department of Mathematics, KU Leuven,\\ Celestijnenlaan 200B bus 2400, B-3001 Leuven, Belgium }
\author[0000-0002-7863-624X]{James A. McLaughlin}
\affiliation{Department of Mathematics, Physics and Electrical Engineering, Northumbria University,\\ Newcastle upon Tyne, NE1 8ST, UK}
\author[0000-0002-5915-697X]{Gert J. J. Botha}
\affiliation{Department of Mathematics, Physics and Electrical Engineering, Northumbria University,\\ Newcastle upon Tyne, NE1 8ST, UK}
\author[0000-0001-8954-4183]{St\'{e}phane R\'{e}gnier}
\affiliation{Department of Mathematics, Physics and Electrical Engineering, Northumbria University,\\ Newcastle upon Tyne, NE1 8ST, UK}

\begin{abstract}
Oscillatory reconnection (a relaxation mechanism with periodic changes in connectivity) has been proposed as a potential physical mechanism underpinning several periodic phenomena in the solar atmosphere including, but not limited to, quasi-periodic pulsations (QPPs). Despite its importance, however, the mechanism has never been studied within a hot, coronal plasma. We investigate oscillatory reconnection in a one million Kelvin plasma by solving the  fully-compressive, resistive  MHD equations for a 2D magnetic X-point under coronal conditions using the PLUTO code. We report on the resulting oscillatory reconnection including its periodicity and decay rate. We observe a more complicated oscillating profile for the current density compared to that found for a cold plasma, due to  mode-conversion at the equipartition layer. We also consider, for the first time, the effect of adding anisotropic thermal conduction to the oscillatory reconnection mechanism, and we find this simplifies the spectrum of the oscillation profile and increases the decay rate. Crucially, the addition of thermal conduction does not prevent the oscillatory reconnection mechanism from manifesting. Finally, we reveal a relationship between the equilibrium magnetic field strength, decay rate, and period of oscillatory reconnection, which opens the tantalising possibility of utilizing oscillatory reconnection as a seismological tool.
\end{abstract}


%%%%%%%%%%%%%%%%%%%%%%%%%%%%%%%%%%%%%%%%%%%%%%%%%%%%

\section{Introduction} \label{sec:introduction}
Null points are magnetic field singularities at which the magnetic field strength and Alfv\'{e}n speed are zero. Although the coronal magnetic field is hard to measure directly \citep{Lin2004ApJ,Gibson2016FrASS}, potential field extrapolations from photospheric magnetograms predict that null points are omnipresent in the solar atmosphere \citep{Galsgaard1997,BrownPriest2001AnA,Longcope2005LRSP,Regnier2008AnA}. The presence of null points in the solar atmosphere drastically changes the behavior of waves and flows around them (e.g. \citealt{Gruszecki2011null,McLaughlin2011SSRv,Santamaria2015,Sabri2018MNRAS,Sabri2019AnA}). In \citet{McLaughlin2004,McLaughlin2005,McLaughlin2006a,McLaughlin2006b}, the phenomenon of wave refraction around various 2D null points was established for both cold ($\beta=0$) plasma and finite-$\beta$ plasma. Those studies  also found that, for finite-$\beta$ plasma, a low-$\beta$ fast wave can generate high-$\beta$ fast and slow waves through mode conversion when crossing the equipartition layer, i.e. the layer where the Alfv\'{e}n speed equals the sound speed. \cite{Thurgood2012,  2013A&A...558A.127T} extended these results to  3D null points.

In addition, simulations involving 2D null points have shown the development of high-frequency wave trains ($\sim 80$\,mHz) in their vicinity \citep{Santamaria2016,Santamaria2017}, while in \citet{Santamaria2018} it was shown that a null point behaves as a resonant cavity that can generate waves at specific frequencies depending on the background equilibrium (the latter is important in the context of coronal seismology; \citealt{Uchida1970, RobertsEdwinBenz1984}). Waves have also been generated via reconnection at 2D null points (e.g. \citealt{2008ApJ...683L..83N, 2009ApJ...705L.217H}) and compared directly with observations. The interaction of fast magnetoacoustic waves with a $2.5$D null point has been shown to produce plasmoids due to the tearing mode instability in a numerical study \citep{Sabri2020ApJ}. Numerical simulations have shown that 3D null points to act as sources of Alfv\'en waves (e.g. \citealt{2014SoPh..289.3043L, 2018ApJ...862....6C}), coronal jets in the form of propagating nonlinear Alfv\'en waves \citep{2017ApJ...834...62K} as well as other sources of fast and slow (and Alfv\'en) waves \citep{Thurgood2017ApJ}, when these 3D null points are subjected to various wave-based driving motions.

Apart from their interaction with the ubiquitous waves in the solar atmosphere \citep[e.g.][]{Nakariakov2005LRSP,depontieu2007,okamoto2007,tomczyk2007,mcintosh2011}, null points also play a key role in highly energetic phenomena such as solar flares (e.g. \citealt{ShibataMagara2011LRSP}). One way to achieve this is by dissipating the currents that are being accumulated at the null point, which leads to plasma heating, anomalous resistivity (via micro-instabilities), and, ultimately, reconnection \citep{2006A&A...452..343N}.

Reconnection occurs when strong currents allow neighboring magnetic field lines to diffuse, leading to a change in connectivity (e.g. \citealt{Parker1957JGR,Sweet1958IAUS,Petschek1964NASSP}) and is expected to occur at null points \citep{PriestForbes2000book}. While considering the relaxation of a 2D X-point that had been disturbed from equilibrium, \citet{CraigMcClymont1991ApJ}  identified oscillatory reconnection as a means of dissipating magnetic energy. As the name suggests, oscillatory reconnection describes a series of reconnection events associated with periodic changes in the  connectivity of the magnetic field. A key characteristic of oscillatory reconnection is that its periodicity is not determined by a periodic external force, rather the periodicity manifests from a relaxation mechanism following from a finite, aperiodic initial perturbation (similar to a decaying harmonic oscillation).

Oscillatory reconnection was further studied in a 2D X-point for finite-$\beta$ and nonlinear effects in \citet{McLaughlin2009}. In that study, an initially cold ($\beta=0$) plasma setup was considered, but the solution included the full compressible resistive MHD equations and thus allowed for plasma heating to take place. This mechanism has also been studied for a 3D null point in \citet{Thurgood2017ApJ}, where the by-product of oscillatory reconnection was the generation of freely propagating MHD waves, escaping the vicinity of the null point. Additional studies have focused on the periodicity of oscillatory reconnection from linear and nonlinear perturbations, and how it is affected by resistivity and the amplitude of the initial perturbation \citep{McLaughlin2012A&A, 2018ApJ...855...50T,
2018PhPl...25g2105T, Thurgood2019A&A}.

As a mechanism, oscillatory reconnection is an appealing explanation of quasi-periodic pulsations (QPPs), which have now been detected in a multitude of solar  flares (e.g. \citealt{Kupriyanova2016SoPh,VanDoorsselaere2016SoPh,Pugh2017AnA,2019ApJ...886L..25Y,2020ApJ...895...50H, 2020A&A...639L...5L,2020ApJ...893....7L,2021ApJ...921..179L, 2021ApJ...910..123C}) as well as detected in a variety of stellar flares (e.g. \citealt{2019A&A...629A.147B,2019A&A...622A.210G,2019MNRAS.482.5553J,2019ApJ...876...58N,2019ApJ...884..160V,2020A&A...636A..96M,2021arXiv210810670R}). QPPs are now understood to be a central feature of the flaring process, and thus understanding the true underpinning mechanism for QPPs holds the key to understanding solar and stellar flares. Detailed reviews of the possible  mechanism(s)  underpinning QPPs can be found in \citet{McLaughlin2018SSRv}, \citet{Kupriyanova2020STP} and \citet{Zimovets2021SSRv}.

Oscillatory reconnection is also a possible mechanism behind phenomena like  periodicities observed in jets (\citealt{2019ApJ...874..146H}), periodic behavior associated with the formation, disappearance, and eruption of magnetic flux ropes (\citealt{2018ApJ...853....1S,2019ApJ...874L..27X}) and quasi-periodic flows associated with spicules in the solar atmosphere (e.g. \citealt{DePontieu2010ApJ,DePontieu2011Sci,2019Sci...366..890S, 2020ApJ...891L..21Y}).  For spicules, a 2D flux emergence model that triggered the oscillatory reconnection mechanism  was able to reproduce such flows with the observed periodicities \citep{McLaughlin2012ApJ}. Oscillatory reconnection may also be a possible explanation of phenomena now being observed by the Parker Solar Probe (e.g. \citealt{2016SSRv..204...49B, 2019Natur.576..237B,2019Natur.576..228K}) including Alfv\'enic spikes/kinks  \citep{2021ApJ...913L..14H} and periodicities correlated with Type III radio bursts \citep{2021A&A...650A...6C}.

Even though oscillatory reconnection has been used within the context of a realistic stratified solar atmosphere (\citealt{2009A&A...494..329M,McLaughlin2012ApJ}) so far there has been no thorough study of the mechanism in hot, coronal plasma. In this paper, we will take that next step by investigating  the oscillatory reconnection mechanism within a hot plasma under coronal conditions. After describing the details of the numerical setups used in our studies (\S\ref{sec:setup}), we will present our results on the initiation and evolution of oscillatory reconnection for hot plasma (\S\ref{sec:OR in 1MK plasma}) while providing a direct comparison to the cold plasma case. We will address the effects of anisotropic thermal conduction (\S\ref{sec:thermal-conduction}) which is a necessary step toward a more realistic coronal environment, and we will present the results of our parameter study regarding the magnetic field strength and its effects on the periodicity and decay rate of oscillatory reconnection (\S\ref{sec:parameter-study}). The conclusions are presented in \S\ref{sec:discussions} and the Appendix \ref{Appendix} reports on a parameter study investigating the temperature dependence.

%%%%%%%%%%%%%%%%%%%%%%%%%%%%%%%%%%%%%%%%%%%%%%%%%%%%%%%%%%

\section{Numerical setup} \label{sec:setup}

\begin{figure}[t]
    \centering
    \includegraphics[trim={0.cm 0.cm 0.cm 0.cm},clip,scale=0.5]{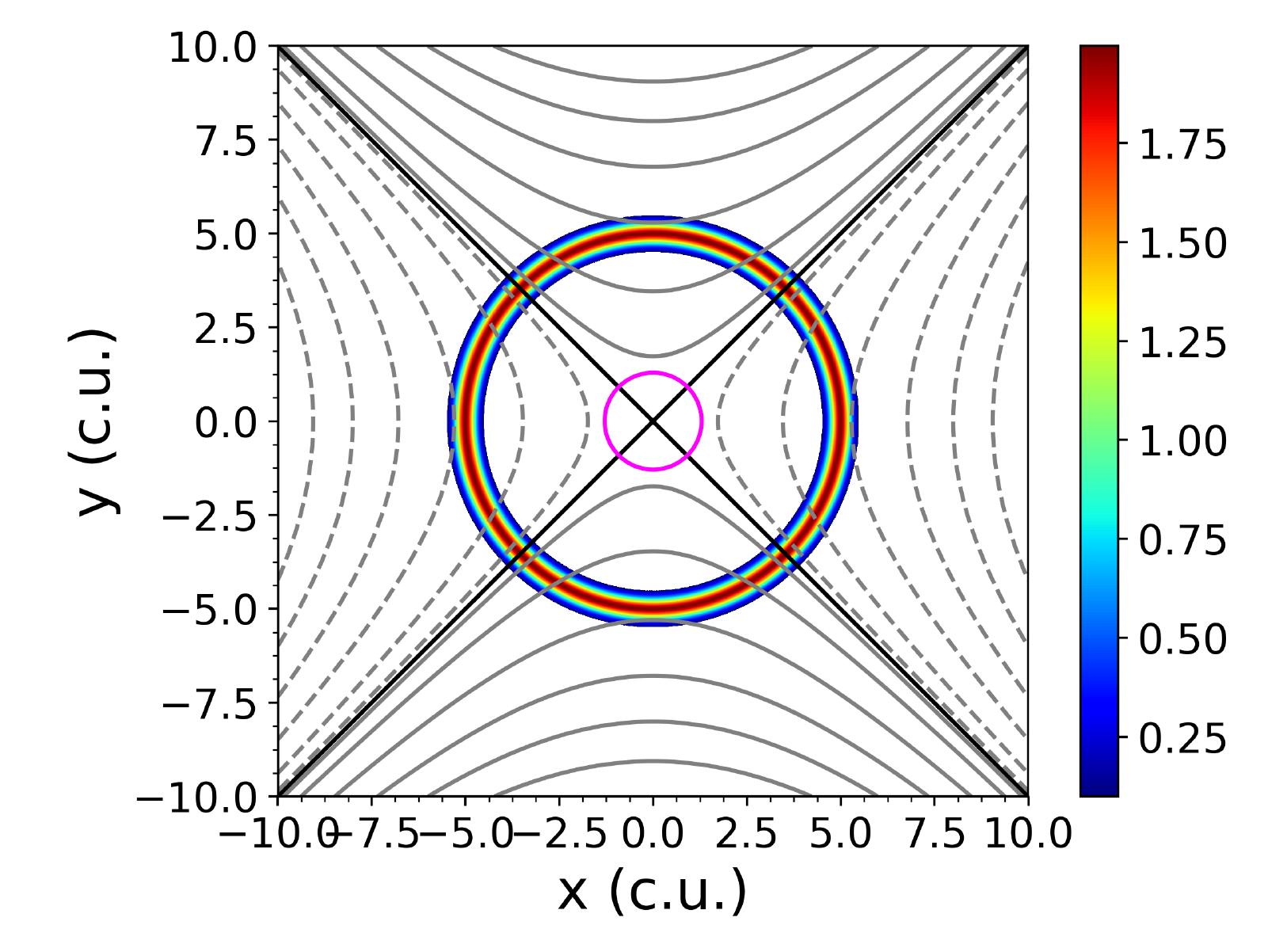}
    \caption{Initial conditions for the X-point setup. The magnetic field is shown with gray solid and dashed lines to highlight the different regions as separated by the separatrices (solid black lines). The initial $v_{\perp}$ is shown here, cropped at its lower values for an easier visualisation. The equipartition layer (magenta line) where $V_A = V_S$ is also shown here.}
    \label{fig:profile}
\end{figure}

\subsection{Numerical scheme}
Our setup closely follows that of \citet{McLaughlin2009}. We solve the 2D compressible resistive MHD equations, also in the presence of numerical resistivity, using the PLUTO code \citep{mignonePLUTO2007, mignonePLUTO2012}, a finite-volume, shock-capturing code, which uses double precision arithmetic for the computations. We employ the third-order Runge-Kutta method to calculate the time step, the fifth-order monotonicity preserving scheme (MP5) for the spatial integration, and the total variation diminishing Lax–Friedrich (TVDLF) solver. To keep the solenoidal constraint on the magnetic field, we employ the Constrained Transport method.

The 2D compressible and resistive MHD equations, in the absence of gravity, written for the primitive variables are:
\begin{equation}
  \frac{\partial\rho}{\partial t} + \nabla \cdot (\rho \mathbf v) = 0\,,
\end{equation}
\begin{equation}
 \rho \left[ \frac{\partial\mathbf v }{\partial t} + \rho(\mathbf v \cdot \nabla)\mathbf v \right] = - \nabla p + \frac{1}{\mu}\left(\nabla\times\mathbf{\mathbf B}\right) \times \mathbf B,
\end{equation}
\begin{equation}
    \rho \left[ \frac{\partial \epsilon}{\partial t} + (\mathbf v \cdot \nabla)\epsilon \right] = -p \nabla \cdot \mathbf v + \frac{1}{\sigma} |\mathbf J|^2 + \Lambda,
\end{equation}
\begin{equation}
  \frac{\partial\mathbf B}{\partial t} = \nabla \times (\mathbf v \times \mathbf B) + \eta\nabla^2 \mathbf B
\end{equation}
where the quantities $\rho$, $p$, and $\mathbf v$ are the density, plasma pressure, and velocity, respectively. The magnetic field $\mathbf B$ satisfies the condition $\nabla\cdot\mathbf B = 0$ and the electric current is defined as $\mathbf J = \nabla\times\mathbf B /\mu$. In addition, $\mu = 4 \pi \times 10^{-7}$\,H m$^{-1}$ is the magnetic permeability, $\sigma$ is the electrical conductivity, $\eta = 1/\mu \sigma$ is the magnetic diffusivity, and $\epsilon = p/\left[\rho (\gamma-1)\right]$ is the specific internal energy density. Finally, $\gamma = 5/3$ is the ratio of the specific heats, and $\Lambda$ represents the volumetric energy gain/loss terms, such as anisotropic thermal conduction. 

We use Cartesian coordinates $(x,y)$ and  express all quantities in code units. A code variable $U_c$ is converted to its physical units $U$ with the help of the normalization unit $U_0$, i.e. $U = U_0U_c$. The constants $U_0$ are specified such that the values of the quantities match the conditions in the solar corona. Specifically, $x_0 = y_0 = L_0 = 10^{6}$\,m is the unit length, $\rho_0 = 10^{-12}$\,kg m$^{-3}$ the unit density, $v_0 \sim 129 \times 10^3$\,m s$^{-1}$ the unit velocity (equal to $c_s / \sqrt{\gamma}$, where $c_s$ the sound speed of coronal plasma at 1 MK), $p_0 = \rho_0 v_0^2$ the unit pressure, $T_0 = 10^6$\,K the unit temperature, $B_0 = \sqrt{\mu p_0} = 1.44$\,G the unit magnetic field, and $t_0 = L_0 / v_0 = 7.78$\,s the unit time. We also let $\nabla = \nabla_c / L_0$. Finally, the magnetic diffusivity takes values $\eta = \eta_0 \, \eta_c = v_0\,L_0 \, R_m^{-1}$, where $R_m$ is the magnetic Reynolds number, assuming that $v_0$ and $L_0$ are the typical velocity and length scales of the system, respectively. In this setup, we take $R_m = 10^5$ and thus $\eta_c = 10^{-5}$. Alongside the explicit magnetic diffusivity, we also have inevitable effects of the `effective' numerical diffusivity present in our code. Through a parameter study, we have estimated the effective numerical diffusion of our lower-resolution setups used in this study to be between $10^{-6}$ and $10^{-5}$ in code units. The effective numerical diffusion of the higher resolution setups is thus lower than this value, since it decreases with an increasing resolution. This numerical dissipation is many orders of magnitude higher than the values for the diffusivity expected in the solar corona, which is why we chose such a relatively small value for $R_m$. We also need to stress that the small value for $R_m$ makes it difficult to make accurate quantitative estimations for the solar corona, but it will still allow us to get qualitative results for the topic under study.

\begin{figure*}[t]
    \centering
    \resizebox{\hsize}{!}{
    \includegraphics[trim={0.cm 0.cm 2.8cm 0.cm},clip,scale=0.5]{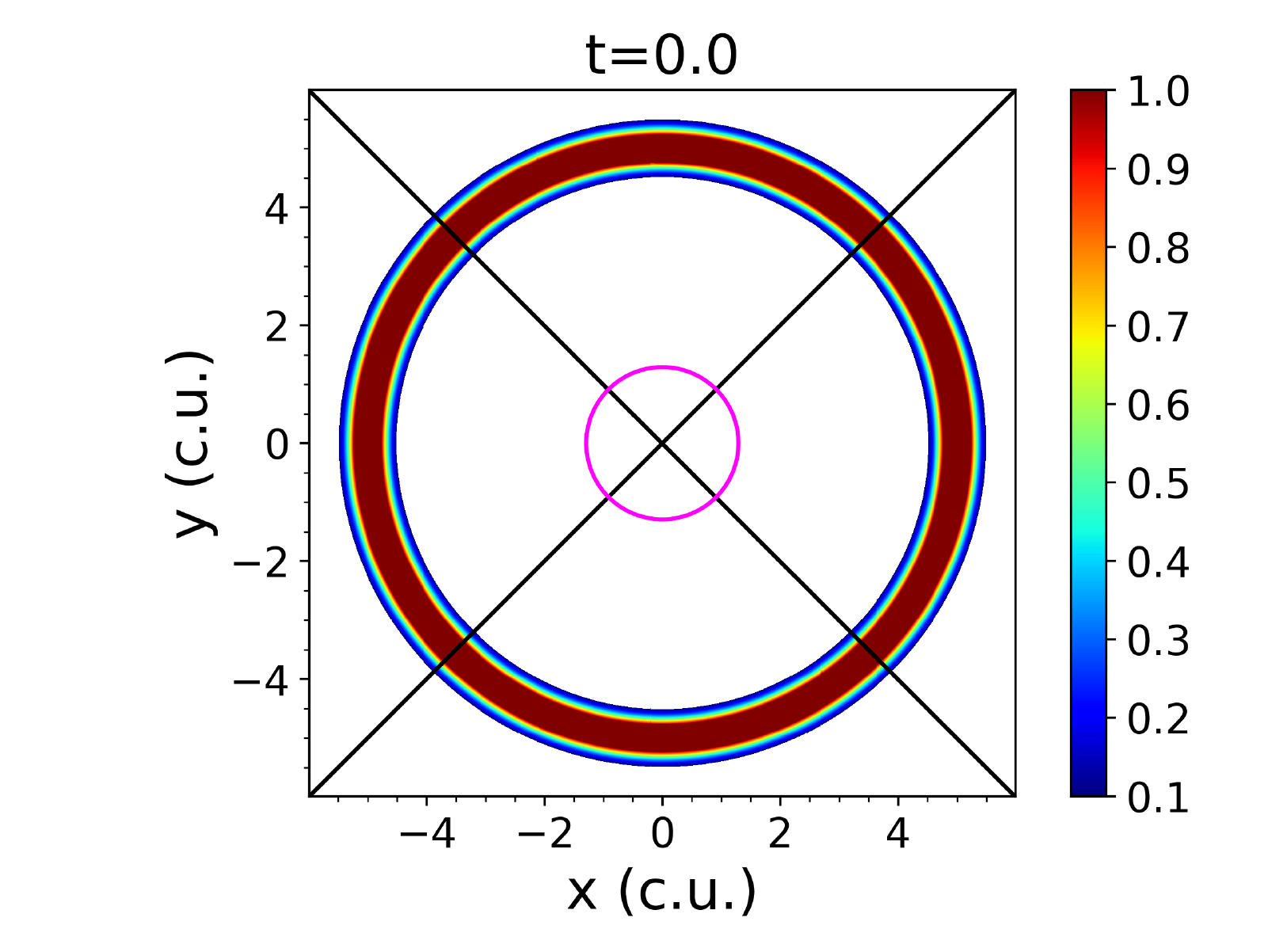}
    \includegraphics[trim={3.8cm 0.cm 2.8cm 0.cm},clip,scale=0.5]{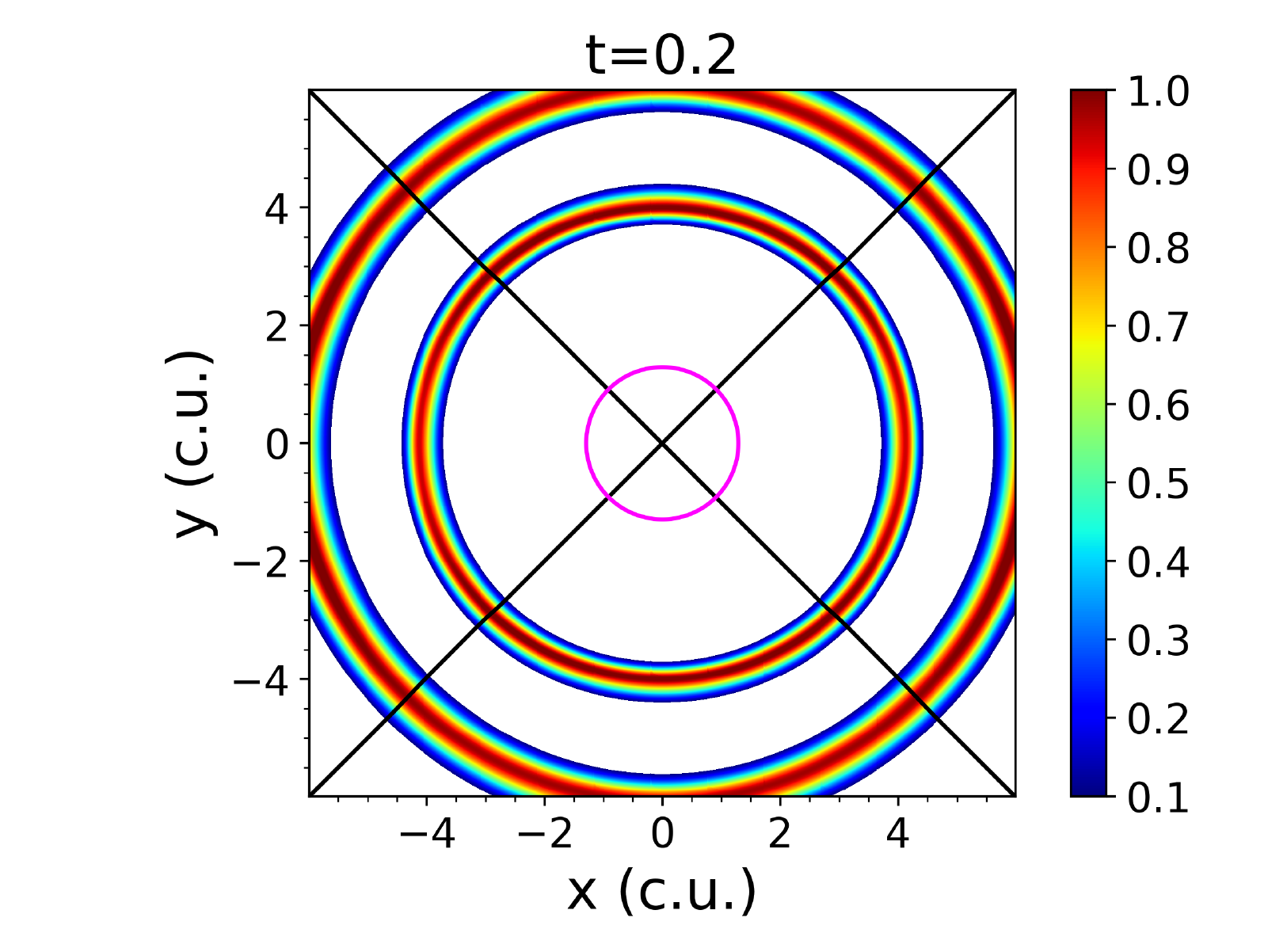}
    \includegraphics[trim={3.8cm 0.cm 2.8cm 0.cm},clip,scale=0.5]{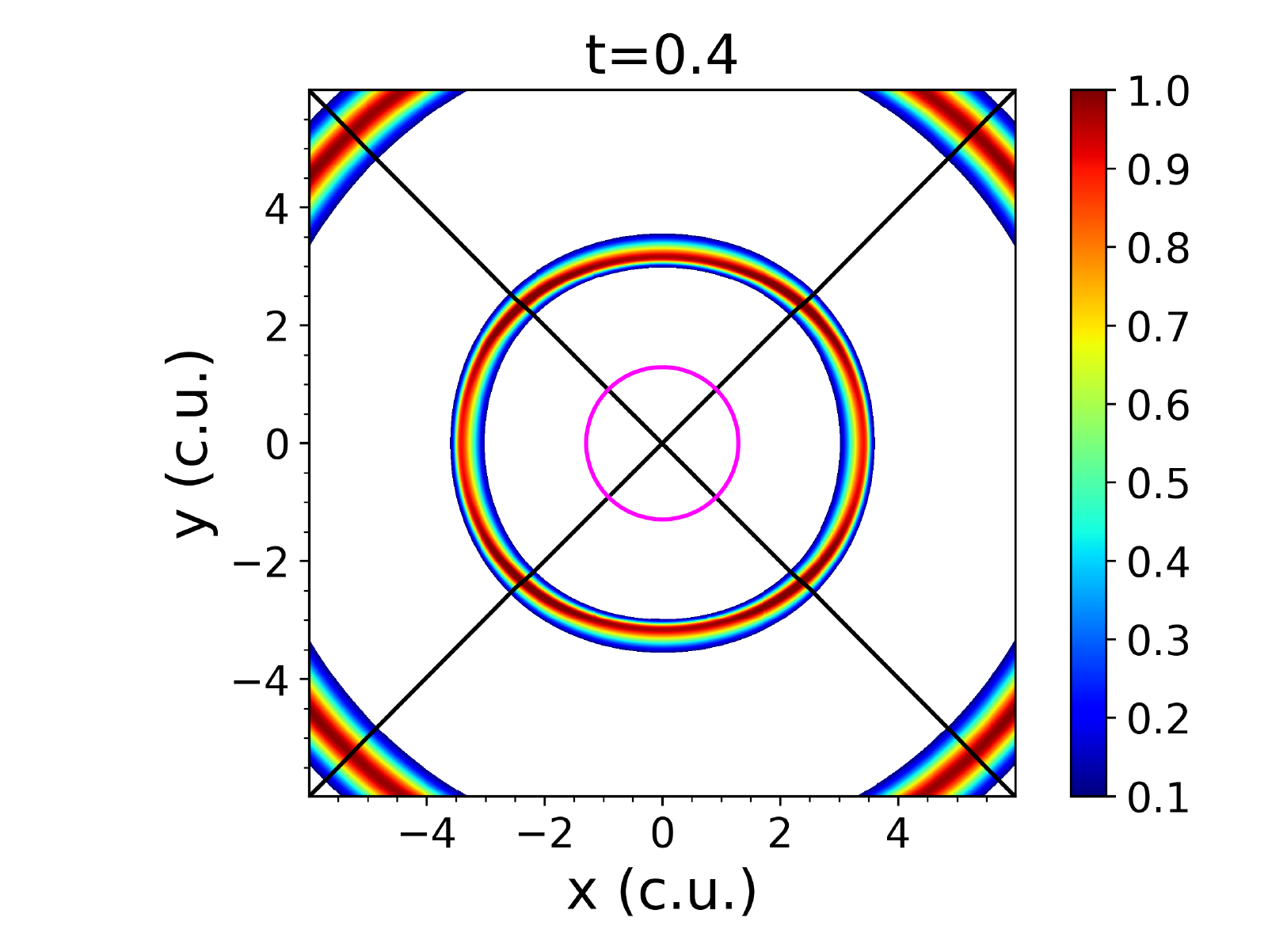}
    \includegraphics[trim={3.8cm 0.cm 0.cm 0.cm},clip,scale=0.5]{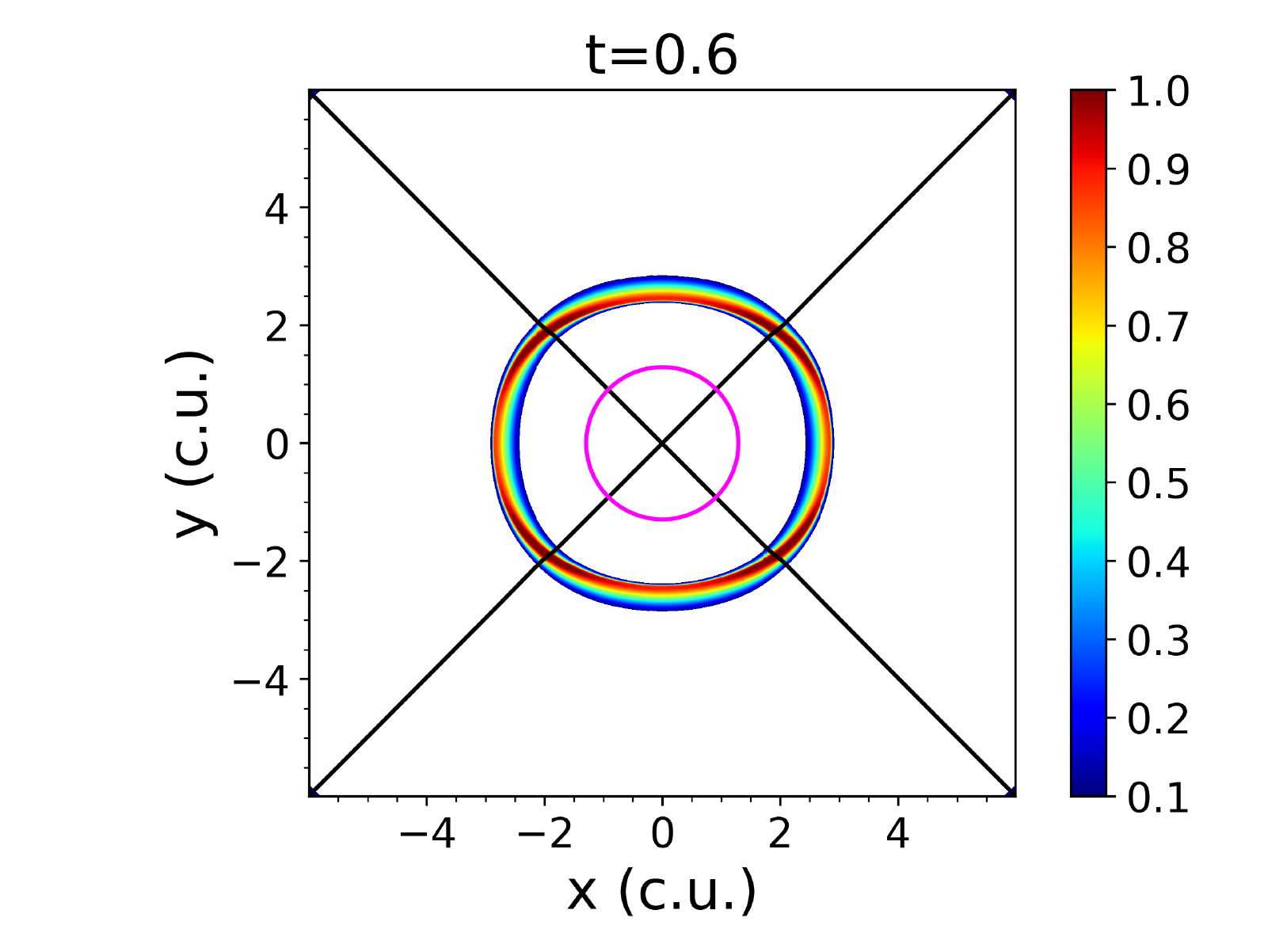}}
    \caption{Zoomed in profiles of the $v_{\perp}$ for the initial states of the simulation. The split of the initial pulse in an inward and outward traveling pulse is shown. The magnetic field separatrices (black solid lines) and the equipartition layer (magenta solid line) are included.}
    \label{fig:initial}
\end{figure*}

When switching on anisotropic thermal conduction (TC) in the PLUTO code, we use values of the parallel and perpendicular coefficients that are derived fro the Spitzer conductivity \citep{Orlando2008ApJ}. The values for the parallel and perpendicular coefficients (in J\,s$^{-1}$\,K$^{-1}$\,m$^{-1}$) are the following:
\begin{eqnarray}
\kappa_{\parallel} &=& 5.6 \times 10^{-12}\, T^{\frac{5}{2}},\label{eq:kpar}\\
\kappa_{\perp} &=& 3.3 \times 10^{-21}\,  T^{\frac{3}{2}} \frac{n_H^2}{\sqrt{T}B^2},\label{eq:kprp}
\end{eqnarray}
where $n_H$ is the hydrogen number density, and $T$ and $B$ are in physical units.

Finally, we note that for the rest of the paper, we will write the quantities for the code variables as $U$ instead of $U_c$ and show them in code units, unless otherwise stated.

%%%%%%%%%%%%%%%%%%%%%%%%%%%%%%%%%%%%%%%%%%%%%%%%%%%%%%%%%%%%%%%%%%%%%%%%

\subsection{Initial and boundary conditions}\label{sec:Initial and boundary conditions}
Just like in \citet{McLaughlin2009}, our setup consists of a square domain containing a 2D magnetic X-point. Our domain consists of a structured uniform grid that extends to $(x,y)\in (-10,10)$\, code units, with a resolution of $2401\times 2401$ grid points (and $1801\times 1801$ grid points for the parameter studies in \S\ref{sec:results} and in the Appendix \ref{Appendix}). 

We consider a uniform density equal to $\rho_0 = 10^{-12}$\,kg m$^{-3}$ and a temperature that is equal to $T_0 = 1$\,MK (with the exception of the parameter study; see the Appendix \ref{Appendix}). We take the equilibrium density to be uniform, as a spatial variation in density can cause phase mixing (e.g. \citealt{heyvaerts1983}). This creates a uniform initial sound speed $V_S$.
 
The initial magnetic field is given by:
\begin{equation}\label{eq:magneticfield}
    \mathbf{B} = \frac{B_0}{L_0}\left(y,x,0\right),
\end{equation}
where $B_0$ is the characteristic field strength (taken equal to the unit magnetic field), and $L_0$ is the length scale for magnetic field variations (taken here as the unit length). This magnetic field takes its highest values at the boundaries and goes to zero at the X-point. The Alfv\'{e}n speed $V_A$ increases the farther we go away from the X-point (the origin in our simulations). The contour where $V_A = V_S$ defines the equipartition layer, and this is initially a circle (it will deform as the system evolves). In most parts of the solar corona, the plasma-$\beta$ is much less than unity but near null points the plasma-$\beta$ becomes large (since the magnetic field strength is itself becoming small). Thus, the equipartition layer defines a system where outside is a low-$\beta$ environment and inside is a high-$\beta$ environment.\footnote{Note that the key parameter here is the location of the equipartition layer as opposed to the $\beta=1$ layer, since the coupling between fast and slow waves is most efficient where $V_A=V_S$. For interest, the radius of the $\beta=1$ layer occurs at a slightly greater radius (specifically a radius $\sqrt{\frac{2}{\gamma}}\approx 1.095$ greater than the equipartition radius).}

\begin{figure*}[t]
    \centering
    \resizebox{\hsize}{!}{
    \includegraphics[trim={0.cm 0.cm 3.cm 0.cm},clip,scale=0.5]{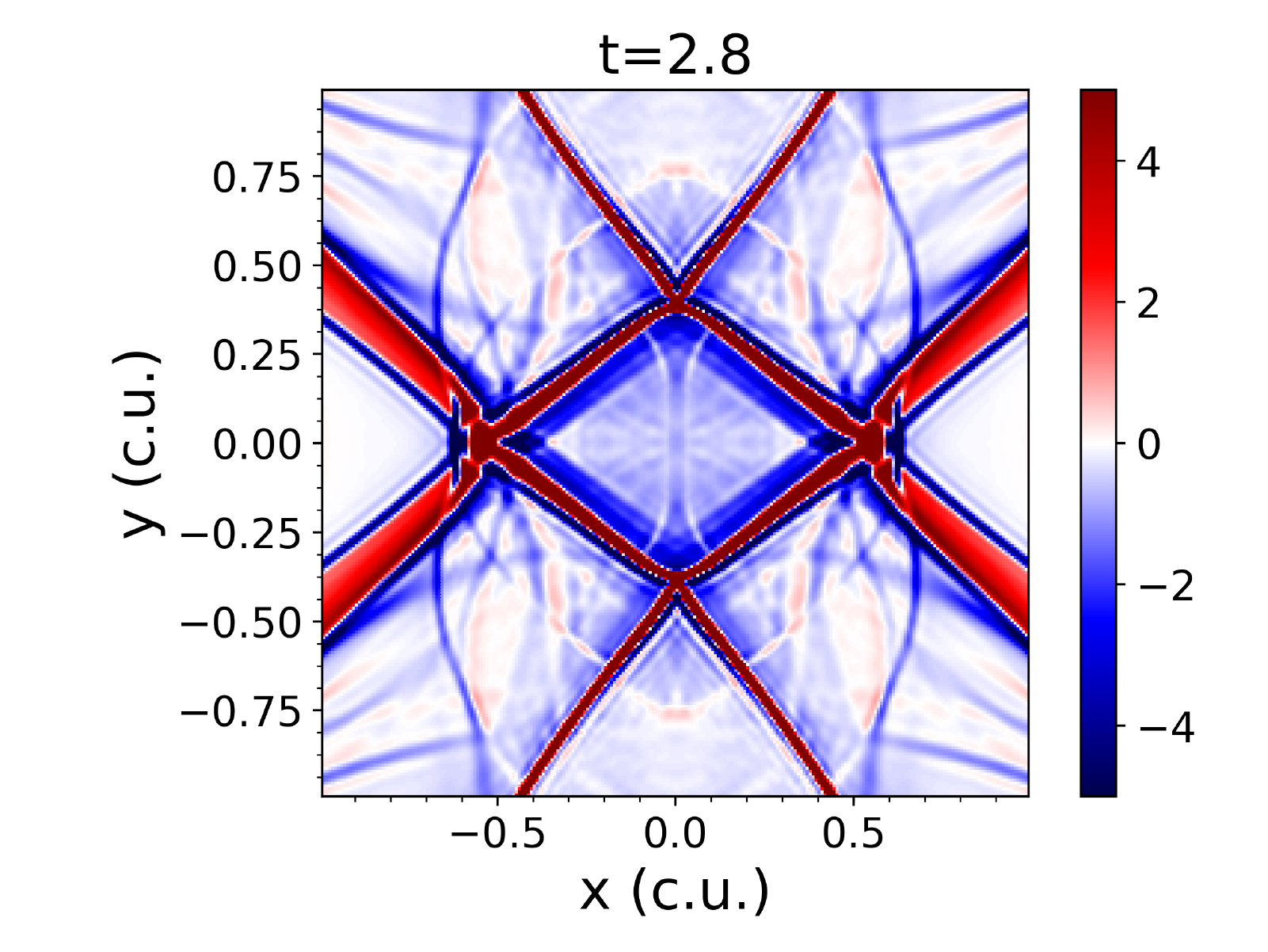}
    \includegraphics[trim={4.cm 0.cm 3.cm 0.cm},clip,scale=0.5]{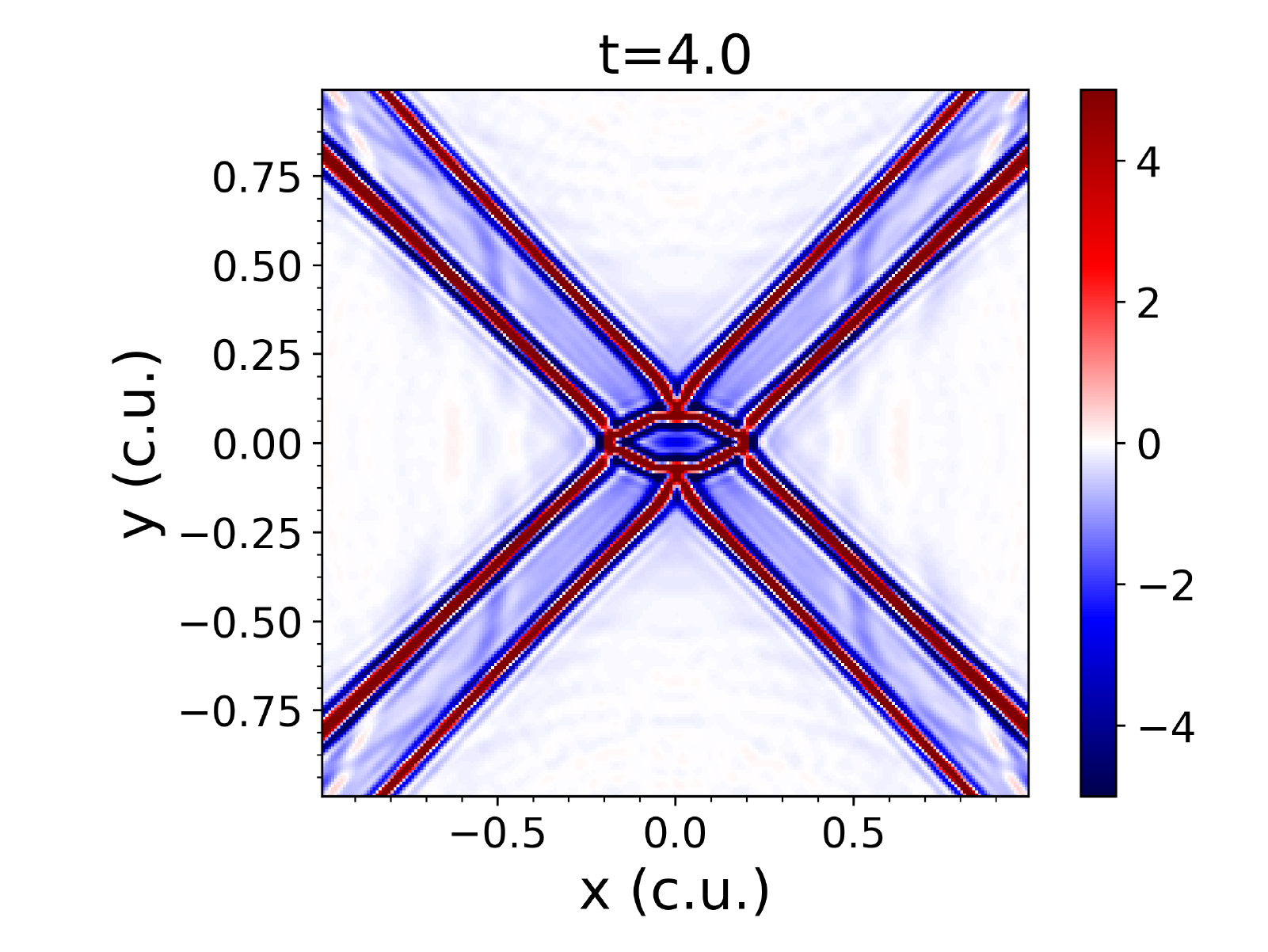}
    \includegraphics[trim={4.cm 0.cm 3.cm 0.cm},clip,scale=0.5]{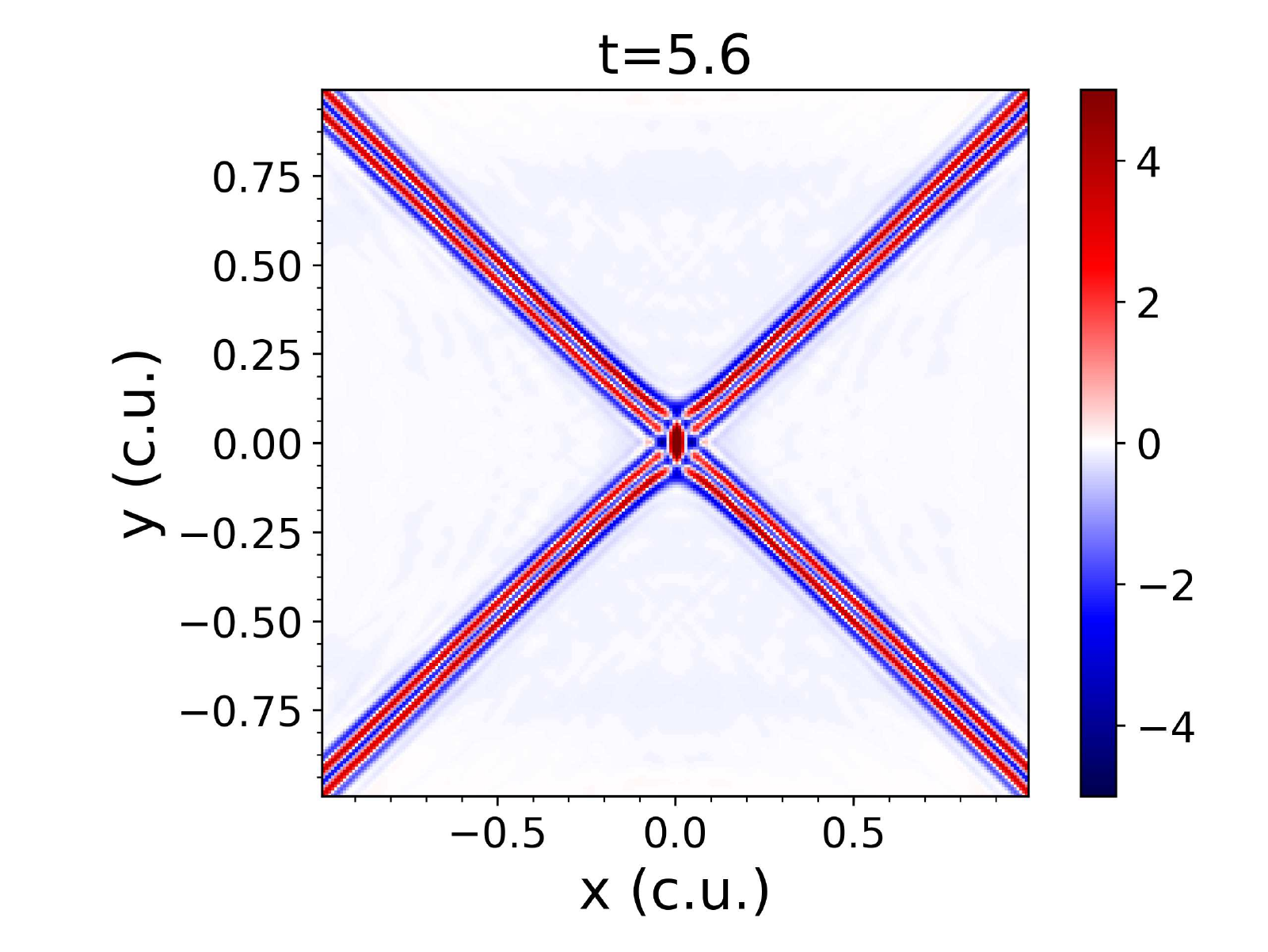}
    \includegraphics[trim={4.cm 0.cm 3.cm 0.cm},clip,scale=0.5]{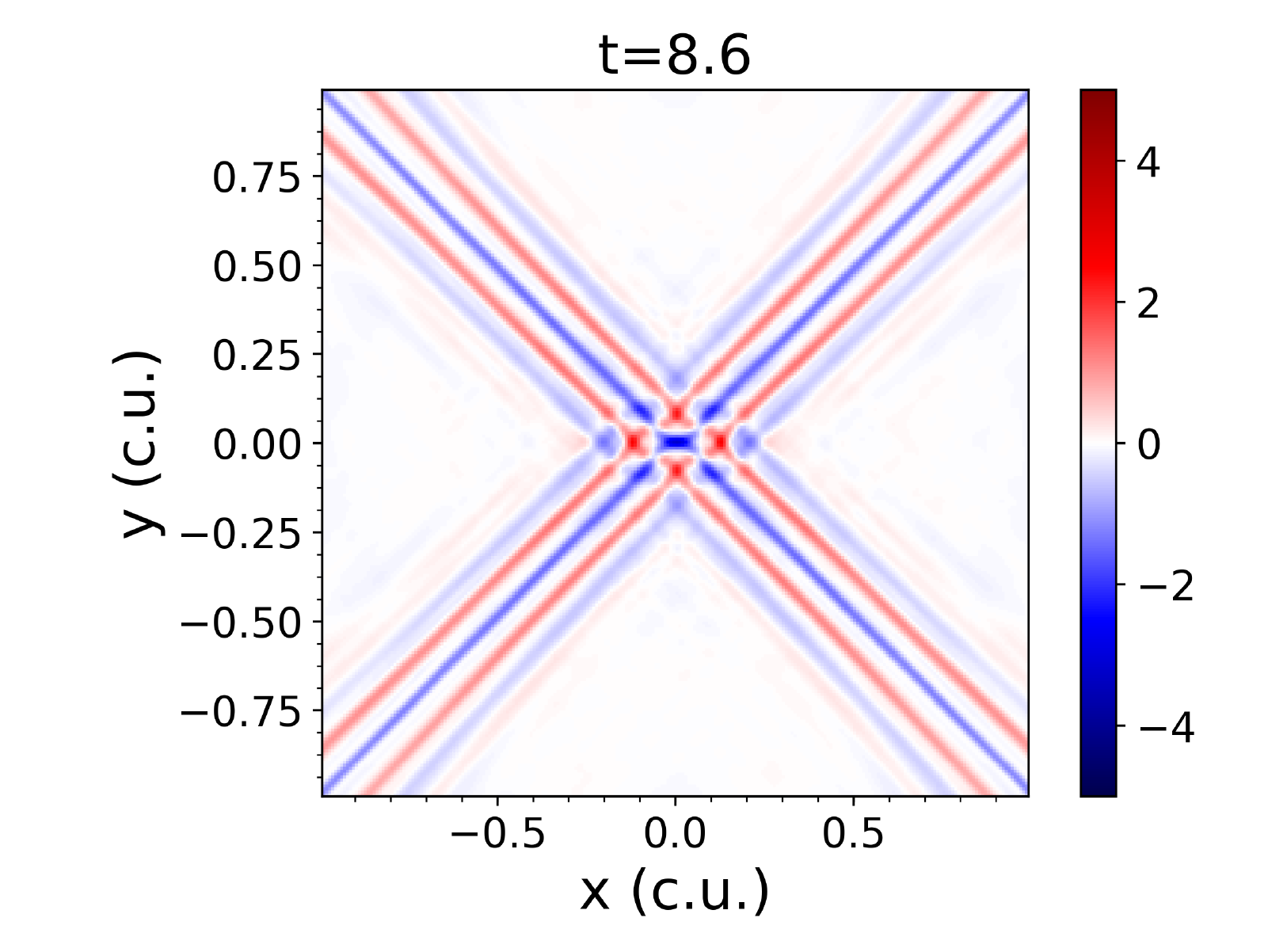}
    \includegraphics[trim={4.cm 0.cm 0.cm 0.cm},clip,scale=0.5]{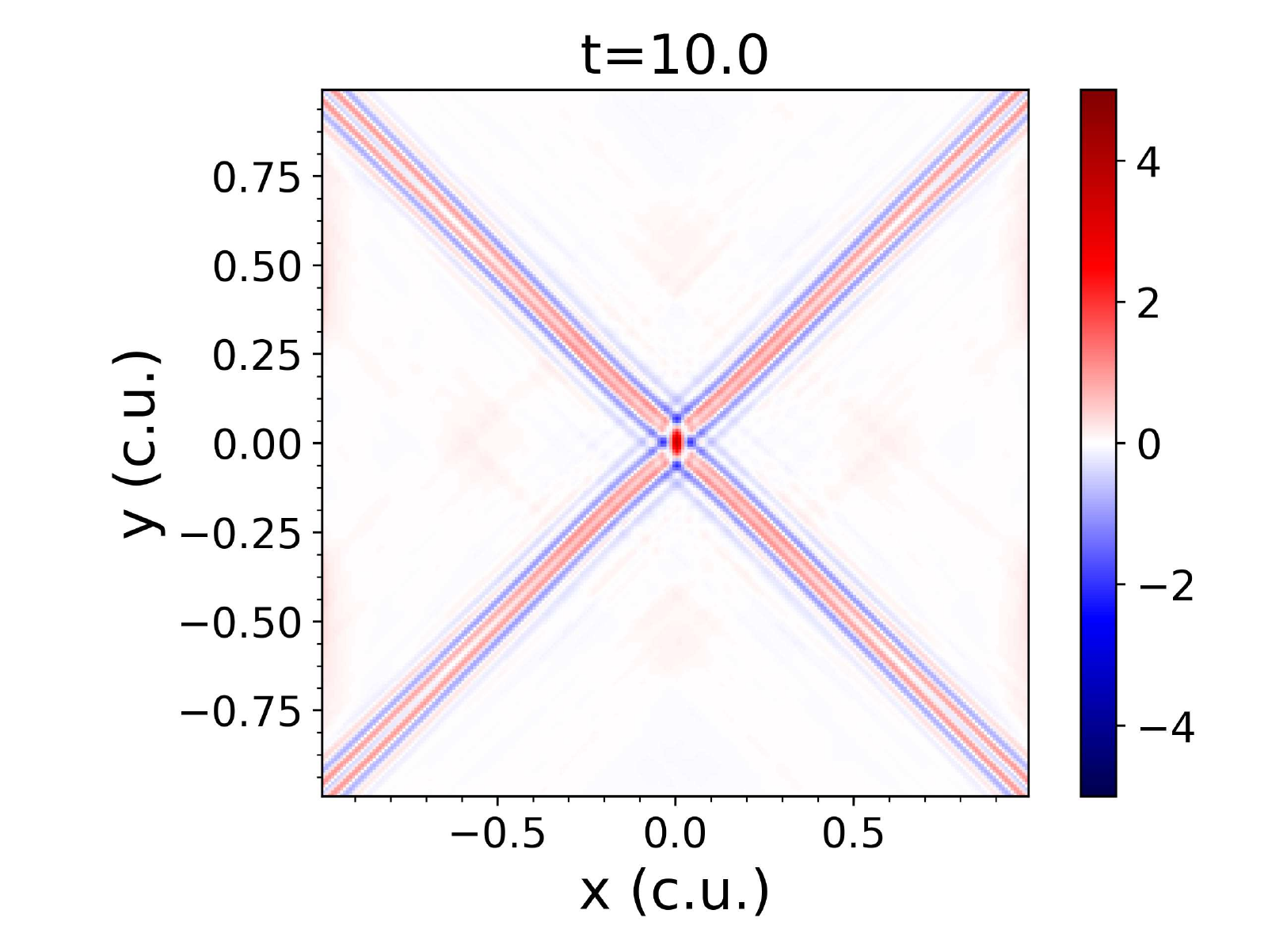}}
    \caption{Zoomed in profile of the $J_z$ current density. Shown here are the horizontal and vertical profiles associated with negative and positive values for the current density at the initial stages of the simulation.} \label{fig:highresjz}
\end{figure*}

Following \citet{McLaughlin2009}, in this paper, we consider an initial condition in velocity, such that:
\begin{equation}\label{eq:vp}
v_{\perp} = \frac{C}{0.2 \sqrt{2\pi}}\exp\left[ -0.5 \frac{(r-5)^2}{0.2}\right],
\end{equation}
\begin{equation}\label{eq:vl}
v_{\parallel}(t=0)=0,
\end{equation}
where $v_{\perp} = (\mathbf{v}\times \mathbf{B})\cdot\hat{\mathbf{z}}$ is related to the velocity perpendicular to the magnetic field lines, and $v_{\parallel}=\mathbf{v}\cdot\mathbf{B}$ is  related to the velocity parallel to the magnetic field lines. We will consider $C=1$, unless otherwise stated, which brings us to the nonlinear regime of the wave--null point interaction, similarly as in \citet{McLaughlin2009}. The magnetic field configuration and the initial pulse are shown in Fig. \ref{fig:profile}. The initial conditions for the velocity describe a circular velocity pulse in $v_{\perp}$, which crosses the magnetic field lines as it propagates, and can thus be identified as an (initially) fast magnetoacoustic wave. At the start of the simulation, the initial velocity pulse splits into two counterpropagating waves, each with half the velocity amplitude of the initial pulse. As shown in Fig. \ref{fig:initial}, the first wave is traveling toward the null point and is the one we will be focusing on for the rest of this study, while the second wave is traveling away from the null point and toward the edges of our domain. 

We take what are effectively reflective boundaries for the velocity across the boundaries, by fixing the velocity components to zero there, and use constant gradient boundary conditions for the magnetic field (of the form $d_i - d_{i-1} = d_{i-1} - d_{i-2}$). This ensures that, for a Cartesian grid, no currents are developed artificially at the boundaries. Finally, we fix the values of the pressure and density at the boundaries to their initial conditions. This allows for heat to freely leave the domain and not accumulate at the boundaries in the case where thermal conduction is used.

\subsection{Additional numerical considerations: Damping regions and the solenoidal constraint}\label{sec:damping region}

Our numerical domain covers $(x,y)\in (-10,10)$\, code units, but our primary interest is in a much smaller region around the null point itself. We  employ a few different numerical strategies to ensure that our reflective boundaries will not result in reflected/returning waves contaminating the solution near the X-point. First, in order to deal with the outgoing pulse generated by our initial conditions, we reset the velocity after a radius of $r\geqslant 7$ at $t=0.6$ (both in code units), removing the initial outward propagating velocity pulse. 

Once this initial perturbation is removed, we then initiate a numerical dissipation scheme for the velocity, for a radius of $r\geqslant 6$ and time $t>0.6$ of the form $v_i = v_i/n_d$. This creates a damping region that removes kinetic energy from the outgoing waves and their reflections from the boundaries, per iteration, before they can reach the X-point and interfere with the solution. The dissipation coefficient $n_d = 1 + 0.007\,\tanh(r - 6)^2$ regulates how fast the weakening will occur.

In addition to the numerical dissipation scheme, we introduce viscosity into our system, with coefficient $n_{visc}= 0.5\tanh(r - 6)^2$, again in code units, for a radius of $r\geqslant 6$ and time $t>0.6$. Both the viscous layer and the numerical damping scheme reduce the contamination of the area of interest from reflected waves.

Finally, we need to point out the importance of keeping the solenoidal constraint. The area of interest in our setups is in the vicinity of the X-point. There, by its definition, the magnetic field goes to zero. Because of that, the numerical solution can be very sensitive to any type of errors associated with a poor treatment of the emerging $\nabla \cdot \mathbf{B} \neq 0$ numerical errors. To that end, we chose to use the Constrained Transport method, which is the most effective treatment provided with the PLUTO code to ensure the solenoidal constraint is satisfied.

%%%%%%%%%%%%%%%%%%%%%%%%%%%%%%

\begin{figure*}[t]
    \centering
    \includegraphics[trim={0.cm 0.cm 0.cm 0.cm},clip,scale=0.5]{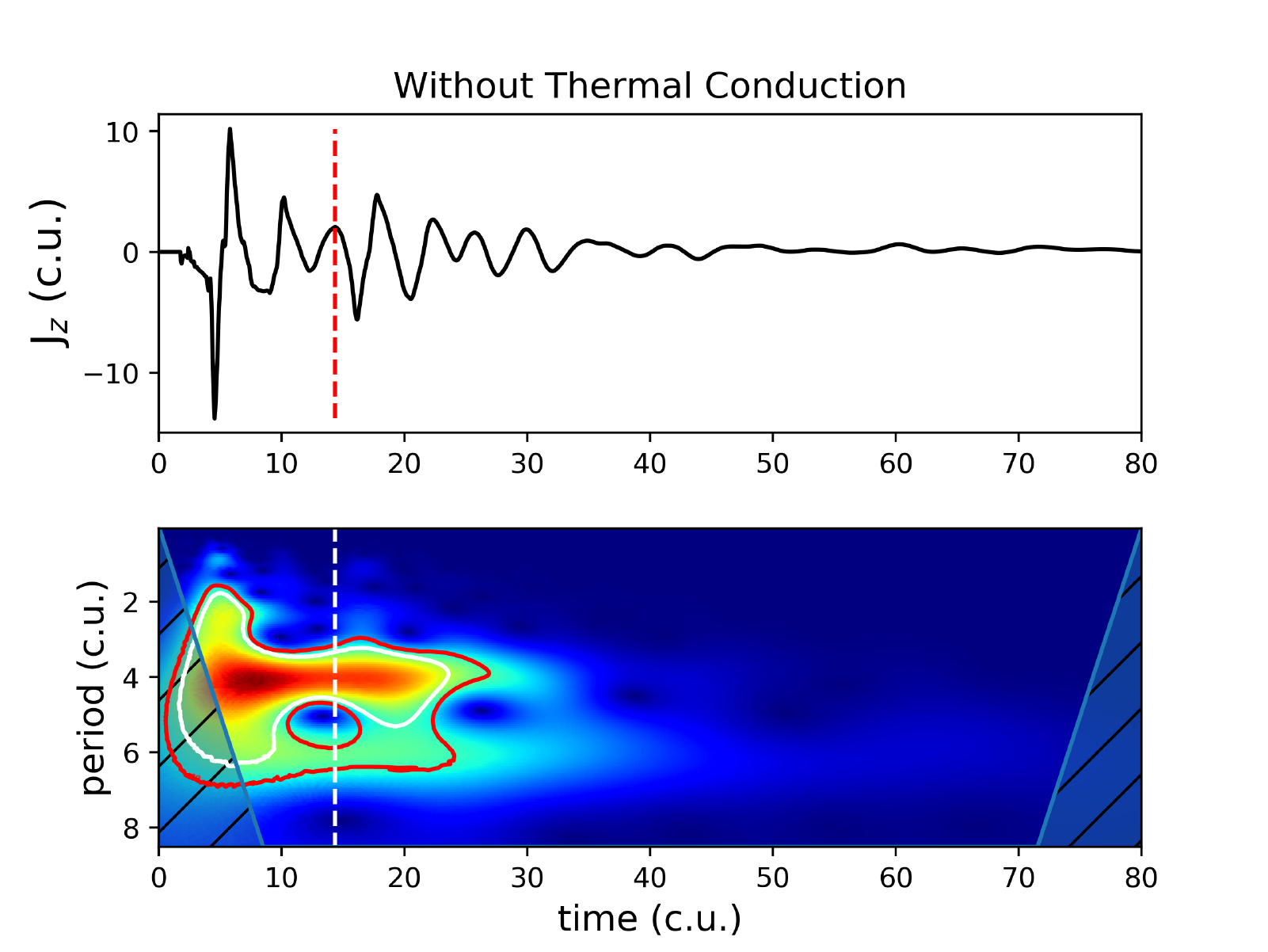}
    \includegraphics[trim={0.cm 0.cm 0.cm 0.cm},clip,scale=0.5]{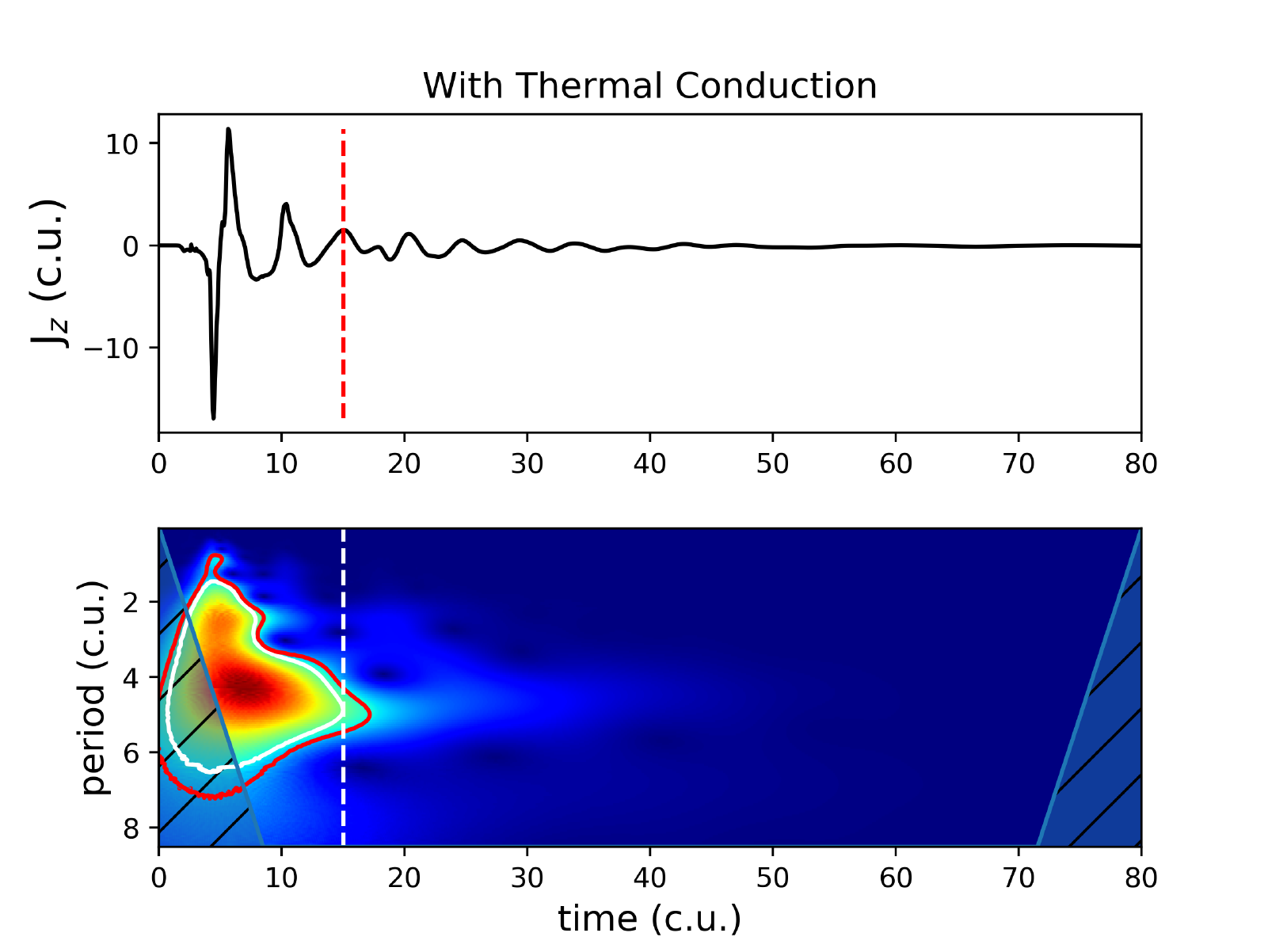}
    \caption{Top row: oscillating profile of the $J_z$ current density at the X-point as a function of time. Bottom row: wavelet profile for the oscillating $J_z$ current density. The vertical axis shows the oscillation period and the horizontal the time. The left panels show the results for the base setup with 1 MK, and the right show the results with anisotropic thermal conduction. The dashed lines show the time at which we estimate the oscillation period, while the solid red and white contours delineate the areas with confidence levels $\geqslant 80\%$ and $\geqslant 90\%$ respectively. The default magnetic field, as described in Eq. \ref{eq:magneticfield}, is equal to $B_0=1.44$\,G.}
    \label{fig:highreswavelet}
\end{figure*}

\section{Results} \label{sec:results}

\subsection{Oscillatory Reconnection in a 1\,MK Coronal Plasma}\label{sec:OR in 1MK plasma}

Our first goal in this study is to explore oscillatory reconnection in the case of a hot coronal plasma; we will first focus on the setup with a resolution of $2401^2$ grid points for an initial hot plasma of 1 MK and in the absence of thermal conduction. 

As was already mentioned, the initial velocity pulse described by Eqns. \ref{eq:vp} and \ref{eq:vl} splits into two counterpropagating waves, each with a velocity amplitude of $0.5\,C$ (see Fig. \ref{fig:initial}). The first wave travels toward the null point and is the one we focus on in this study. The other travels away from the null point and toward the boundaries of our computational domain, where it is removed by resetting the velocity components, as described in section \ref{sec:damping region}. Due to refraction, the incoming pulse gradually focuses at the X-point and, because of the decreasing Alfv\'en speed profile, the length scales between the leading and trailing edges of the wave pulse contract  \citep[e.g.][]{McLaughlin2004,McLaughlin2006b,McLaughlin2011SSRv}. In agreement with the findings of \citet{McLaughlin2009}, the incoming wave develops an asymmetry, with the wave peak at the trailing edge catching up with the leading footpoint of the pulse in the $y$-direction, while in the $x$-direction the footpoint of the pulse is catching up with the leading wave peak. This asymmetry was studied in terms of rarefaction and compression pulses respectively in \citet{Gruszecki2011null}, where it was found that only smooth, low-amplitude pulses can reach a null point without overturning.

Once the incoming wave reaches the vicinity of the null point, it perturbs it from its initial (X-point) equilibrium. Following this initial perturbation, the system starts the process of oscillatory reconnection, where the initial collapse leads to reconnection, with the resulting restoring force causing an overshoot beyond its equilibrium state. As in \citet{McLaughlin2009}, the system exhibits a series of reconnection events, with a periodically alternating connectivity. This periodicity can be traced by the formation of vertical/horizontal current sheets or their equivalent positive/negative values of the $J_z$ current density at the null point, as can be seen in Fig. \ref{fig:highresjz}. Also visible in the same figure is the accumulation of current density along what is the essentially the location of the magnetic field separatrices.

\begin{figure*}[t]
    \centering  
    \includegraphics[trim={0.cm 0.cm 0.cm 0.cm},clip,scale=0.5]{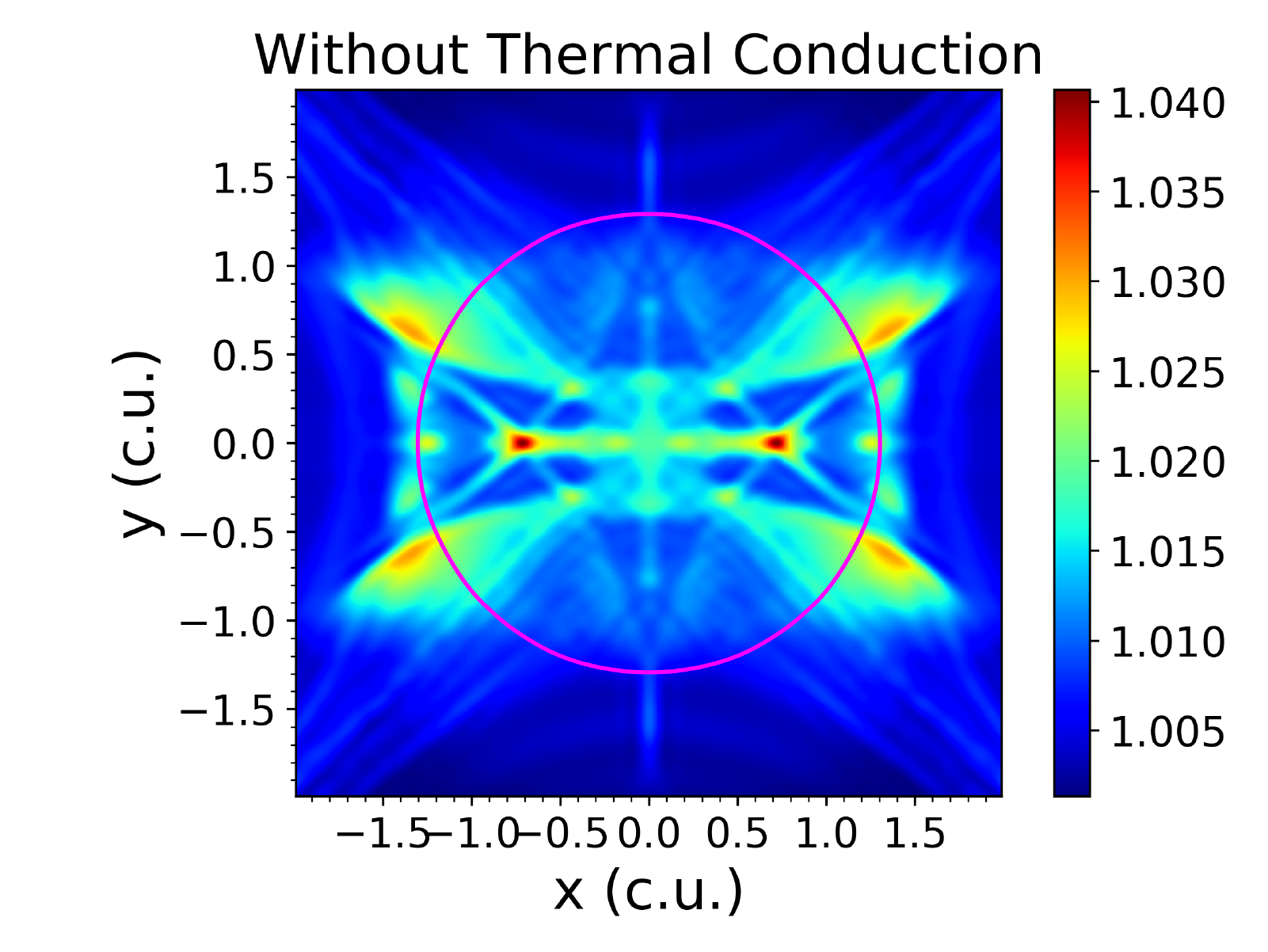}
    \includegraphics[trim={0.cm 0.cm 0.cm 0.cm},clip,scale=0.5]{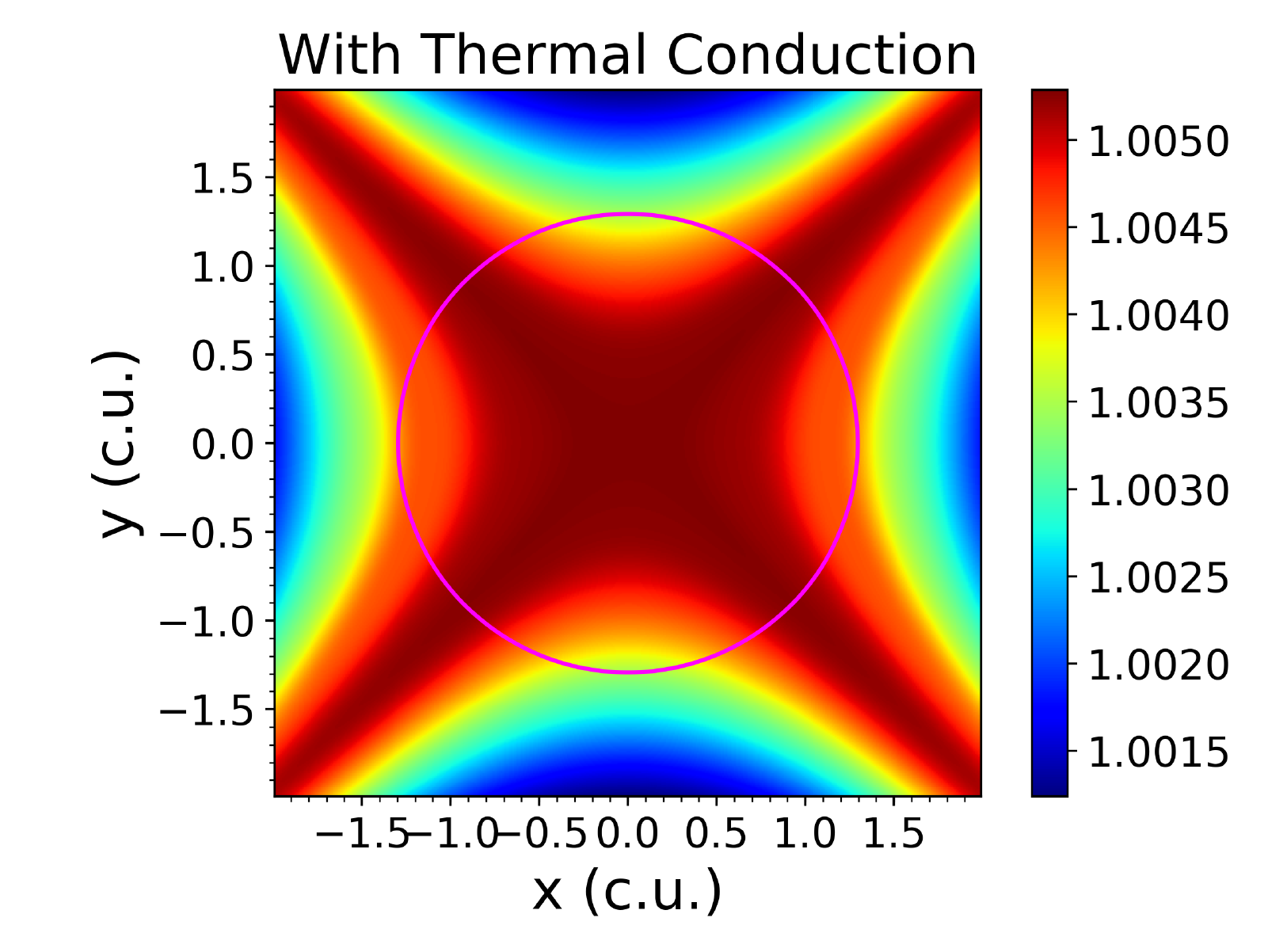}
    \caption{Left panel: temperature profile near the X-point for the setup with a base temperature of 1 MK, without thermal conduction. Right panel: the same temperature profile but for a simulation with anisotropic thermal conduction included. The equipartition layer (magenta line) is overplotted in both panels. The profiles are showing snapshots at $t=80$, in code units. All the quantities are in code units.}
    \label{fig:highrestempe}
\end{figure*}

As shown in the top left panel of Fig. \ref{fig:highreswavelet}, the profile of the $J_z$ current density at the null point, following the initial perturbation, clearly exhibits an oscillatory behavior, with an amplitude which decays over time. This is in agreement with the results of \citet{McLaughlin2009} for 2D oscillatory reconnection in a cold plasma. This oscillatory behavior, which is a signature of oscillatory reconnection, persists over multiple periods, and can be roughly divided into two regimes: a regime where the amplitude decays over time, and a later regime, at $t\gtrsim 40$, where the profile fluctuates around a near-zero but positive asymptotic value for the $J_z$ current density, equal to $J_z=0.19$.
The existence of an asymptotic nonzero value for the current density has been previously addressed for cold plasma and is associated with the (slightly) increased thermal pressure force on the left and right from the null point compared to above and below, due to an asymmetry in the produced heating. For a hot plasma, this asymmetry in the heating around the null point can be seen in Fig. \ref{fig:highrestempe}. In the left panel of this figure, we can see that plasma temperature increase is not  symmetrical around the null point, but is oriented closer along the horizontal ($x$-axis). Note that there is nothing significant about the horizontal direction compared to the vertical; the orientation of the final asymmetry is dictated by the choice of incoming pulse asymmetry described previously. 

The $J_z(0,0,t)$ profile seen in the top left panel of Fig. \ref{fig:highreswavelet} shows a more complex behavior than that of a cold plasma. In particular, we see that the current density peak amplitudes fluctuate significantly, which hints toward the existence of additional waves perturbing the null point after the initial pulse perturbation. After performing a parameter study for setups of different base temperatures (see the Appendix \ref{Appendix} for details), we attribute these additional waves to the effects of mode conversion at the equipartition layer.

\begin{figure*}[t]
    \centering
    \resizebox{\hsize}{!}{
    \includegraphics[trim={0.cm 1.1cm 3.5cm 0.cm}, clip,scale=0.25]{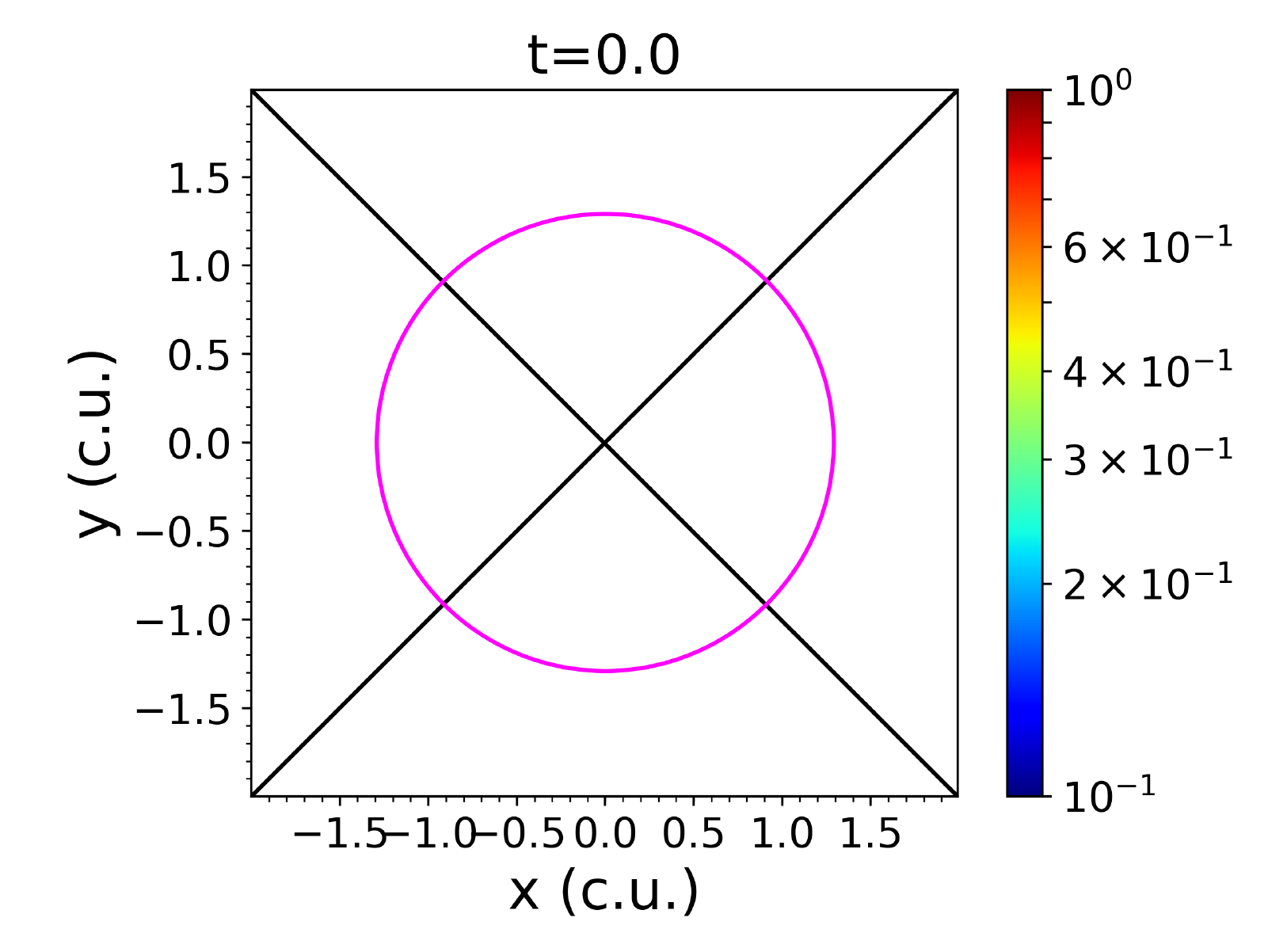}
    \includegraphics[trim={3.cm 1.1cm 3.5cm 0.cm}, clip,scale=0.25]{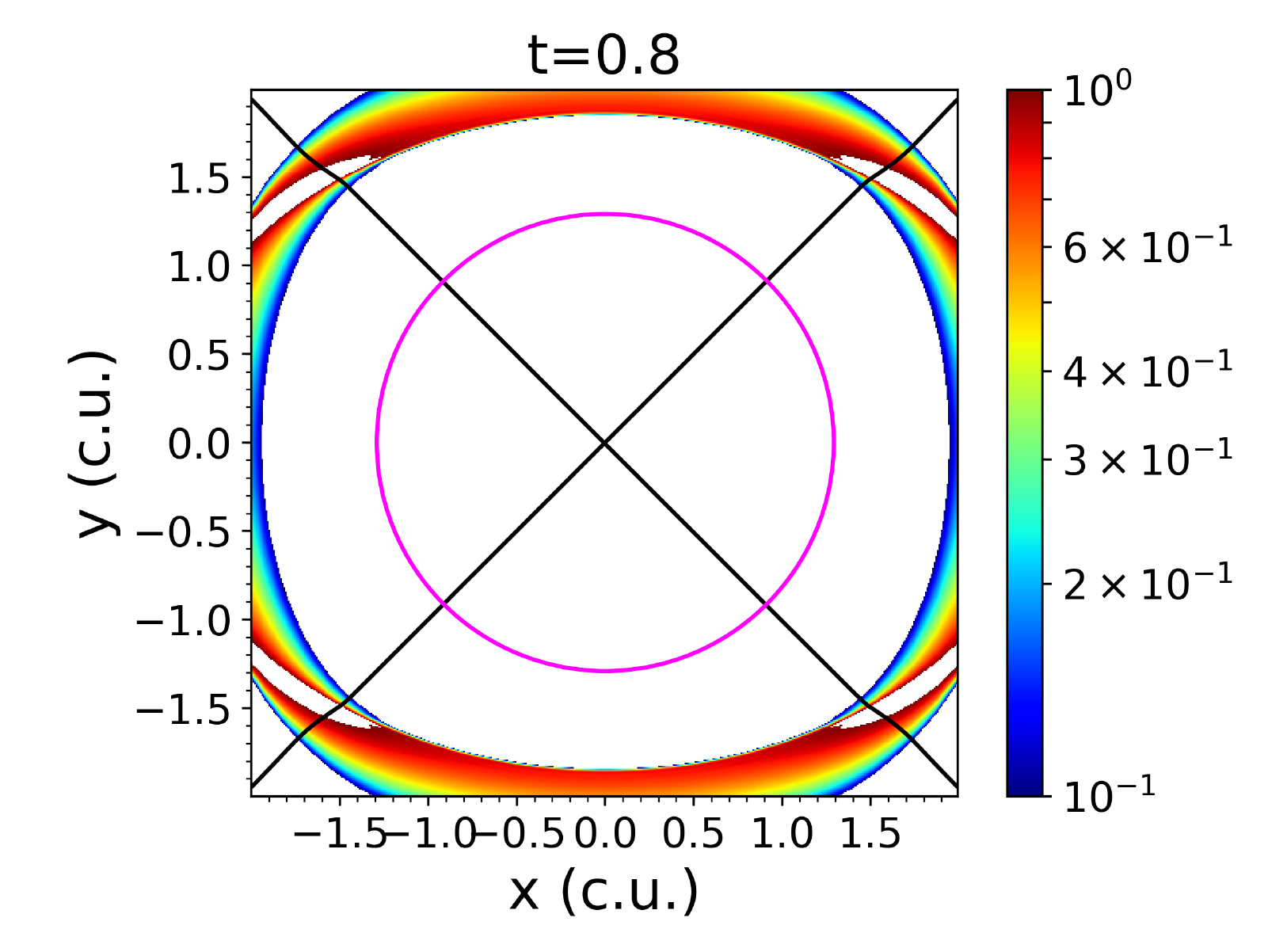}
    \includegraphics[trim={3.cm 1.1cm 3.5cm 0.cm}, clip,scale=0.25]{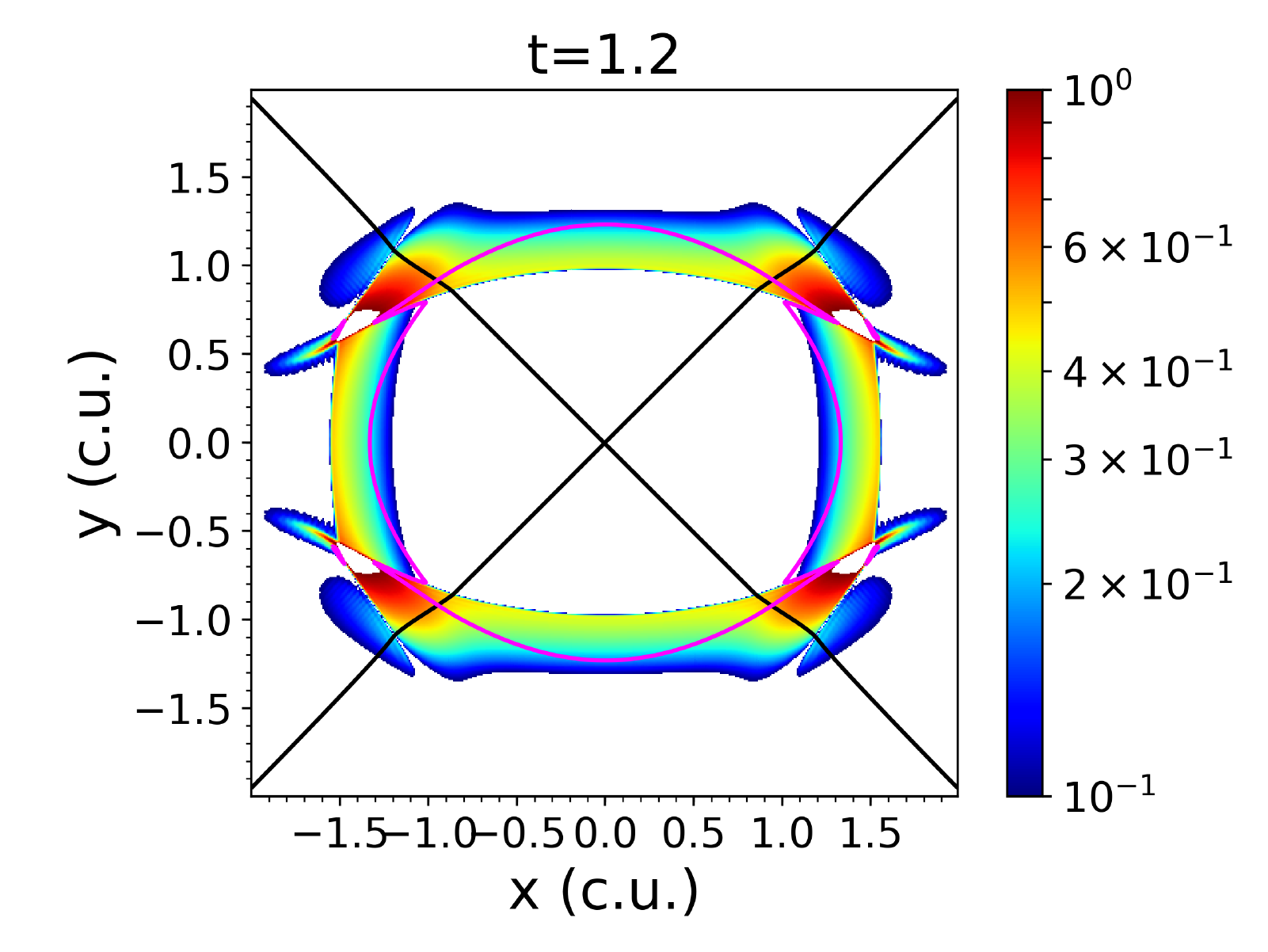}
    \includegraphics[trim={3.cm 1.1cm 0.cm 0.cm}, clip,scale=0.25]{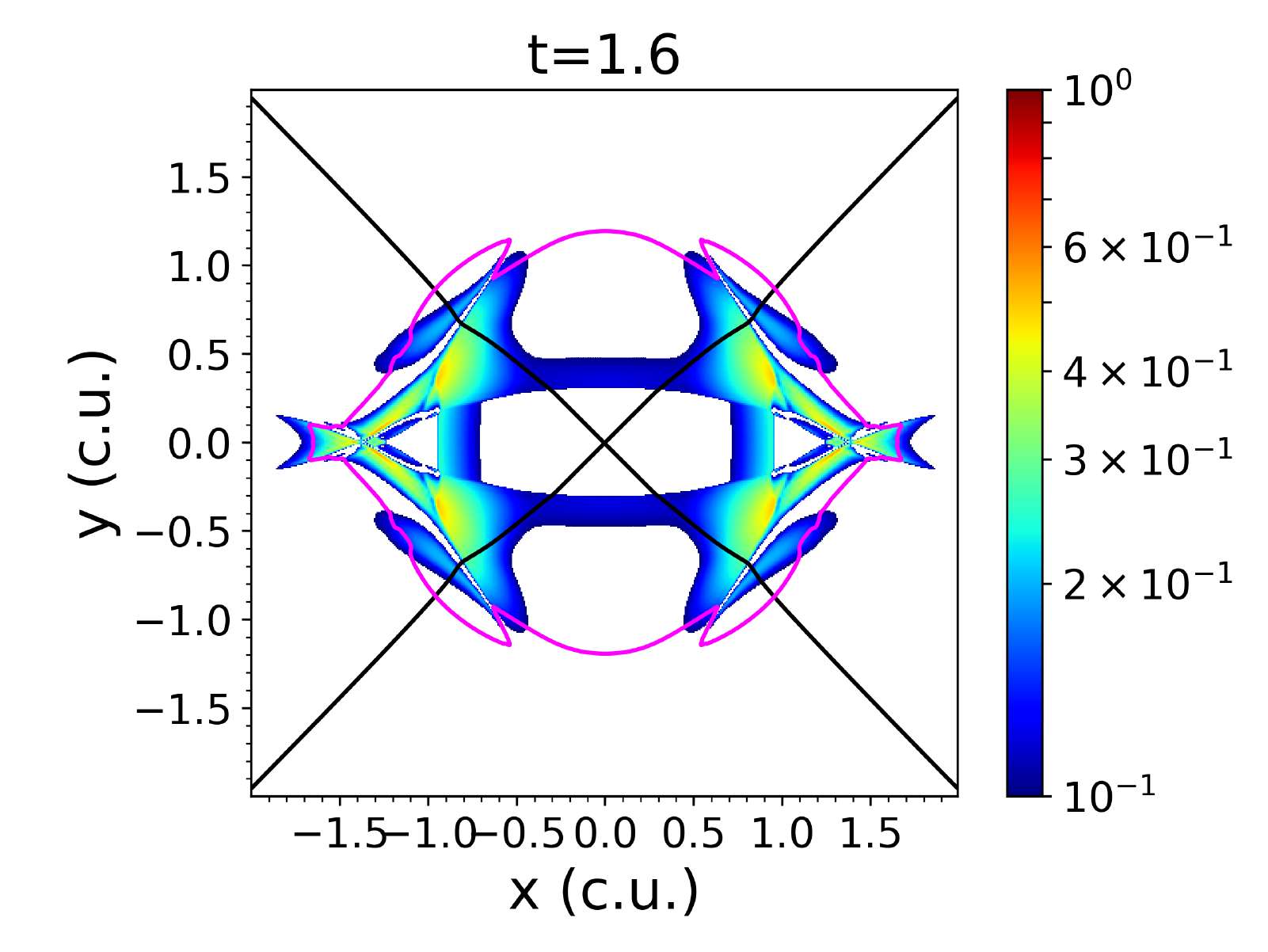}}
    \resizebox{\hsize}{!}{
    \includegraphics[trim={0.cm 0.cm 3.5cm 0.97cm},clip,scale=0.25]{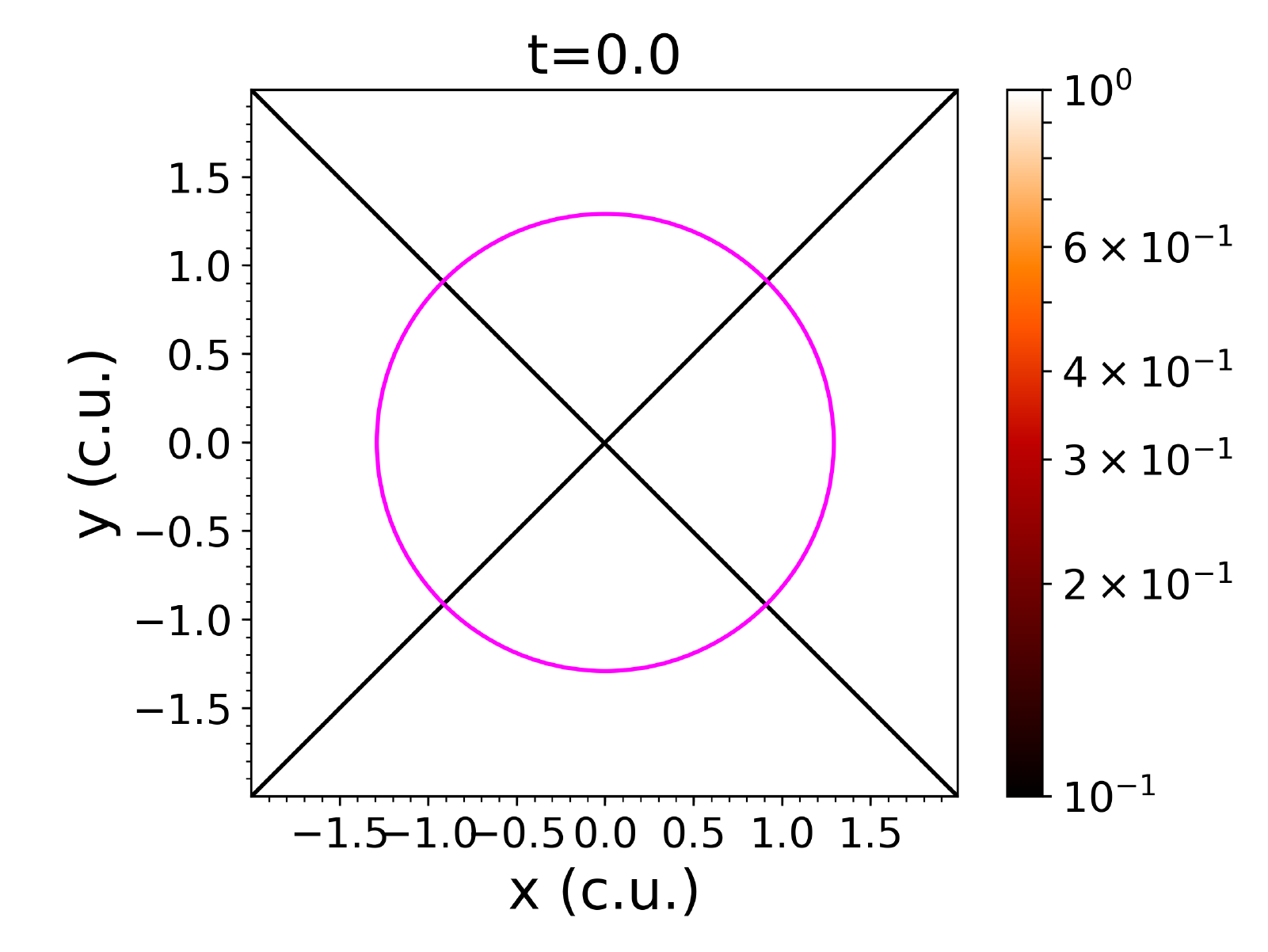}
    \includegraphics[trim={3.cm 0.cm 3.5cm 0.97cm},clip,scale=0.25]{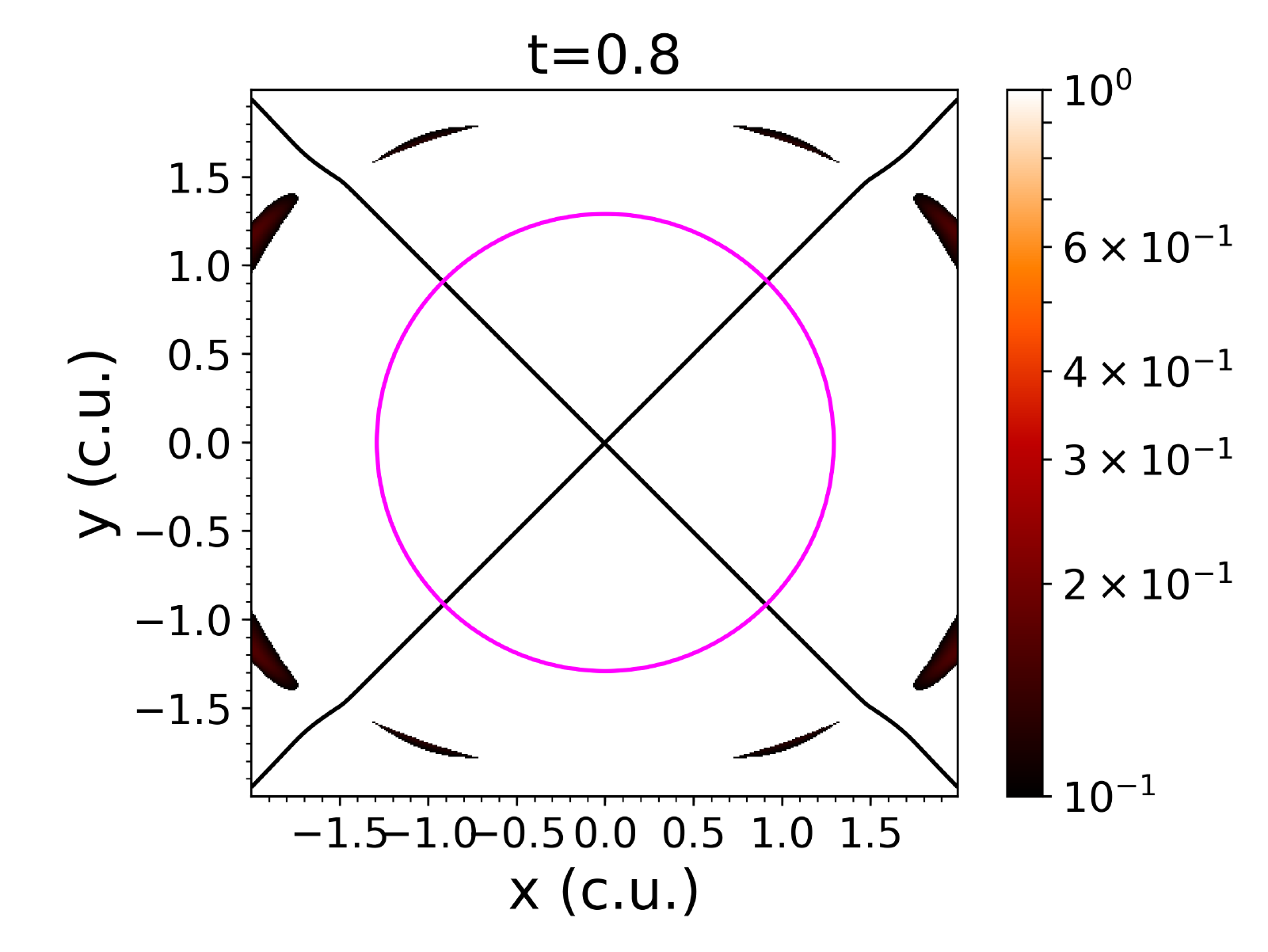}
    \includegraphics[trim={3.cm 0.cm 3.5cm 0.97cm},clip,scale=0.25]{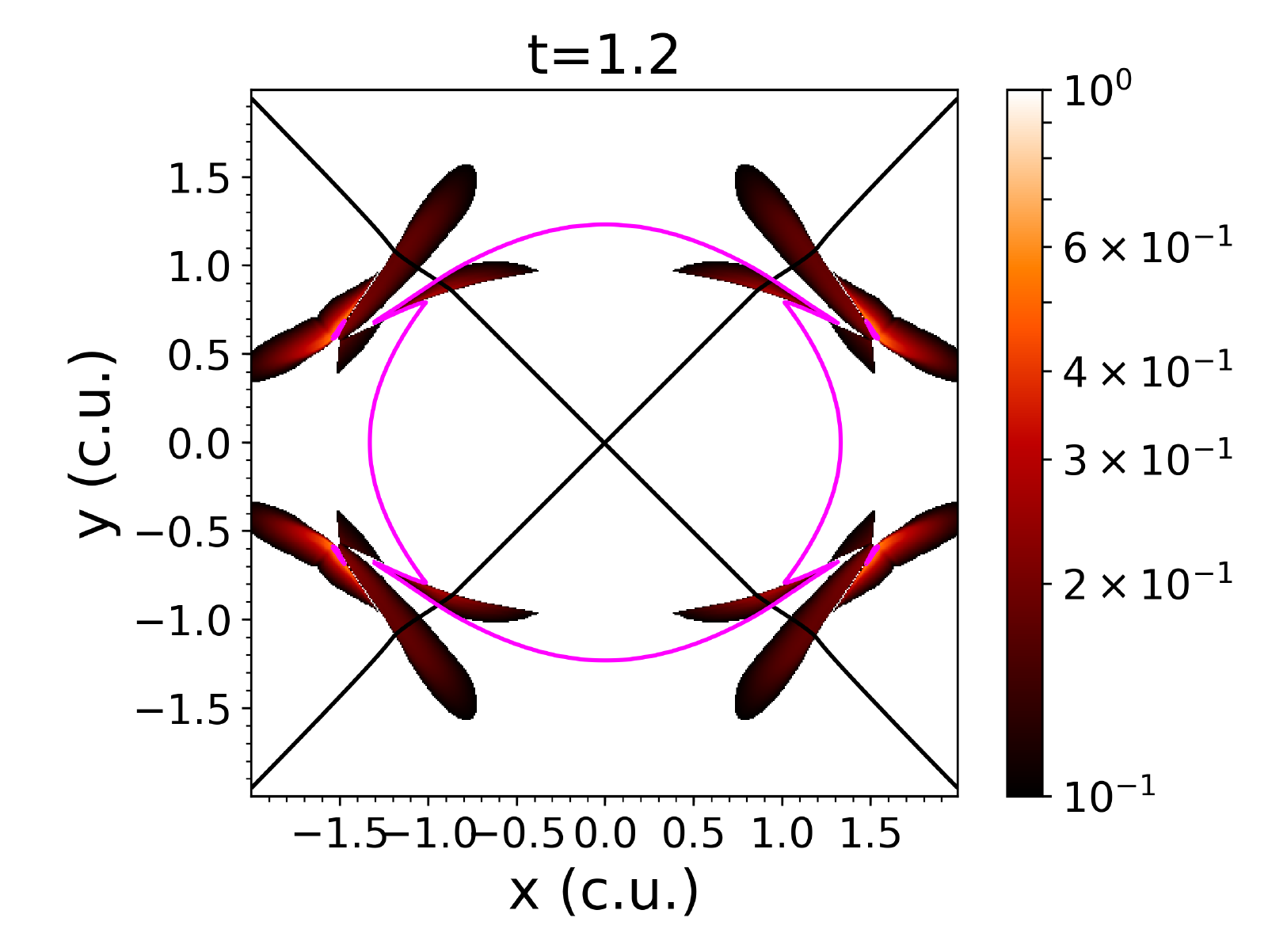}
    \includegraphics[trim={3.cm 0.cm 0.cm 0.97cm},clip,scale=0.25]{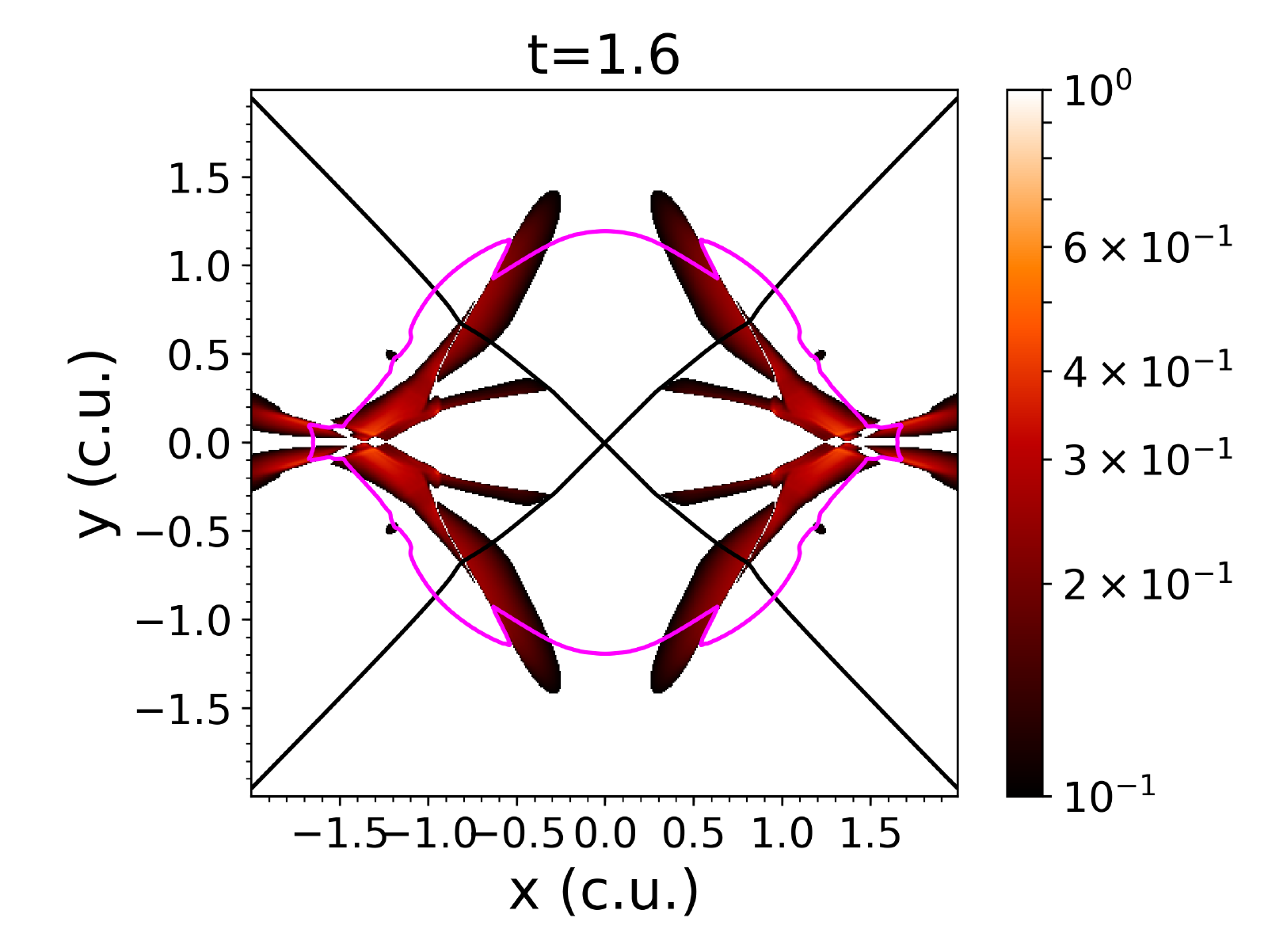}}
    \caption{Zoomed in profile of the incoming pulse for the initial state of the simulation. The $v_{\perp}$ velocity (top row) and $v_{\parallel}$ velocity (bottom row) and are shown, alongside the equipartition layer (magenta solid line) and the magnetic field separatrices (black solid lines).}
    \label{fig:highresmode}
\end{figure*}

Between the initial pulse location (radius $r=5$) and the null point lies the equipartition layer, which is the layer where the Alfv\'{e}n $V_A$ speed equals the sound speed $V_S$. In our setup, due to the initial uniform plasma density and temperature, the sound speed is uniform everywhere, while the Alfv\'{e}n speed increases further away from the null point. Thus the temperature of the setup and the strength of the magnetic field will affect the location of the equipartition layer. As the $v_{\perp}$ pulse approaches this layer, shown with the magenta line in the panels of Fig. \ref{fig:highresmode}, it starts to generate a velocity signal longitudinal to the magnetic field ($v_{\parallel}$). In \citet{McLaughlin2004,McLaughlin2005,McLaughlin2006a,McLaughlin2006b} and in \citet{Thurgood2012}, the behavior of fast and slow MHD waves was investigated in the neighborhood of various 2D and 3D null points. It was found that, for finite plasma-$\beta$, a low-$\beta$ fast wave can generate high-$\beta$ fast and slow waves through mode conversion when crossing the equipartition layer. Here, the velocity component perpendicular to the magnetic field corresponds to the (two) fast wave(s), while the component parallel to the field roughly corresponds to that generated slow MHD wave. As we see, there is extensive mode conversion happening not only along the equipartition layer but also inside the area defined by the layer. Given that, for hot plasma, in the context of our setup, this layer will include a larger area, we expect the mode conversion to generate the additional waves that perturb the $J_z$ profile after the initiation of oscillatory reconnection.

Coming back to the top left panel of Fig. \ref{fig:highreswavelet}, we have performed a wavelet analysis of the oscillation profile in order to identify its periodicity. This wavelet profile can be found on the bottom left panel of Fig. \ref{fig:highreswavelet}. Once the initial perturbation takes place and the oscillation is set up, we can visually identify two different period bands, persisting for the larger part of the oscillation, and we calculate the respective periods. The first period is $4t_0 = 31.1$\,s and is present up to approximately 35$t_0$ with a confidence level $\geqslant 90\%$, while the second period is $6.2t_0 = 48.2$\,s, with a confidence level $\geqslant 80\%$. Although the second period has a lower confidence level than the first, we believe that it is an important feature of the simulation, due to its persistence for the entirety of the simulation, even if its power and confidence level drop, especially after approximately 40$t_0$. The values for the periods were calculated by using a Python script to find the maximum values along the dashed lines, located at the third positive $J_z$ peak, as shown in Fig. \ref{fig:highreswavelet}. We choose the value at the third peak to be consistent with the measurements in the following subsections, where this choice is further explained. Here we need to point out that, since we expect the oscillation periods to be dependent to an extent on the the intensity of the reconnection procedure, these estimated values for the periods will be affected by the small values for the magnetic Reynolds number.

The presence of two periodic signals is something new that was not present in the previously studied system of oscillatory reconnection in a cold plasma. It is likely that this can be attributed to the extensive mode conversion taking place within the larger equipartition layer. The latter periodic signal ($ 6.2t_0 = 48.2$\,s) can be attributed to the oscillatory reconnection process, due to its persistence for the entirety of the simulation, while the former periodic signal ($ 4t_0 = 31.1$\,s) can be attributed to additional wave motions in the system, where those waves are generated by mode conversion as the fast wave initial condition crosses the equipartition layer (see also the Appendix \ref{Appendix} for a direct comparison of the oscillating profiles for $J_z$ for setups at different temperatures). It is the presence of these two signals that leads to a superimposed signal. 
 
%%%%%%%%%%%%%%%%%%%%%%%%%%%%%%%%%%%%%%%%%%%%%%%%%
\subsection{The Effect of Adding Thermal Conduction to the System}\label{sec:thermal-conduction}
%Our next step is to add anisotropic thermal conduction in the previously studied high resolution setup
Our next step is to add anisotropic thermal conduction in the setup from $\S$\ref{sec:OR in 1MK plasma}. At coronal conditions for the temperature, the density and the magnetic field, the ratio of the parallel to perpendicular conduction coefficients (Eq. \ref{eq:kpar} and \ref{eq:kprp}) is $\kappa_{\parallel} \gg \kappa_{\perp}$. That means that anisotropic thermal conduction is expected to dissipate heat predominantly along the magnetic field.

In the right panel of Fig. \ref{fig:highreswavelet}, we see the $J_z(0,0,t)$ profile and the corresponding wavelet profile for the setup with the anisotropic thermal conduction. From the profile for the current density, we see that oscillatory reconnection manifests as expected, and it exhibits a similar qualitative behavior as in the case without thermal conduction. However, the two profiles show a few clear differences. First of all, the wavelet analysis now exhibits only one dominant period, at $ 4.9t_0 = 38.1$\,s, persisting for the entirety of the simulation, but with a faster decreasing signal compared to the system without thermal conduction. This dominant period is associated with the process of oscillatory reconnection, given that the secondary wavefronts generated through mode conversion are now getting dissipated faster, since anisotropic thermal conduction removes energy from the system in a very efficient way. The value for this period was calculated by finding the maximum value along the dashed lines, located at the third positive $J_z$ peak, as shown in Fig. \ref{fig:highreswavelet}. The third peak is chosen, since by then the additional the periods from the initial perturbation have decayed, while we are bordering the line with confidence level of $90\%$, as shown in Fig. \ref{fig:highreswavelet}.

Secondly, the current density profile shows smaller fluctuations to its peak values. As stated before, the incoming wavefronts generated through mode conversion, which are the cause of these fluctuations, are now dissipated faster than before, due to anisotropic thermal conduction removing energy from the system in a very efficient way. If we compare the temperature profiles for the two different simulations near the X-point in  Fig. \ref{fig:highrestempe}, we can see that the produced heating is not confined in small areas around the X-point any more. Instead, it is spread/conducted along the magnetic field, creating a more uniform heating profile.

As it was previously mentioned, the profiles for the current density clearly indicate that, by adding thermal conduction, the decay rate of oscillatory reconnection has increased. This is to be expected, since one of the restoring forces of the oscillation, alongside magnetic tension, is the thermal pressure force from the generated heating \citep{McLaughlin2009}. However, the heat dissipation by thermal conduction leads to a more spatially extended and uniform, as well as less prominent temperature increase, which leads to a smaller thermal pressure force and thus as weaker overall oscillation.

\begin{figure*}[t]
    \centering
    \includegraphics[trim={0.cm 0.cm 0.cm 0.cm},clip,scale=0.5]{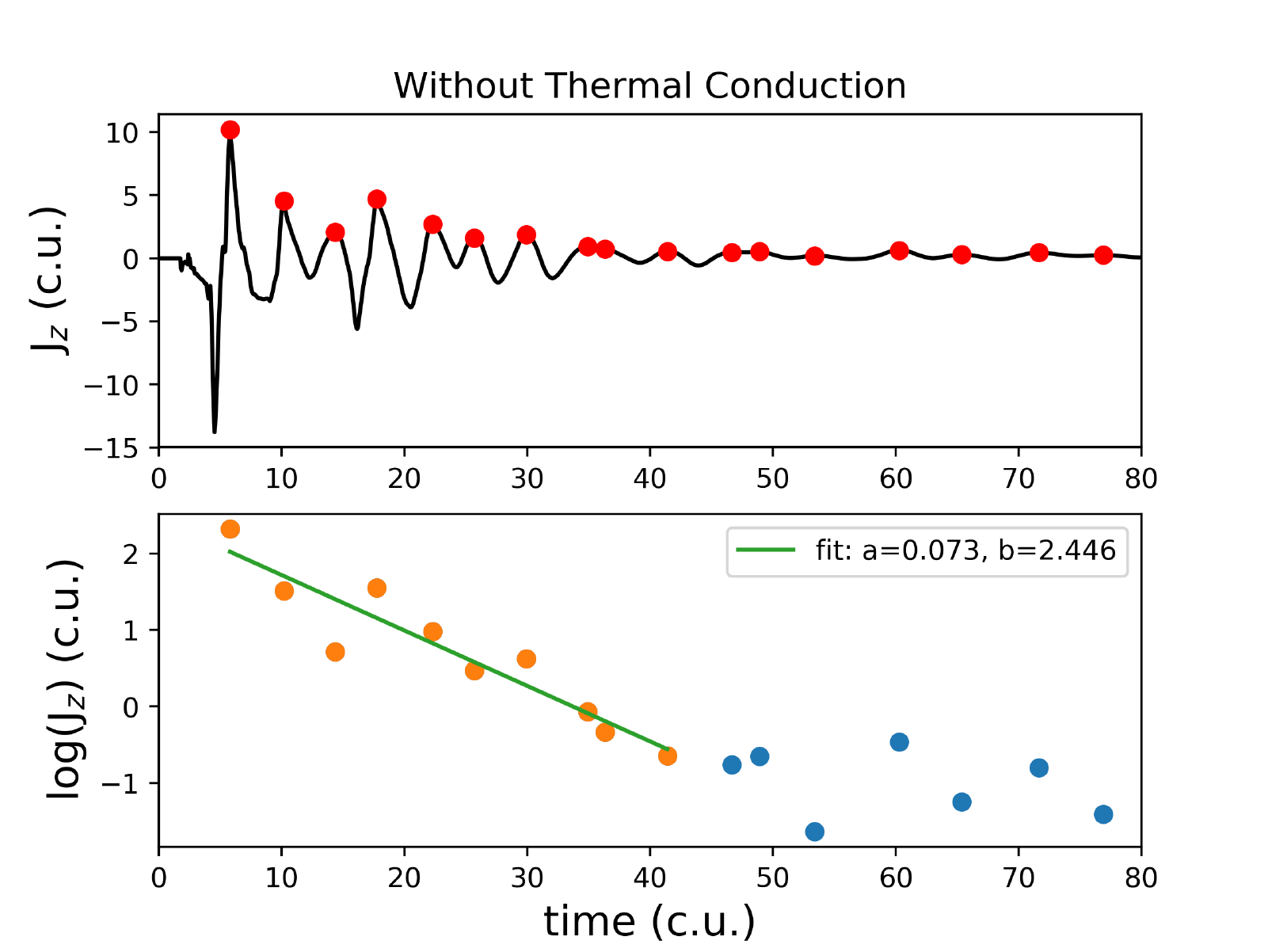}
    \includegraphics[trim={0.cm 0.cm 0.cm 0.cm},clip,scale=0.5]{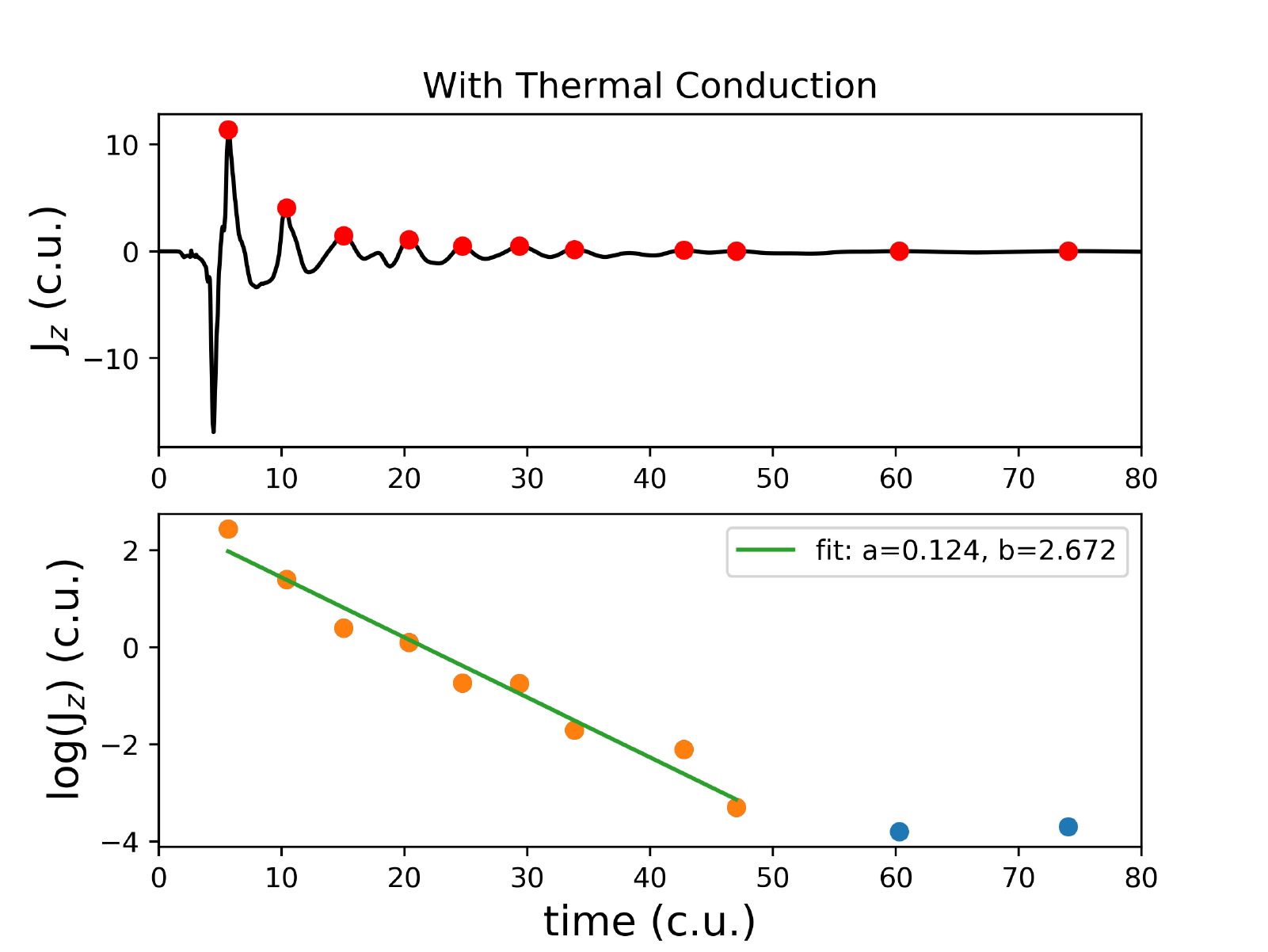}
    \caption{Top row: oscillating profile of the $J_z$ current density at the X-point as a function of time. The red dots show the local maxima of the profile used for estimating the decay rate of the oscillation. Bottom row: logarithmic plot of the local maxima of the $J_z$ profile. The different colored dots show the two different regions of the profile. The line fit for the decay rate is overplotted. The left panels show the results for the base setup with 1 MK, and the right panels show the results for anisotropic thermal conduction.}
    \label{fig:highresdecay}
\end{figure*}

In order to quantify this result, we have studied the decay rate of the oscillations \lq{envelope}\rq{} for each profile. As seen in Fig. \ref{fig:highresdecay}, we have considered the local maxima of the two oscillating profiles that can define the trend in the evolution of the oscillation amplitude. These are highlighted with red dots for the two current density profiles. By plotting the logarithm of these values as a function of time, we can divide the values in two distinct regions, as a first-order approximation. We will focus here on the first region, which is clearly associated with an exponential decay. Indeed, by fitting a linear function $f(t)=-at+b$ to the logarithmic values, we can estimate the decay rate for the two cases, $a=0.073$ and $a=0.124$ for the setup with and without thermal conduction, respectively. From this result, it becomes clear that, by adding thermal conduction, we end up with a faster decay for oscillatory reconnection. 

%%%%%%%%%%%%%%%%%%%%%%%%%%%%%%%%%%%%%%%%%%%

\subsection{The Sensitivity of Oscillatory Reconnection to Magnetic Field Strength}\label{sec:parameter-study}
Studying oscillatory reconnection for hot coronal plasma and adding the effects of thermal conduction has so far provided us with a very useful insight into the behavior of this mechanism in the solar corona and has put older results into perspective, as will be discussed in $\S\ref{sec:discussions}$. However, our ultimate goal, which is beyond the scope of the current work, is to associate the mechanism of oscillatory reconnection with direct observations from the solar atmosphere and use our findings in the context of coronal seismology (e.g. \citealt{Uchida1970,RobertsEdwinBenz1984}). To that end, we want to understand the connection - if any - between the periodicity and decay rate of oscillatory reconnection at a null point, compared to the strength of the (local) magnetic field.

\begin{figure}[b]
    \centering
    \includegraphics[trim={0.4cm 0.cm 0.cm 0.cm},clip,scale=0.5]{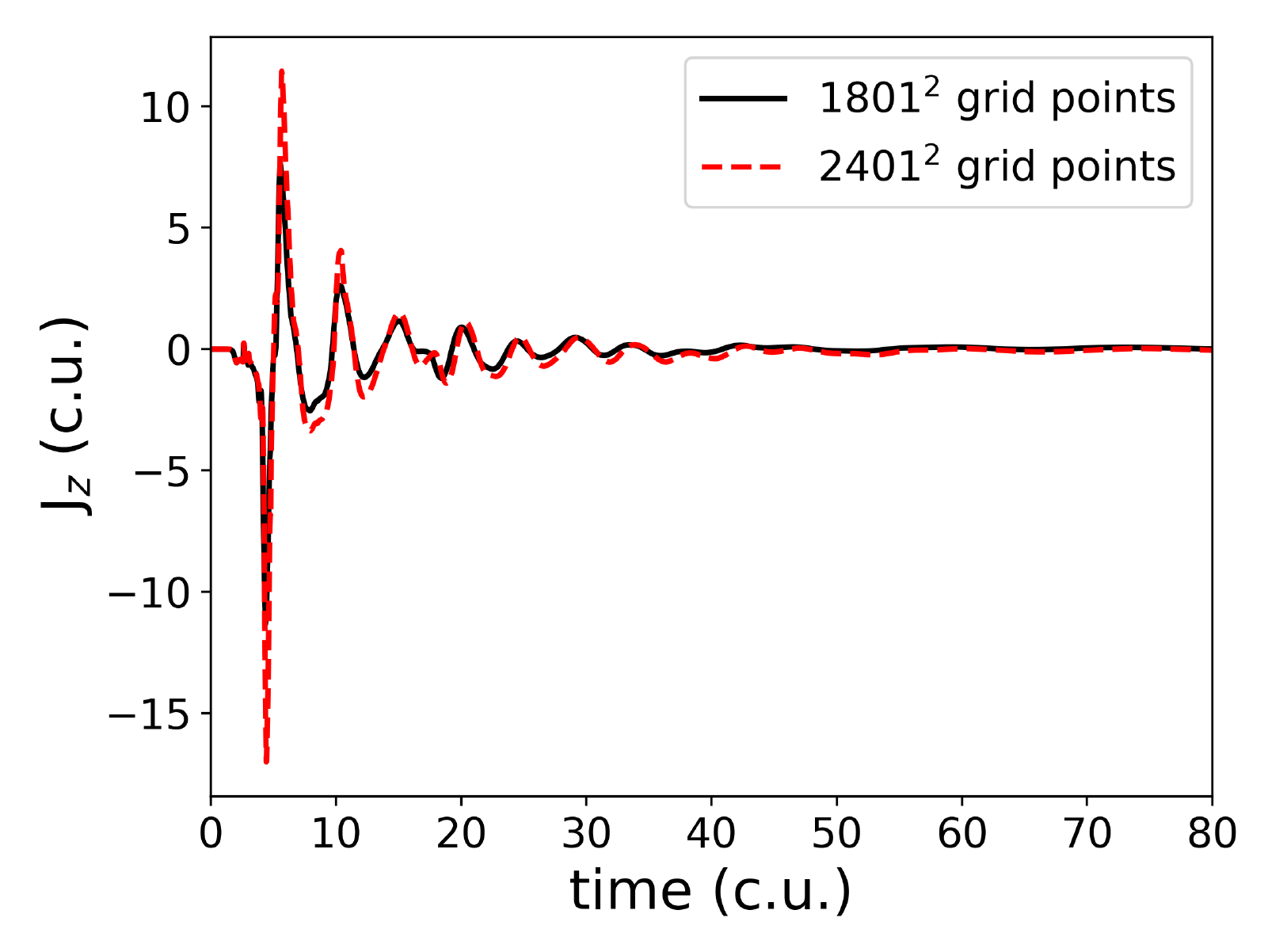}
    \caption{The $J_z$ current density profiles at the X-point for two setups at 1 MK base temperature and anisotropic thermal conduction are overplotted. The black line corresponds to the lower-resolution setup with $1801^2$ grid points, and the red dashed line to the higher resolution setup with $2401^2$ grid points.}
    \label{fig:rescompare}
\end{figure}

The first step is taken in this study, where we use our theoretical model of a magnetic field X-point configuration. We start from our basic setup, with added anisotropic thermal conduction and with a 1 MK base temperature (see $\S\ref{sec:Initial and boundary conditions}$). We will consider four cases of different equilibrium magnetic fields. One will have the \lq{base}\rq{} magnetic field ($1\times B_0=1.44$\,G) described in Eq. \ref{eq:magneticfield} (note that this is the same as that already studied in $\S\ref{sec:OR in 1MK plasma}$, albeit with a coarser resolution - see below), one with half the base magnetic field strength ($0.5\times B_0=0.72$\,G), one with double the base magnetic field ($2.0\times B_0=2.88$\,G), and one with triple the magnetic field strength ($3.0\times B_0=4.32$\,G). At the same time, we will alter the value of $C$ in Eq. \ref{eq:vp} for the velocity pulse to ensure that we inject the same amount of kinetic energy in each setup. In particular, we will have:
\begin{itemize}
    \item $C=1$, for the case with $1\times B_0$,
    \item $C=2$, for the case with $0.5 \times B_0$,
    \item $C=1/2$, for the case with $2.0\times B_0$ and
    \item $C=1/3$, for the case with $3.0\times B_0$.
\end{itemize}

In this parameter study, we will use a coarser grid for our simulations ($1801^2$ points) in order to benefit from the faster computations. Therefore, we first need to compare with our higher resolution run (from $\S\ref{sec:OR in 1MK plasma}$) to make certain that our results are in agreement and not significantly affected by the decreased resolution. In particular, we compare the $J_z$ current density profiles for the two resolutions, with  thermal conduction present in the system. The results are shown in Fig. \ref{fig:rescompare}. As we can see, the two profiles are in good agreement with respect to the oscillation frequency and decaying behavior. The only slight difference present is the amplitudes of the current density. In particular, they have higher absolute values for the setup with a higher resolution, since the current sheets are better resolved in that case.

\begin{figure*}[t]
    \centering  
    \includegraphics[trim={0.cm 0.cm 0.cm 0.cm},clip,scale=0.5]{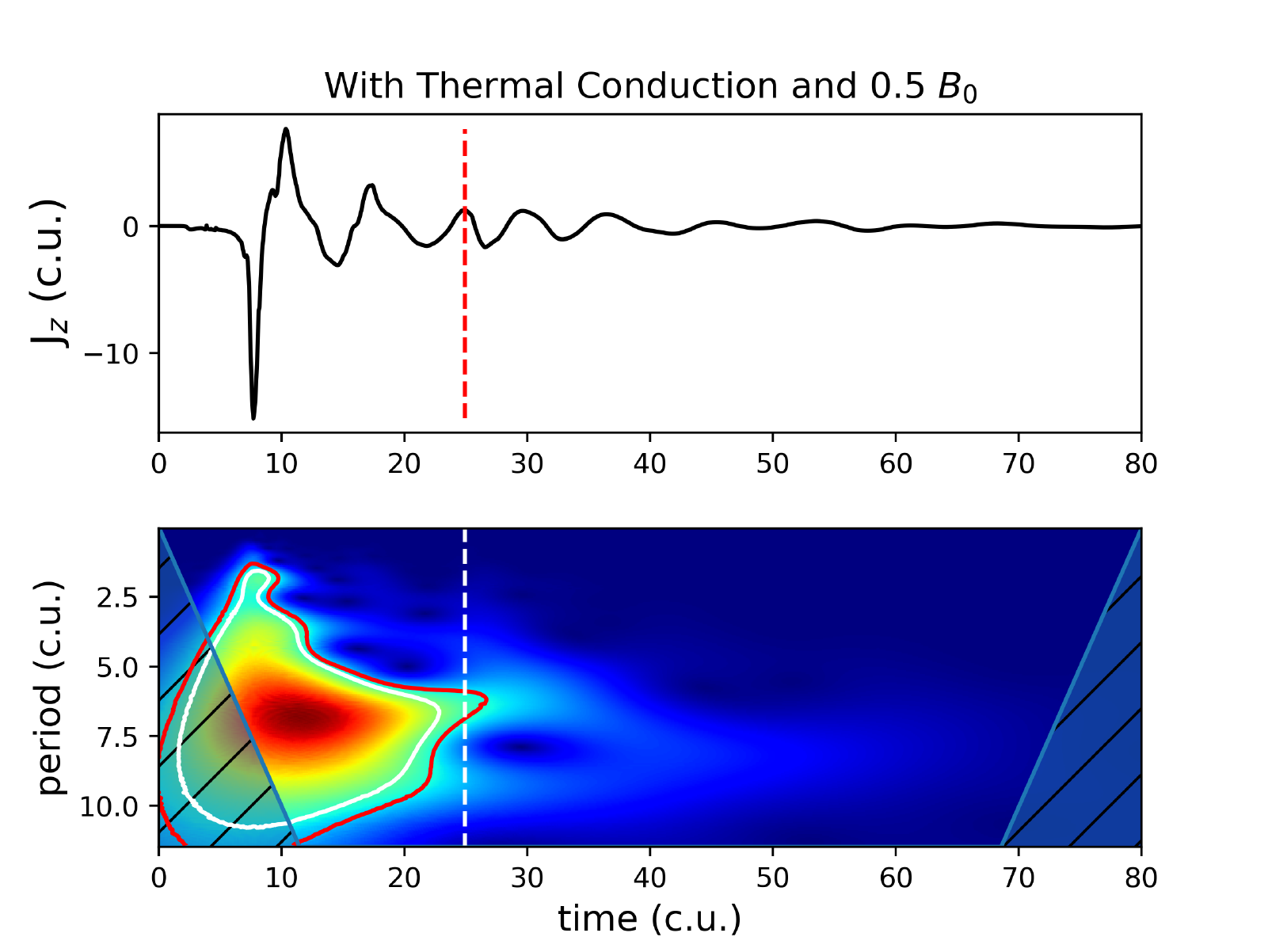}
    \includegraphics[trim={0.cm 0.cm 0.cm 0.cm},clip,scale=0.5]{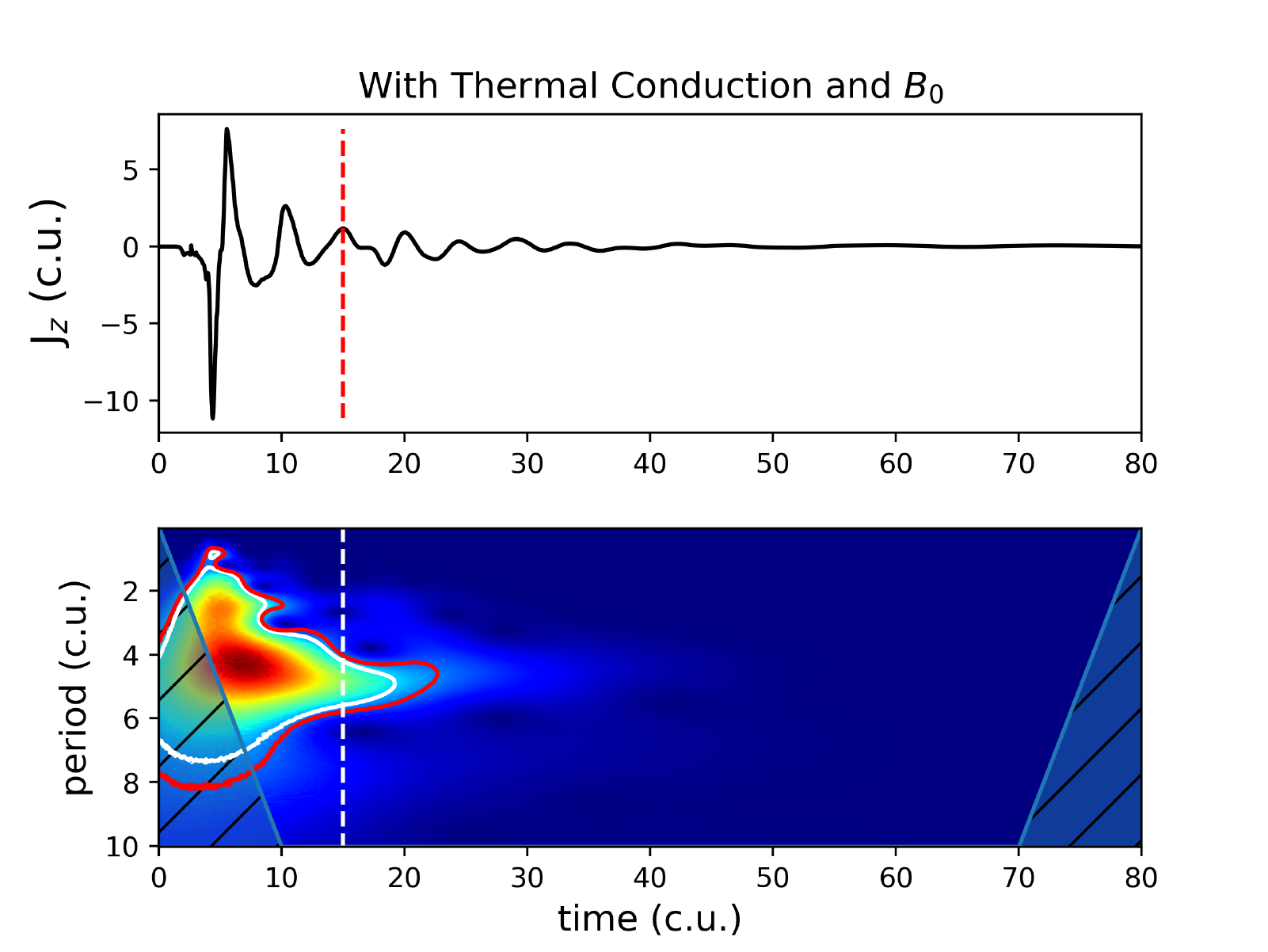}
    \includegraphics[trim={0.cm 0.cm 0.cm 0.cm},clip,scale=0.5]{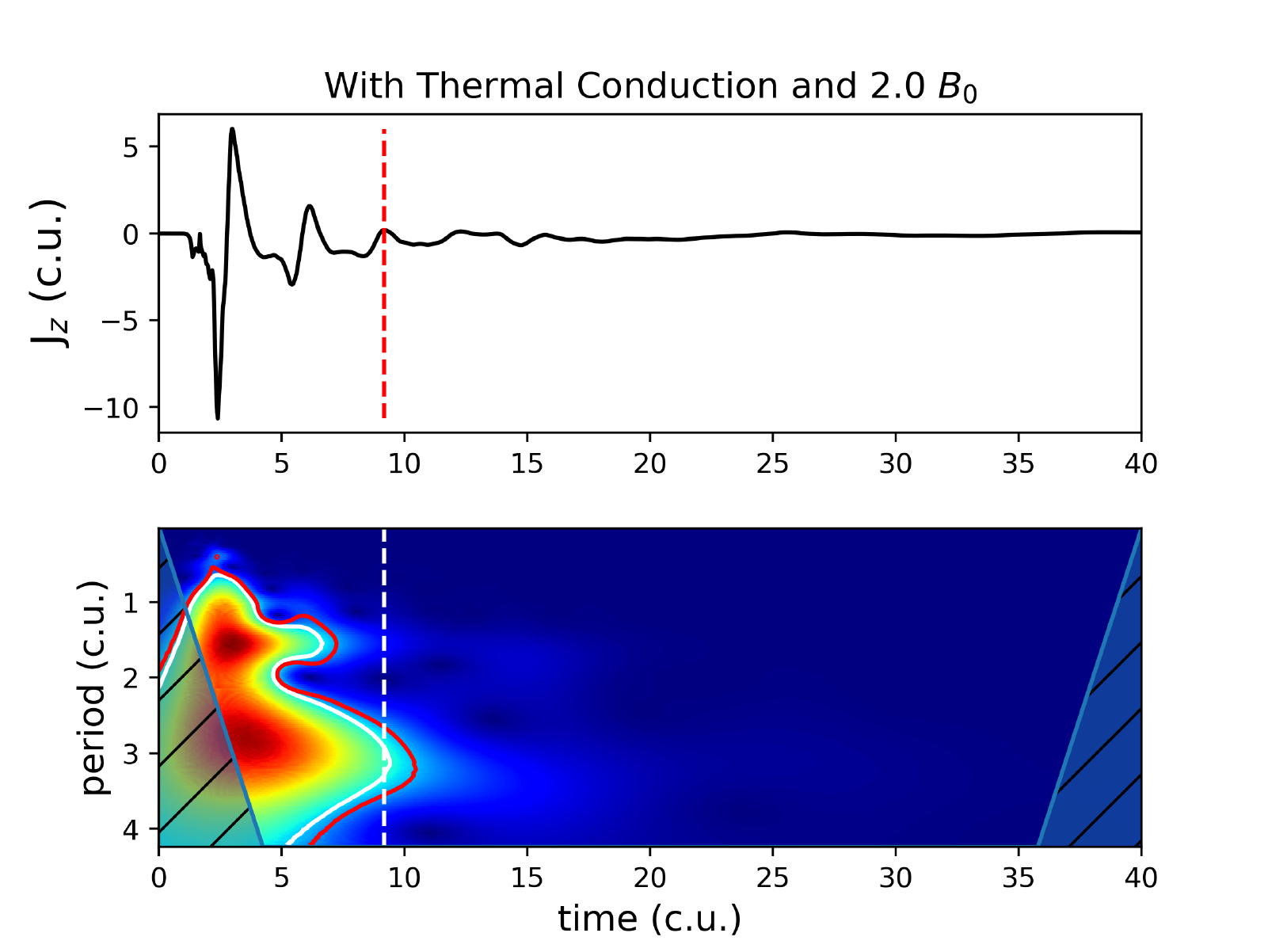}
    \includegraphics[trim={0.cm 0.cm 0.cm 0.cm},clip,scale=0.5]{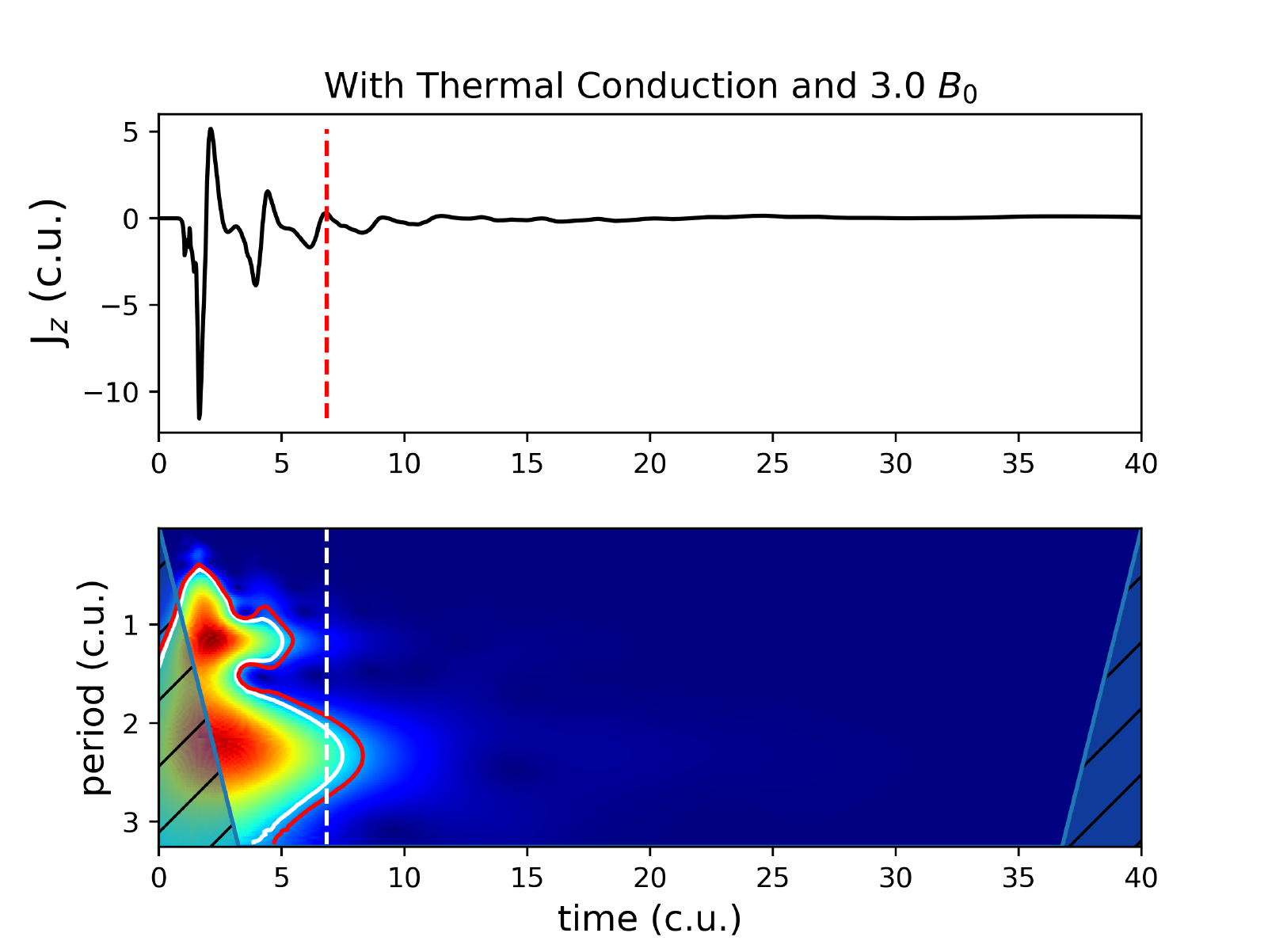}
    \caption{Same as Figure \ref{fig:highreswavelet}, but for the lower-resolution setups with anisotropic thermal conduction. The cases with different initial magnetic fields (half - $0.5B_0$, regular - $B_0$, double - $2.0B_0$ and triple - $3.0B_0$) are shown. Note that the profiles for double and triple the magnetic field strength show data only up to $t=40$.}
    \label{fig:paramwavelet}
\end{figure*}

The first thing to do is to study the spectra of the different $J_z$ profiles at each null point. As seen in Fig. \ref{fig:paramwavelet}, the expected oscillatory form of the profile is acquired in all the different cases, with the initial perturbation followed by the decaying oscillation. It is clear from both the profiles and the wavelet analysis that the oscillation period changes for the different magnetic field strengths. In particular, starting from half the base magnetic field strength ($0.5B_0$) and heading to higher values, the dominant period of the oscillation (as was identified for the setup studied in Section \ref{sec:thermal-conduction}) has approximate values of $ 6.3\,t_0 = 49$\,s, $ 4.9\,t_0 = 38.1$\, s, $ 3.1\,t_0 = 24.1$\, s, and finally $ 2.3\,t_0 = 17.9$\, s. For the reasons explained in the previous section, these values for the period were calculated by using a Python script to find the maximum values along the dashed lines, located at the third positive $J_z$ peak, as shown in Fig. \ref{fig:paramwavelet}. We use this method to calculate the periods for all cases, in order to be consistent. The confidence levels for these periods are above $90\%$ with the exception of the $0.5B_0$ which shows a confidence level between $80\%$ and $90\%$. Note that, for magnetic fields of $2B_0$ and $3B_0$, we only show the results up to $t=40\,t_0$, for a more convenient visualisation.

Here we need to point out that all the wavelet profiles show a peak that is associated with the initial perturbation and which has a lower period. This is usually confined in the very initial part of the oscillation and does not affect the results as time progresses. However, this peak starts to be increasingly more important for the cases with the stronger magnetic field ($2.0\times B_0$ and $3.0\times B_0$), where the oscillation decays rapidly. To remain consistent with the other setups, and with the higher resolution run, we will only focus on the second peak, which is clearly associated with the decaying oscillation for those other setups.  

\begin{figure*}[t]
    \centering
    \includegraphics[trim={0.cm 0.cm 0.cm 0.cm},clip,scale=0.5]{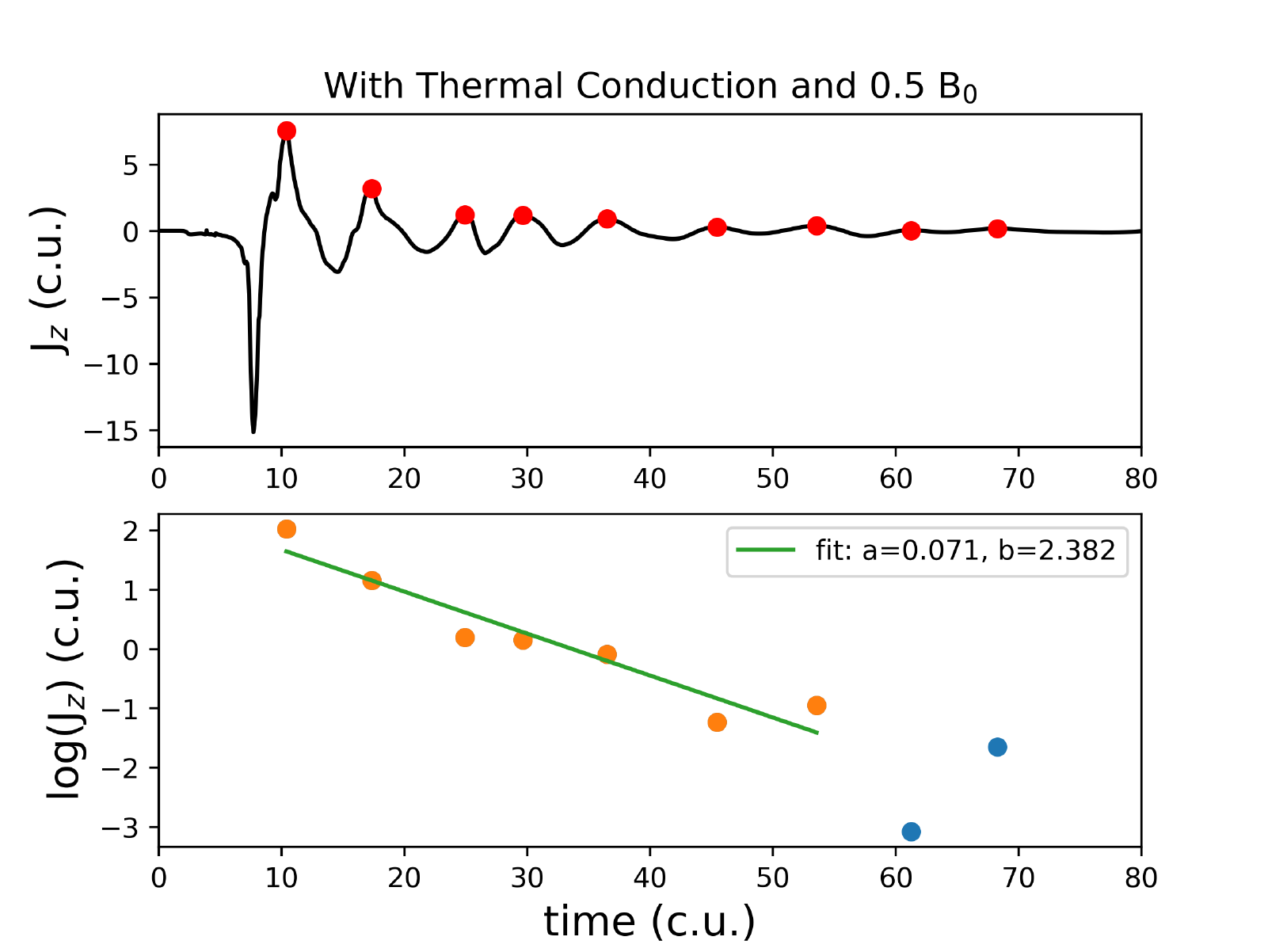}
    \includegraphics[trim={0.cm 0.cm 0.cm 0.cm},clip,scale=0.5]{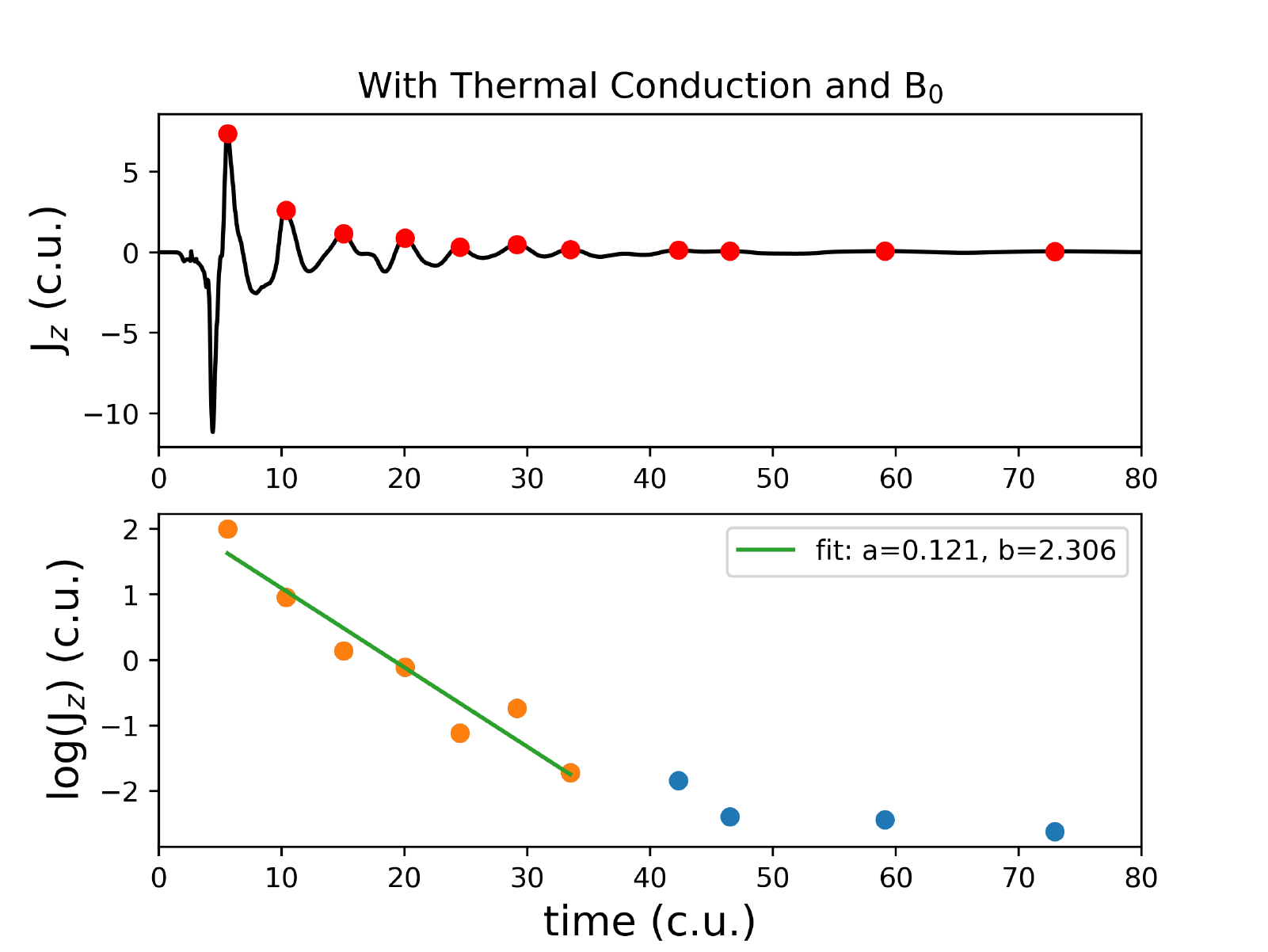}
    \includegraphics[trim={0.cm 0.cm 0.cm 0.cm},clip,scale=0.5]{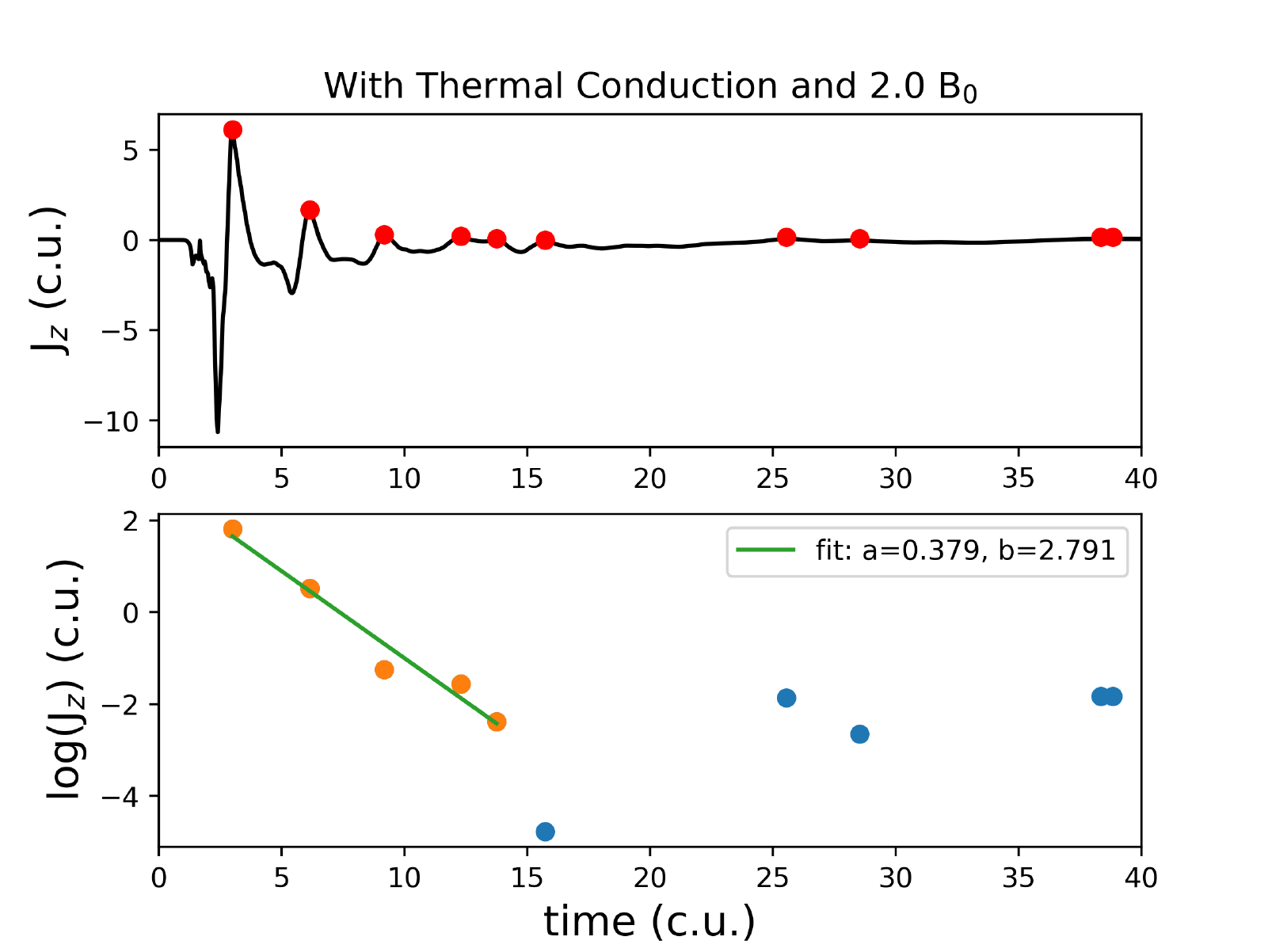}
    \includegraphics[trim={0.cm 0.cm 0.cm 0.cm},clip,scale=0.5]{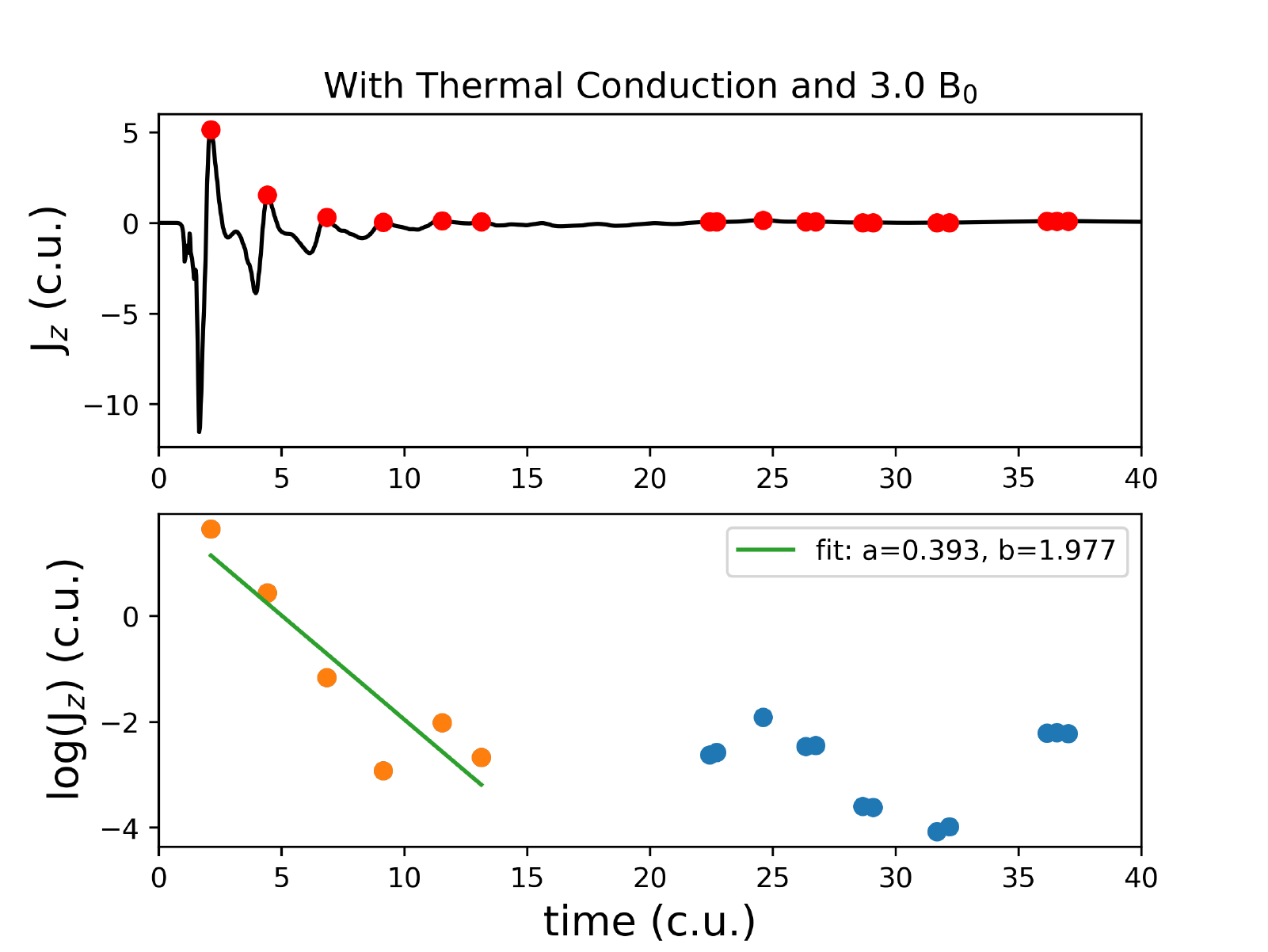}
    \caption{Same as Figure \ref{fig:highresdecay}, but for the lower-resolution setups with anisotropic thermal conduction. The cases with different initial magnetic fields (half - $0.5B_0$, regular - $B_0$, double - $2.0B_0$ and triple - $3.0B_0$) are shown. Note that the profiles for double and triple the magnetic field strength show data only up to $t=40\,t_0$.}
    \label{fig:paramdecay}
\end{figure*}

\begin{figure*}[t]
    \centering
    \resizebox{\hsize}{!}{
    \includegraphics[trim={0.cm 0.cm 0.cm 0.cm},clip,scale=0.5]{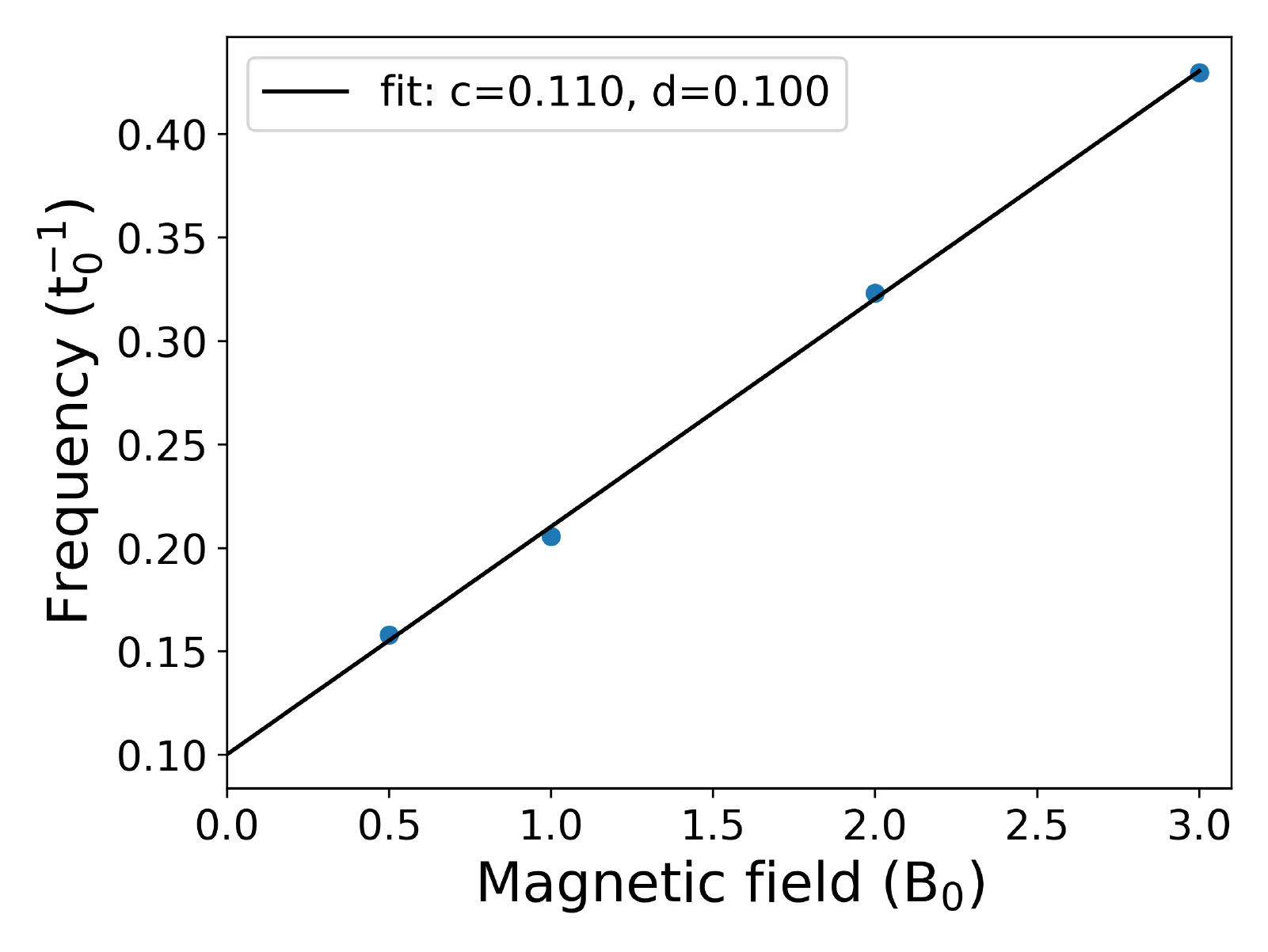}
    \includegraphics[trim={0.cm 0.cm 0.cm 0.cm},clip,scale=0.5]{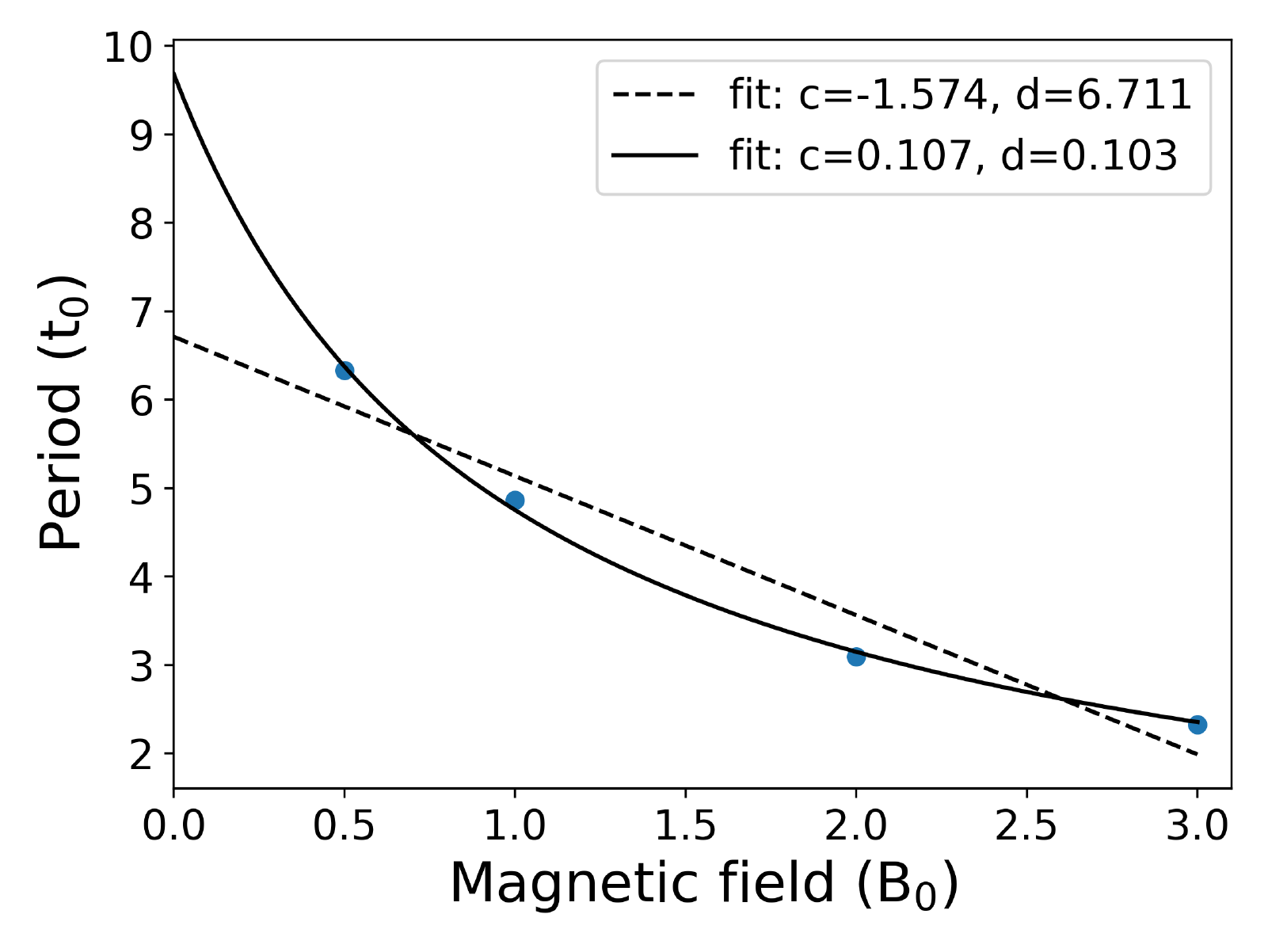}
    \includegraphics[trim={0.cm 0.cm 0.cm 0.cm},clip,scale=0.5]{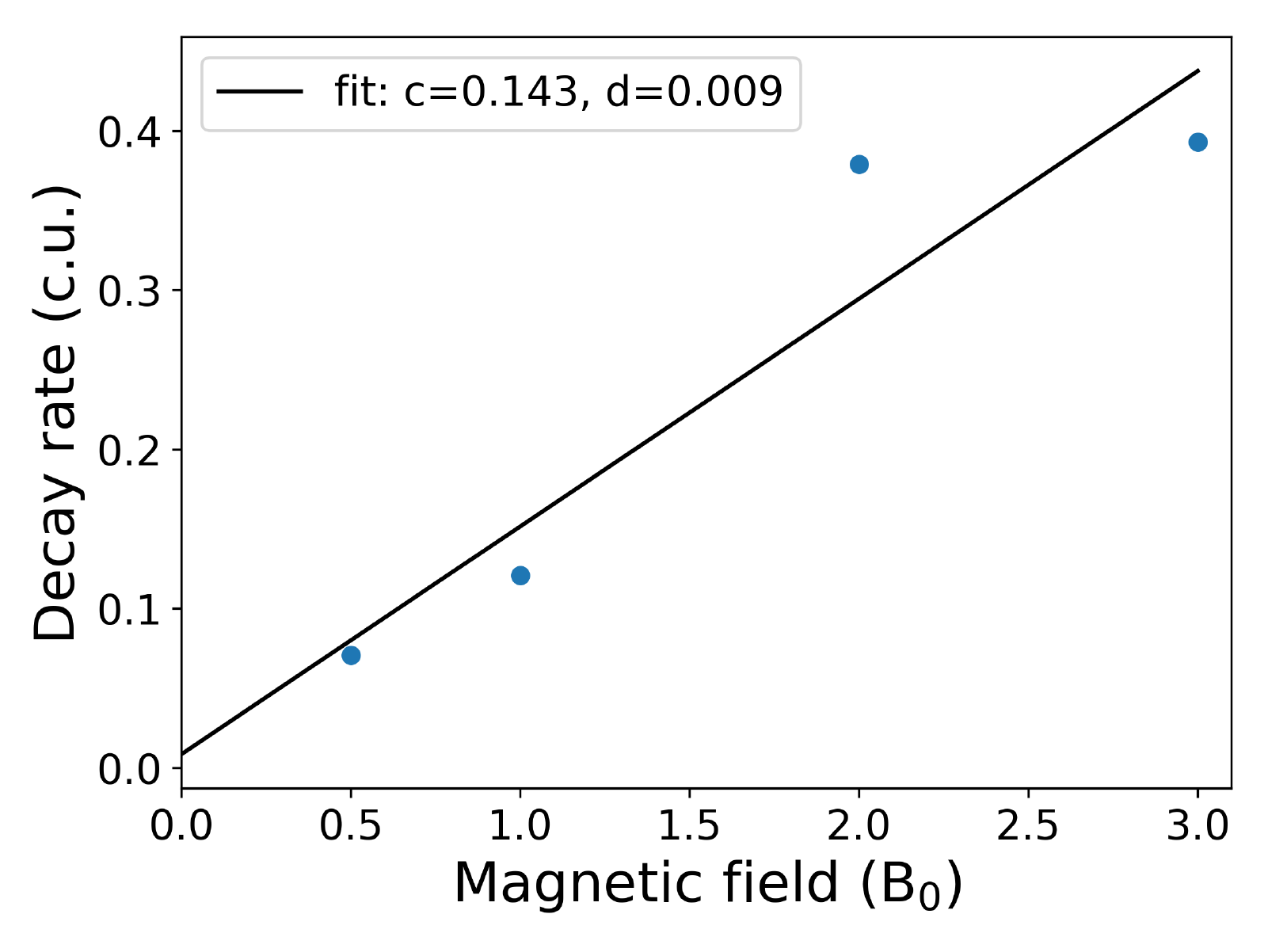}
    }
    \caption{Left panel: frequency of the $J_z$ oscillation profile with respect to the magnetic field strength (times the initial $B_0$), and a fit of the $c\,B_0 + d$ function. Middle panel: period of the $J_z$ oscillation profile with respect to the magnetic field strength (times the initial $B_0$), and a fit of the $1/(c\,B_0 + d)$ (solid line) and $c\,B_0 + d$ (dashed line) functions. Right panel: decay rate of the $J_z$ oscillation profile with respect to the magnetic field strength (times the initial $B_0$) and a fit of the linear $c\,B_0 + d$ function.}
    \label{fig:slopes}
\end{figure*}

To analyze the decay rate, we repeated the analysis with the local maxim (from $\S\ref{sec:thermal-conduction}$, in order to estimate the rates of the exponential decay, as seen in Fig. \ref{fig:paramdecay}. Plots of the frequency and period versus the initial magnetic field strength can be seen in the left and middle panels of Fig. \ref{fig:slopes}. From the plot of the frequency, we can fit the linear function $Frequency(B_0) = c\,B_0 + d$, showing that the frequency is proportional to the magnetic field strength. From the plot of the period, we can fit the inverse function $Period(B_0) = 1/Frequency(B_0)$, showing that the period seems to be inversely proportional to the magnetic field strength. This good fit for the periods and their respective frequencies is a strong indication that our calculated periods are real features of the signal and not the result of noise. For comparison, we also include a linear fit for the same plot ($Period(B_0) = c\,B_0 + d$, dashed line). Plotting the decay rate values for each case versus the magnetic field, we find a clear linear relation between the two, as seen in the right panel of Fig. \ref{fig:slopes}. Although more data points are required for to confirm the exact functional relationship of this result, there is a clear trend found by this parameter study, i.e. that the period of the oscillation decreases compared to increasing magnetic field strength, and that the decay rate increases as the magnetic field strength increases.

%%%%%%%%%%%%%%%%%%%%%%%%%%%%%%%%%%

\section{Discussion and Conclusions} \label{sec:discussions}
This paper presents an investigation of the phenomenon of oscillatory reconnection in the neighborhood of a 2D magnetic X-point for hot coronal plasma. In previous papers, this fundamental plasma mechanism has only been studied for cold plasma environments, which makes this current study an essential step  for revealing a more complete insight on the nature of this mechanism and its possible role in observable events in the solar atmosphere (see $\S\ref{sec:introduction}$ for the plethora of observables that have attributed their underpinning physical mechanism to oscillatory reconnection). 

We have solved the compressible and resistive 2D MHD equations using the PLUTO code. We follow the work of \citet{McLaughlin2009}, but chose to focus on hot coronal plasma instead of cold plasma. Using an initially circular velocity pulse, we initiate a fast wave that reaches the X-point, perturbing it from its equilibrium and leading to the manifestation of oscillatory reconnection. Note that all other previous studies detailing the oscillatory reconnection mechanism, have utilized the LARE code \citep{2001JCoPh.171..151A}, and so this study represents the first time that oscillatory reconnection has been studied in isolation, in the absence of other effects or instabilities, using a different code, i.e. the PLUTO code. This is important since it shows that the mechanism is not some sort of artifact associated with one particular code.

The first aspect of our study focused on oscillatory reconnection in a 1\,MK hot coronal plasma. Focusing on the current density profiles at the null point, we have identified the oscillatory behavior of the process, which is connected to the periodic manifestation of horizontal and vertical current sheets. We have estimated the decay rate for the peak amplitudes of the current density, and through a wavelet analysis we have revealed a complicated spectrum for the oscillating signal. The complexity of the $J_z$ oscillating profile and spectrum is attributed to extensive mode conversion, taking place as the $v_{\perp}$ pulse approaches the layer where the Alfv\'{e}n speed $V_A$  equals the sound speed $V_S$, generating a fast and slow MHD wave in the region with high plasma-$\beta$. This phenomenon, which has already been investigated for various 2D and 3D null points \citep{McLaughlin2004,McLaughlin2005,McLaughlin2006a,McLaughlin2006b,Thurgood2012} leads to an additional periodic signal superimposed on top of the oscillatory reconnection signal. This layer has a larger radius for this setup of a hot plasma, resulting in more extensive mode conversion in this case (see also the Appendix \ref{Appendix} for a direct comparison of $J_z$ profiles for setups with different equilibrium temperatures). 

The second aspect of our study investigated the effect of adding anisotropic thermal conduction to our system. We found that adding anisotropic thermal conduction to our model did not prevent the development of oscillatory reconnection but led to a faster decay of the oscillation while also simplifying the spectrum of the profile. We also observed a \lq{simpler}\rq{} oscillatory profile than that of the case without thermal conduction; in particular, we report one dominant period (from the oscillatory reconnection) as opposed to two (in the absence of thermal conduction), and thus the effect of thermal conduction is twofold: it removes the significance of the generated mode-converted perturbations (since thermal conduction leads to a more effective energy dissipation from these generated perturbations), and, secondly, thermal conduction leads to a faster decay of the oscillation. This second point is intuitive when considering the physical driving and restoring forces at work in oscillatory reconnection: the thermal-pressure gradients (primarily from the reconnection jets) act to sustain/prolong the oscillation, whereas the magnetic forces act to damp the oscillation and restore the system to equilibrium. Thus, the addition of thermal conduction weakens the thermal-pressure gradients (the heat is transported away from the region of interest via the $\kappa_\parallel$), and consequently the system loses one of the forces sustaining the oscillation. Thus, the system is restored to equilibrium at a faster rate, i.e. the oscillation decays at a faster rate with the addition of thermal conduction.

For the last part of this work, we performed a parameter study of hot coronal plasma with anisotropic thermal conduction present. We have found a linear relation between the magnetic field strength in the vicinity of an X-point and the decay rate of the oscillating $J_z$ current density profile (one of the key signatures of oscillatory reconnection). This result is intuitive: magnetic forces act to restore the perturbed X-point to its equilibrium, and so stronger magnetic fields will restore the system at a faster rate, i.e. an increase in equilibrium magnetic field corresponds to an increase in decay rate.  Note that, in our parameter study, we altered the initial wave amplitude to ensure we injected the same amount of kinetic energy in each setup.

We have also reported an inverse relationship between the equilibrium magnetic field and the period of oscillatory reconnection, with the respective periods ranging between $17.9$\,s (for a magnetic field of $3B_0 = 4.32G$) up to $49$\,s (for a magnetic field of $0.5B_0 = 0.72G$).  In other words, stronger equilibrium magnetic fields correspond to shorter periods of oscillation (or conversely, weaker magnetic fields oscillate for longer periods). These values of the period are expected to be dependent upon the intensity of magnetic reconnection and by extension the values of the magnetic Reynolds number. Further studies are needed in order to establish a more accurate quantitative relation between the oscillation period and the magnetic field in the context of coronal plasma.

Thus, we have shown that stronger magnetic fields (relative to weaker fields) have shorter periods of oscillation as well as stronger decay rates. This demonstration opens the tantalising possibility of utilizing oscillatory reconnection as a seismological tool, and this will be the focus of future study. However, we can speculate on what aspects such a tool may want to consider: in \citet{McLaughlin2012ApJ}, it was found that the period of the oscillatory reconnection was linked to the length of the initial collapsed current sheet, i.e. the longer/stronger the first current sheet (equivalent to the first 'horizontal' current sheet), then the longer the resulting period.{\footnote{Note that \citet{McLaughlin2012ApJ} also found a positive relationship between the magnetic field strength of a subphotospheric, initially buoyant flux tube and the resultant period, and thus at first sight this result appears to disagree with our findings here. However, \citet{McLaughlin2012ApJ} found that the stronger the magnetic field strength of the emerging flux tube, the longer the initial current sheet formed (i.e. formed when the flux emerged into a preexisting magnetic environment) and those longer initial current sheets corresponded to longer periods. In other words, \citet{McLaughlin2012ApJ} found that it is the length of the initial current sheet that is important to understand resultant periods (and that, in their system, stronger flux emergence corresponded to generating longer initial current sheets).}} We may be seeing the same phenomena here: weaker magnetic fields form longer/stronger first current sheets and consequently have longer periods of relaxation. Conversely, strong magnetic fields form shorter first current sheets and thus have shorter periods of relaxation.  Both of these results, i.e. stronger magnetic fields correspond to shorter periods of oscillation as well as stronger decay rates, can be understood by considering the \lq{stiffness}\rq{} of the magnetic field: stronger fields are more resistant to deformation and, once deformed, relax back to equilibrium at a faster rate. Again, this requires further study.

With this paper, we aim toward a better understanding of nonlinear MHD wave behavior in inhomogeneous plasma and especially in the context of the solar atmosphere. Oscillatory reconnection, in particular, is a type of fundamental time-dependent reconnection mechanism with possible applications in several observed phenomena, including, but not limited, to QPPs (see $\S\ref{sec:introduction}$ for several examples). To further our understanding, we need to focus on the effects of different initial conditions, such as the density distribution, temperature, and the magnetic field strength of this oscillatory process. Expanding the hot plasma approach to 3D null points will provide additional insight into the wave$-$null point interaction. Finally, forward modeling of our results, especially for more complex magnetic field configurations, will aid us in combining these effects with the overarching aim of creating an oscillatory-reconnection-based seismological tool.

%%%%%%%%%%%%%%%%%%%%%%%%%%%%%%%%%%%%%%%%%%%%%%

\begin{acknowledgments}
All authors acknowledge UK Science and Technology Facilities Council (STFC) support from grant ST/T000384/1. K.K. also acknowledges support by an FWO (Fonds voor Wetenschappelijk Onderzoek – Vlaanderen) postdoctoral fellowship (1273221N). This work used the Oswald High Performance Computing facility operated by Northumbria University (UK). 
\end{acknowledgments}

%%%%%%%%%%%%%%%%%%%%%%%%%%%%%%%%%%%%%%%%%%%%%%%%

\appendix
\section{Equilibrium Temperature Parameter study} \label{Appendix}
In the main part of the paper, we have presented our analysis on the evolution of oscillatory reconnection for a hot coronal plasma. However, in order to better understand the  results, we need a more direct comparison with the cold plasma case. To that end, we present here a parameter study that will investigate the differences of oscillatory reconnection for different initial equilibrium temperatures. We will consider a basic setup similar to the one described in $\S\ref{sec:setup}$. For this study, we will use ideal MHD in the presence of numerical resistivity, without the use of anisotropic thermal conduction. We will be using the same Runge-Kutta method for the time step, the same solver (TVDLF), and the Constrained Transport method for keeping the solenoidal constraint. Since this is a parameter study and we want to benefit from faster computations, we use a  resolution of $1801^2$ grid points. 

Another difference from the (higher-resolution) base setup that we have focused on, is the use of different spatial integration schemes. For  all the other cases previously considered, we had included the fifth-order monotonicity preserving scheme (MP5) for the spatial integration, which provides the highest-accuracy results for a given resolution from all the methods provided with the PLUTO code. However, this scheme is unstable for simulations with cold plasma, in the context of the current setup. Therefore, for this parameter study, we have used a slightly more stable, and currently the second most accurate method provided by the PLUTO code: the parabolic reconstruction method.

\begin{figure}[t]
    \centering
    \includegraphics[trim={0.cm 0.cm 0.cm 0.cm}, clip,scale=0.5]{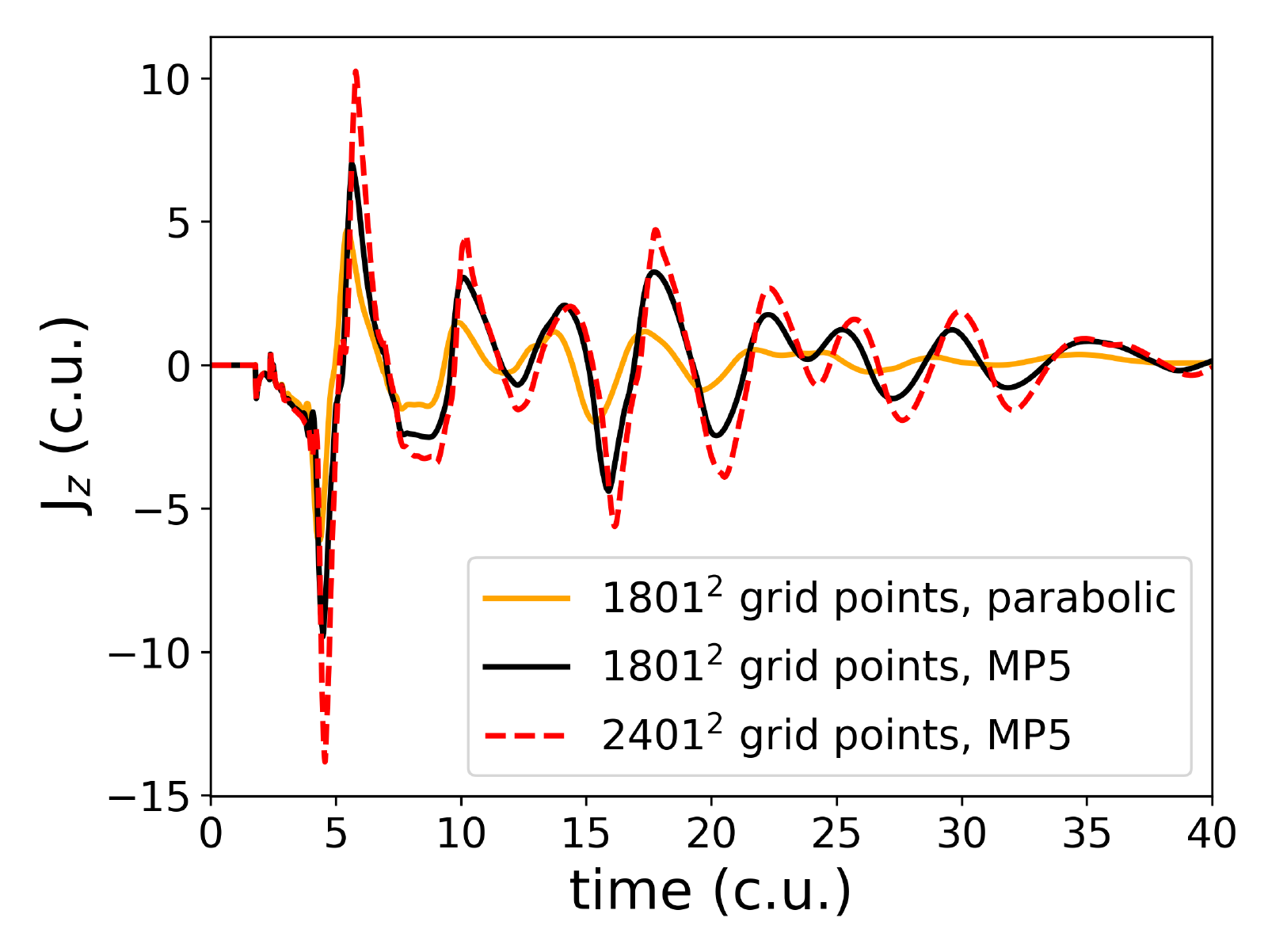}
    \includegraphics[trim={0.cm 0.cm 0.cm 0.cm}, clip,scale=0.5]{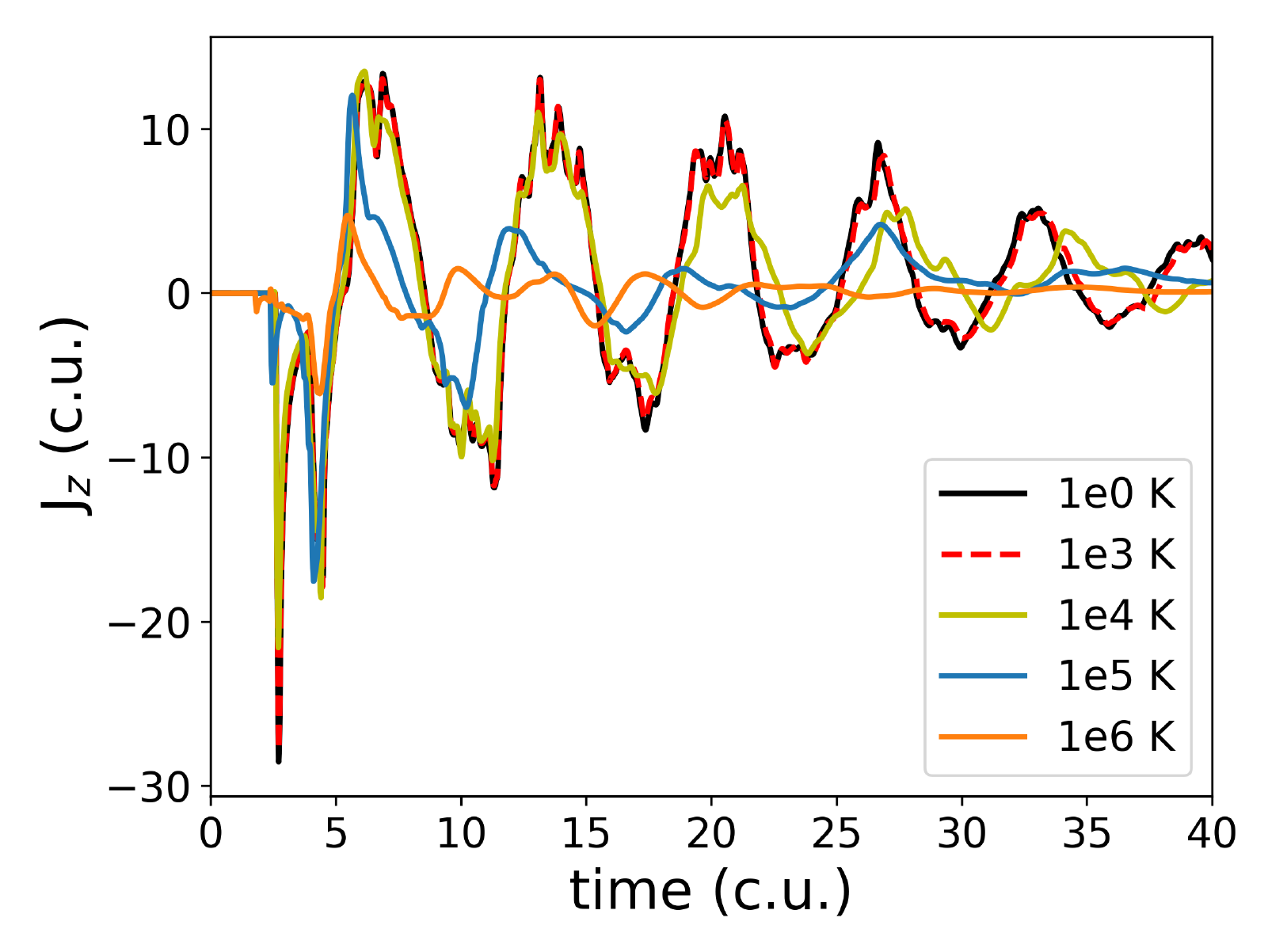}
    \caption{Left panel: a comparison plot of different setups with 1 MK base temperature without thermal conduction. Two of the profiles are for the lower-resolution setups but for different spatial reconstruction methods (MP5 and Parabolic), and the third is for the high resolution setup with the MP5 spatial reconstruction method. This plot connects the results of the temperature parameter study to the high-resolution base setup on which we focused in this study. Right panel: plot of the $J_z$ current density profile for all the setups with the different base temperatures that were used in this parameter study. For these setups, we have not used thermal conduction.}
    \label{fig:lowresparamcomp}
\end{figure}

Our first step is to compare the results from the two different reconstruction methods for a known solution. In the left panel of Fig. \ref{fig:lowresparamcomp} we take the high-resolution run without thermal conduction, which was studied in $\S\ref{sec:OR in 1MK plasma}$, and we compare it with its equivalent setup at the decreased resolution. Then both of these are compared to the decreased resolution setup with the parabolic reconstruction method. From Fig. \ref{fig:lowresparamcomp}, we can see that comparison of the two profiles for the MP5 method are very similar in frequency and overall behavior. In fact, just like in Fig. \ref{fig:rescompare}, the only difference between the two (red dashed line and black solid line) are the amplitudes of the $J_z$ current density. When we then compare the two lower-resolution runs (black solid line and orange solid line), we again see a similar oscillatory behavior and periodicity. The MP5 method yields stronger values for the peaks of the current density at the same resolution, which also better resolves the oscillation at the later part of the simulation. However, the accuracy of the parabolic reconstruction is adequate in order to employ it for the parameter study.

In the right panel of Fig. \ref{fig:lowresparamcomp}, we see a combined plot of the $J_z$ profiles of all the cases. We consider setups with base temperatures of $1$, $10^3$, $10^4$, $10^5$ and $10^6$\,K. In all these cases, the same magnetic field, density, and initial velocity pulse are considered. The first obvious result is that, in all cases, we have oscillatory reconnection manifesting. Each setup exhibits a different initial perturbation amplitude, for the same initial condition from Eq. \ref{eq:vp} and \ref{eq:vl}, with the hotter setups showing smaller overall amplitudes and relatively smaller oscillation periods. This seems counterintuitive, since in \citet{Thurgood2019A&A} it was found that, for the nonlinear regime, stronger initial perturbations would lead to smaller oscillation periods. However, that study did not take into account differences that may occur from varying initial conditions, as in the temperature in our cases. 

\begin{figure*}[t]
    \centering  
    \includegraphics[trim={0.cm 0.cm 0.cm 0.cm},clip,scale=0.5]{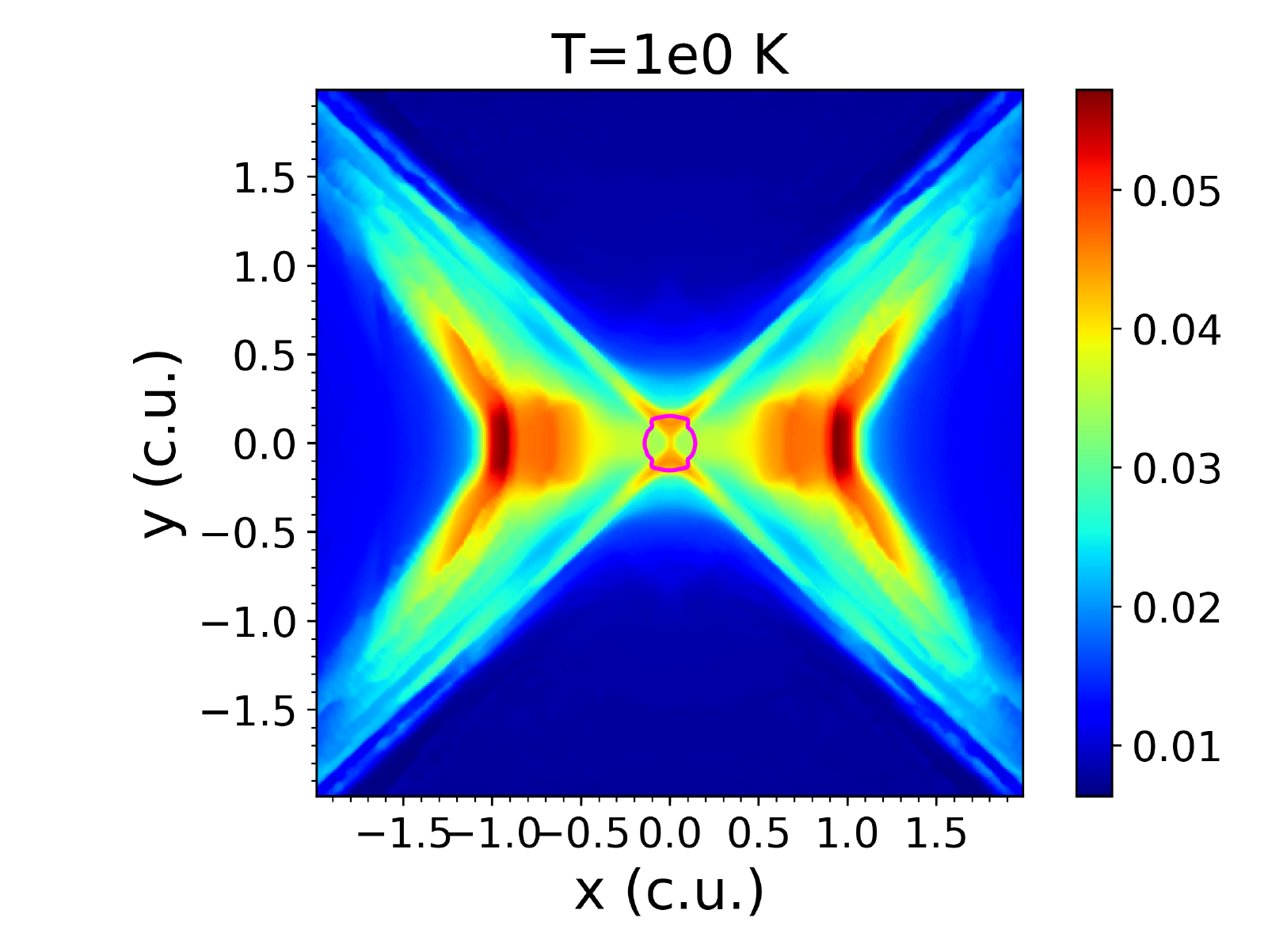}
    \includegraphics[trim={0.cm 0.cm 0.cm 0.cm},clip,scale=0.5]{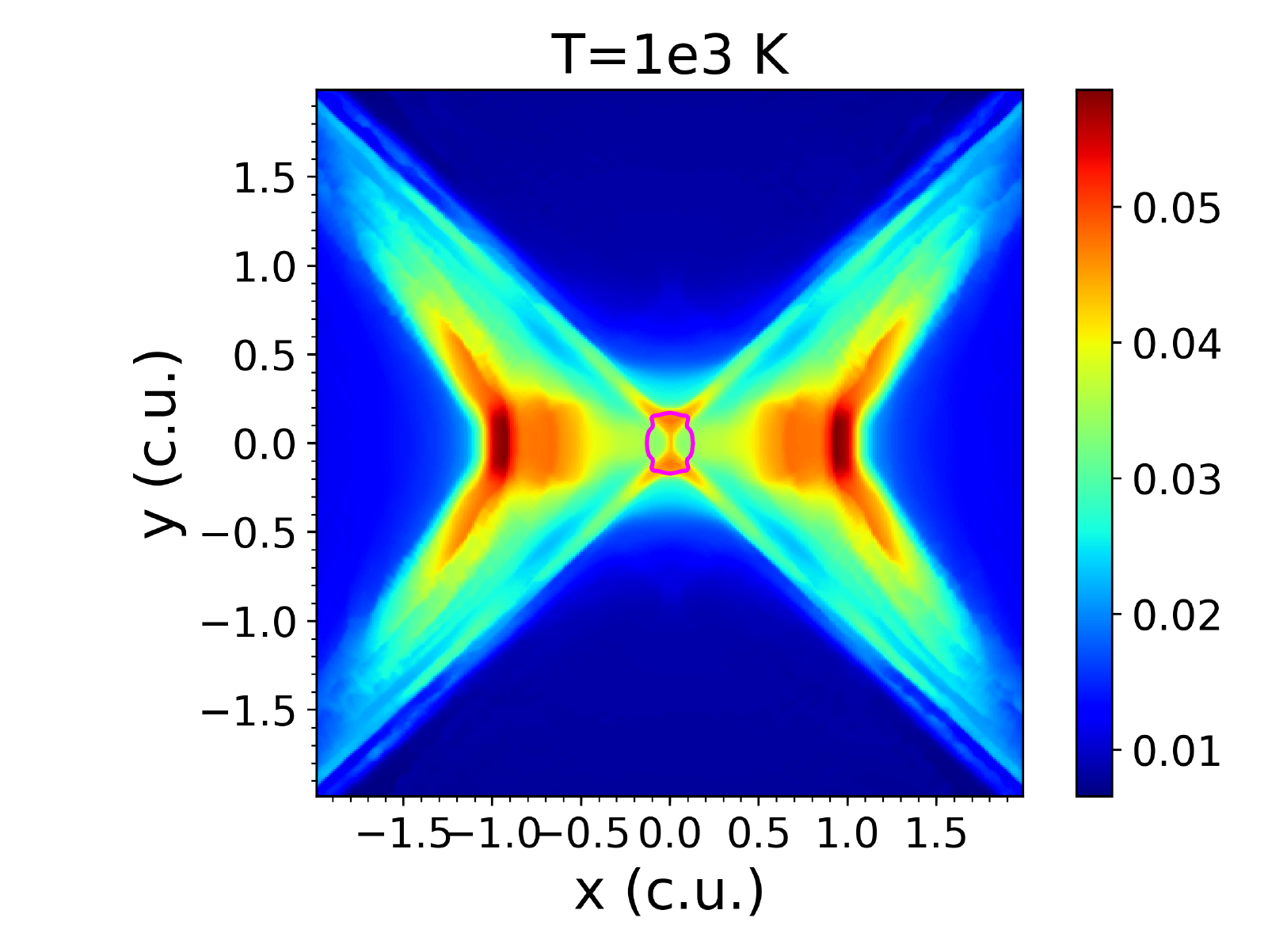}
    \includegraphics[trim={0.cm 0.cm 0.cm 0.cm},clip,scale=0.5]{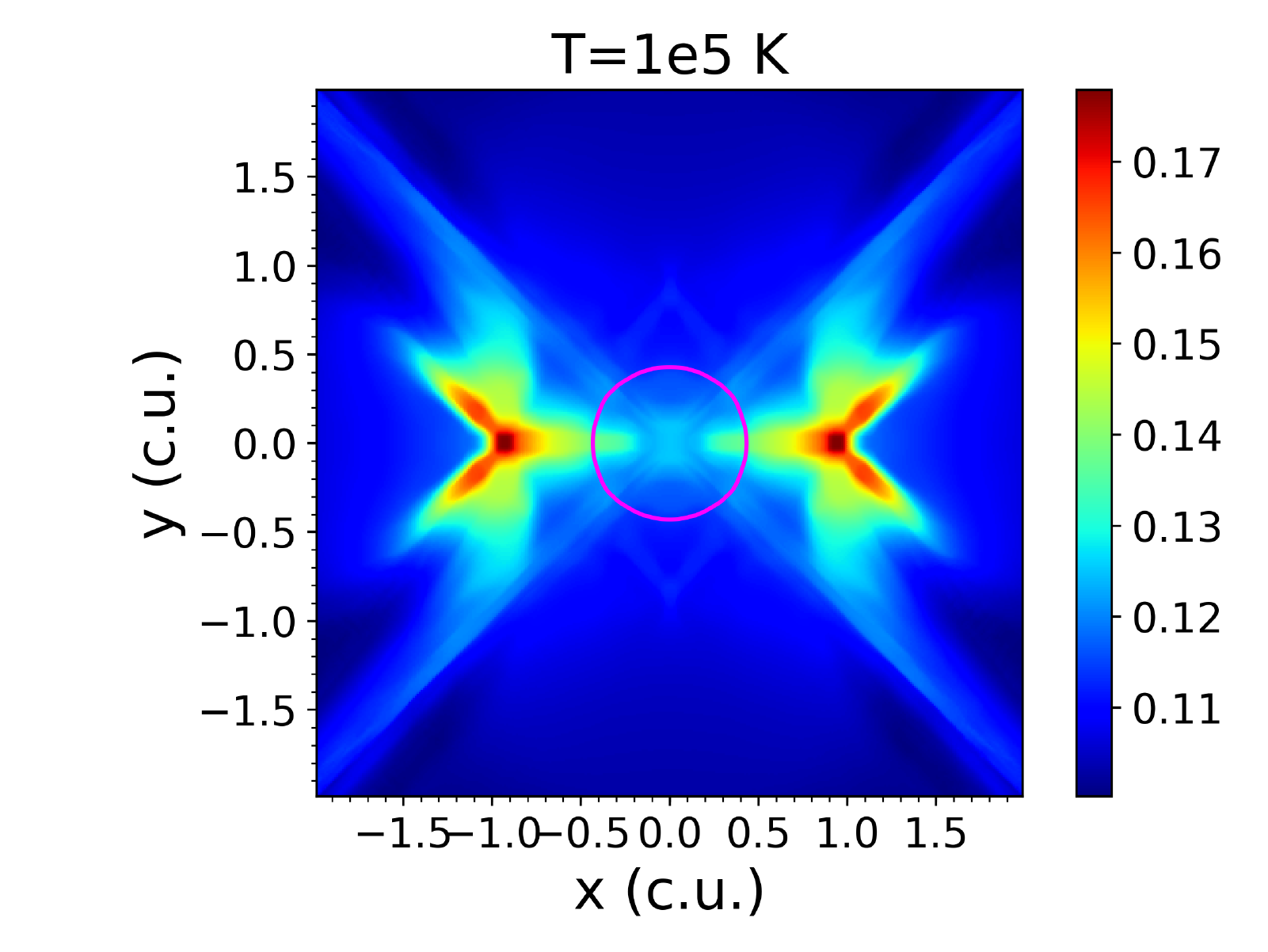}
    \includegraphics[trim={0.cm 0.cm 0.cm 0.cm},clip,scale=0.5]{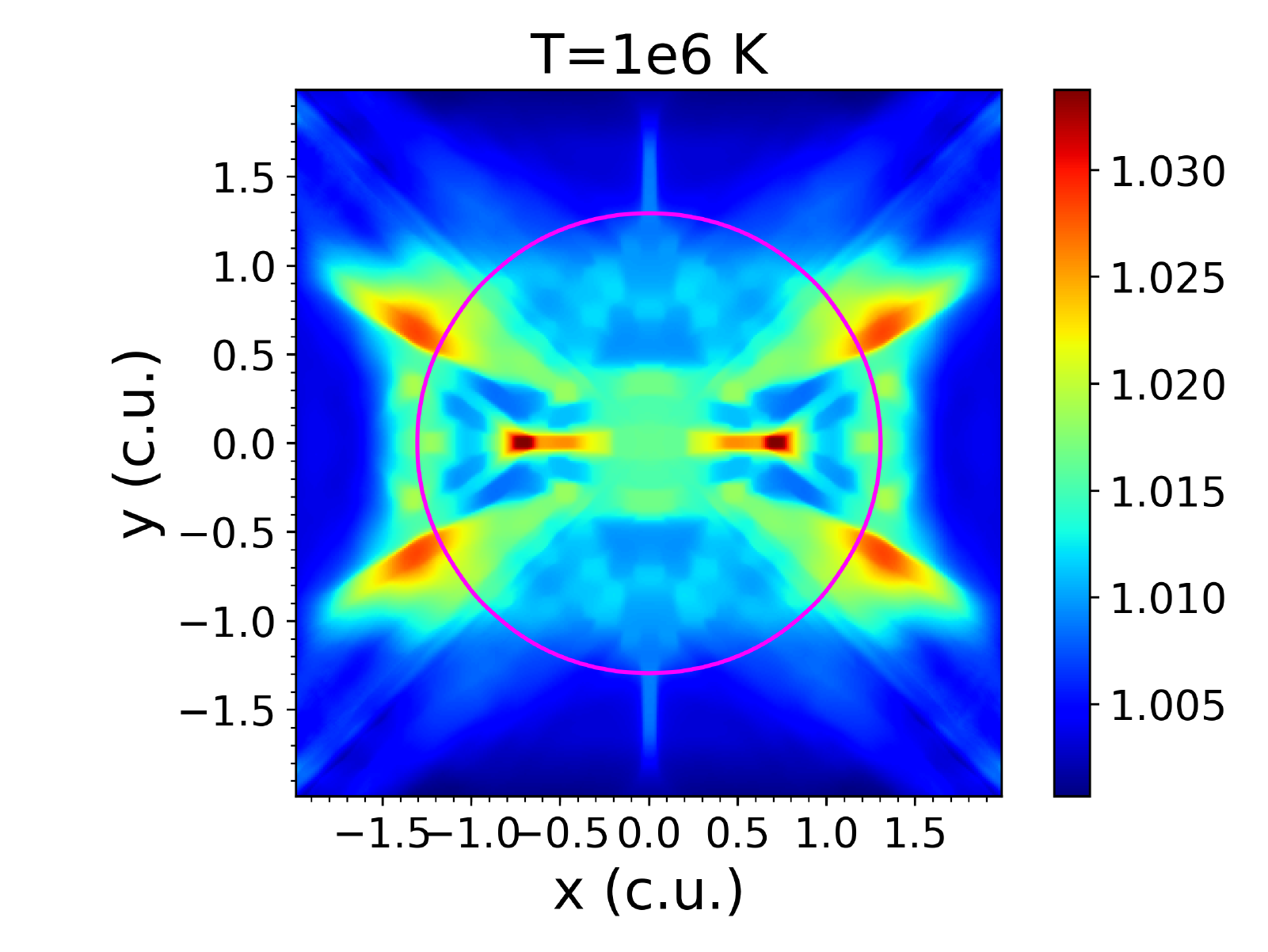}
    \caption{Temperature profiles for four setups with different initial base temperatures (at $1$\,K, $10^4$\,K, $10^5$\,K, and $10^6$\,K). The profiles are showing snapshots at $t=40$, in code units. The equipartition layer for each profile is also plotted. Note that the color scale is different for each profile.}
    \label{fig:lowresparamtempeprofiles}
\end{figure*}

We also see that the lower temperature runs (with $1$ and $10^3$\,K) exhibit an almost identical evolution for the oscillation of the current density profiles. To understand why, one needs to see the temperature profile panels of Fig. \ref{fig:lowresparamtempeprofiles}. These profiles show the temperature near the X-point at time $t=40t_0$. Due to the very low base temperature of the first two profiles ($1$ and $10^3$\,K), the heating from the oscillatory reconnection heats up the plasma near the null point to practically the same temperature. In fact, the only notable difference between the two profiles shown here is the shape of the equipartition layer. 

Once we start increasing the base temperature of our setup, for the same base magnetic field, we start seeing some of the aforementioned additional, superimposed periodic signals in the oscillatory profiles. As we have explained (\S\ref{sec:OR in 1MK plasma}), these  additional periodic signals can be attributed to the more extensive mode conversion taking place near the equipartition layer and inside the area that is included. Indeed, the more prominent  additional periodic signals are seen for the two hotter setups, which also show the equipartition layers with the largest radii. Thus, we can now compare the results of our main study with those for cold plasma and obtain a better understanding of our previous findings.

\bibliography{paper}{}
\bibliographystyle{aasjournal}

\end{document}